\documentclass[aps,prm,floatfix,notitlepage,preprint,showpacs,superscriptaddress,longbibliography]{revtex4-2}
\usepackage{mathtools,amssymb,graphicx,units}\usepackage[plainpages=false,pdfpagelabels,colorlinks=true,linkcolor=red,urlcolor=blue,citecolor=blue,pdftitle={Title},pdfauthor={},pdfdisplaydoctitle=true,pdfduplex=DuplexFlipLongEdge]{hyperref}
\usepackage{soul} 
\usepackage[normalem]{ulem}
\usepackage{xcolor}
\usepackage{bbold, verbatim} 
\usepackage{float}
\usepackage[font=small,justification=justified]{subcaption} 
\emergencystretch=\maxdimen
\usepackage[colorinlistoftodos,prependcaption,textsize=tiny]{todonotes}
\usepackage{cprotect}
\usepackage{ulem}
\usepackage{siunitx}
\usepackage{xr}
\graphicspath{ {./images/} }

\newcommand{\red}[1]{\textcolor{black}{\rm{#1}}}
\newcommand{\blue}[1]{\textcolor{black}{\rm{#1}}}
\newcommand{\rred}[1]{\textcolor{black}{\rm{#1}}}

\usepackage{etoolbox}

\makeatletter
\patchcmd{\frontmatter@abstract@produce}
  {\vskip200\p@\@plus1fil
   \penalty-200\relax
   \vskip-200\p@\@plus-1fil}
  {}
  {}
  {}
\makeatother

\begin{document}
\title{Topological data analysis for revealing structural origin of density anomalies in silica glass}

\author{Andrea Tirelli} 
\affiliation{International School for Advanced Studies (SISSA), Via Bonomea 265, 34136 Trieste, Italy}

\author{Kousuke Nakano} 
\email{kousuke\_1123@icloud.com}
\affiliation{International School for Advanced Studies (SISSA), Via Bonomea 265, 34136 Trieste, Italy}
\affiliation{School of Information Science, JAIST, Asahidai 1-1, Nomi, Ishikawa 923-1292, Japan}

\begin{abstract} 
Topological data analysis (TDA) is a new emerging and powerful tool to understand the medium range structure ordering of multi-scale data. This study investigates the density anomalies observed {\rred{during cooling of liquid silica}} from topological point of view using TDA. The density of {\rred{liquid}} silica does not monotonically increase during cooling; it instead shows a maximum and minimum. Despite tremendous efforts, the structural origin of these density anomalies is not clearly understood.
%
%
Our approach reveals that the one-dimensional topology of the -Si-Si- network changes at the temperatures at which the maximum and minimum densities are observed {\rred{in our MD simulations}}, while those of the -O-O- and -Si-O- networks change at lower temperatures. These results are also supported by conventional ring analysis.
%
%
Our work demonstrates the value of new topological techniques in understanding the transitions in glassy materials and sheds light on the characterization of glass--liquid transitions.

\end{abstract}
\date{\today}
\maketitle


\section{\label{sec:intro}Introduction}
\vspace{2mm}
The characterization of the transitions that occur in glassy materials is one of the most challenging and long-standing problems in materials science and solid-state physics~{\cite{1995AND}}.
Glasses have been practically used for various purposes before the Common Era. However, questions such as ``What is the difference between a regular fluid--solid transition and fluid--glass transition?" are still under debate.
The \textit{density anomaly} in tetrahedral liquids, such as water and {\rred{silica}}~{\cite{2019TAN}}, is a widely investigated transition. In this work, we focus on the density anomalies {\rred{observed during cooling of liquid silica}} because structural understanding of silica glass is lacking even though it is one of the most fundamental glasses. 
It has been experimentally observed that the density of {\rred{liquid silica}} does not monotonically increase during cooling. Instead, it shows a maximum and minimum at two temperatures~{\cite{1960BAC,1970BRU,2003KAK}}.
%
%
In the past few decades, the structural origin of {\rred{such}} anomalies {\rred{observed in silica}}, {\rred{and in}} ``tetrahedral liquids" more in general, have been intensively investigated through experiments ~{\cite{2003KAK, 2013SKI}} and simulations~{\cite{2001YAM,2011SOU,2019SHI,2021TAN}}. Computational studies have played an important role in understanding the density anomalies from an atomistic point of view. Researchers have attempted to characterize the density anomalies in tetrahedral liquids using local and global descriptors, such as tetrahedral ordering and pair translation ordering parameters~{\cite{1998LUI,2000TAN,2001JEF,2002TAN,2002JEF,2006JEF,2006VAL,2006RUC,2007AGA,2007AGA2,2008RUC,2009MAN,2009AGA,2010SHA,2011HUJ,2011AGA}}. 
{\rred{
Specifically for silica, the medium-range structural ordering of the silica network is regarded as the origin of its density anomalies. For instance, Soules et al.~{\cite{2011SOU}} speculated the stronger vibrations of O atoms located at the corner of silica tetrahedra, which collapse the silica network to higher density amorphous structure, are the origin of the density anomalies.
Yamahara et al.~{\cite{2001YAM}} claimed that the density anomalies of the silica melt are caused by two opposing factors in the density variation with decreasing temperature: densification due to the increase in number of bridging bonds and opening of the tetrahedral network. 
Shin et al.~{\cite{2020SHI}} found a similar behavior of the ring statistics. They showed that the population of 6-membered rings is the most dominant during the cooling process of silica liquid, which compensates for the regular volume shrinkage in the cooling process.
Skinner et al.~{\cite{2013SKI}} revealed that the density maximum corresponds to the first sharp diffusion peak (FSDP) height, and thus, they claimed that the density maximum in liquid silica is correlated with the onset of slightly reduced intermediate range order and coherence between the rings and cage structures.
Sen et al.~{\cite{2004SEN}} proposed that the system could be considered as a low density amorphous (LDA) phase at temperatures below the density minimum, while the liquid forms a high density liquid (HDL) phase above the density maximum. They claimed that the anomalous density behavior between extrema can be interpreted as a smearing of the HDL $\leftrightarrow$ LDA transitions. However, only subtle changes were observed in the temperature dependent X-ray diffraction measurement performed by Skinner et al.~{\cite{2013SKI}}, which implies that the HDL and LDA structures would be very similar even if such a transition occurs.
}}
New descriptors are also being investigated to better understand the structural ordering of glassy materials~{\cite{2001VOL,2009TAK,2011MEG,2013GAL,2014RUS,2014JON,2016HIR,2016YIC,2016YIC2,2016PAT,2018TAK,2020VIC,2021FOF,2021TAN}}.
{\rred{Despite the tremendous efforts}}, the origin of density anomalies {\rred{observed in silica}} is still a matter of debate.

\vspace{2mm}
Topological data analysis (TDA)~{\cite{2018LAR, 2021CHA}} is an emerging and powerful tool for understanding the medium-range structure ordering of multiscale data. The possible applications of TDA range widely from cosmology~{\cite{2015CHE}} to condensed matter physics~{\cite{tirelli2021a,2022SUN}}. Furthermore, in recent years, topological concepts have played an important role in materials science{~\cite{2020TOR, 2021SCO, 2022ANA}} and chemical engineering~{\cite{2021ALE}}.
Persistence diagrams (PDs) are particularly important tools in TDA. PDs are sets of points in a two-dimensional plane that encode the topological information \blue{in arbitrary dimensions} of a certain point cloud.
One of the pioneering works on the application of TDA to glassy materials via PDs is that of Hiraoka et al.~{\cite{2016HIR}}. They have shown that PDs are qualitatively distinguishable between crystalline and glassy silica oxides~{\cite{2016HIR}}. Their results clearly suggest that PDs are \blue{fundamental} tools for extracting detailed geometric and topological information from amorphous structures. PDs have been \blue{used} not only for the simple silica glass but also for alkaline silicate glasses~{\cite{2019ONO, 2020SOR}}. One of the drawbacks of the raw topological features encoded in PDs is the difficulty in determining the {\it quantitative} differences among PDs; however, there has been progress in this area in recent years~{\cite{2016KUS,2020HIR}}. This issue can potentially be  problematic when studying two similar amorphous structures, for instance, glass--liquid or glass--glass transitions in amorphous materials.

\vspace{2mm}
In this study, we used a technique that was first reported in Ref.~{\onlinecite{tirelli2021a}} to investigate the structural origin of the density anomalies in {\rred{silica}}. This technique combined TDA with an unsupervised machine learning (ML) tool, namely, fuzzy spectral clustering~\cite{jimenez2008fuzzy, bezdek1984fcm}. 
To our best knowledge, this is the first application of TDA in combination with fuzzy spectral clustering for studying structural transitions of materials.
The unsupervised analysis of PDs enabled us to precisely quantify their differences. We performed molecular dynamics (MD) simulations with 5000 SiO$_2$ molecules (15000 atoms) and analysed the trajectories during a {\rred{liquid}} cooling process. \blue{Our workflow} was individually performed \blue{on} -Si-Si-, -O-O-, and -Si-O- networks. The -Si-Si- and -O-O- networks refer to those composed of only Si and O atoms, respectively, and the -Si-O- network refers to the entire {\rred{silica}} network. Our analysis revealed that in the cooling process of {\rred{liquid silica}}, the one-dimensional topology~{\footnote{{\it One-dimensional topology} refers to all the topological invariants that generate the first persistent homology group of the point clouds on which we perform TDA. Nontrivial (noncontractible) loops are an example of topological invariants at the first homology group level.}} of the -Si-Si- network changed at the temperatures of the density maximum and minimum, whereas those of the -O-O- and -Si-O- networks changed at lower temperatures.
We also \red{performed} conventional \red{ring analysis and confirmed that \blue{the corresponding results are} consistent with \blue{the findings generated with our approach}.}


\section{Methods}\label{sec:methods}

\subsection{Modelling of silica glass}\label{sec:modeling}
%
{\rred{In this study, the structure models of silica were}} obtained using classical MD simulations implemented in the LAMMPS package~{\cite{1995Plimpton}} with the Beest Kramer van Santen \red{(BKS)} interatomic potential~{\cite{1955VAN}}, which has been widely used to investigate the structure, dynamics, and thermodynamics of silica~{\cite{1996VOL,2015LAN}}. On the basis of previous studies, we {\rred{used}} cutoffs of 5.5 and 10.0 \AA\ for short-range and long-range Coulombic interactions, respectively~{\cite{1996VOL}}.
All simulations {\rred{were}} conducted in an isothermal--isobaric ensemble (NPT) using the Nose–-Hoover thermostat~{\cite{1989NOSE}} and barostat~{\cite{2006Tuckerman}}. 
The glass models {\rred{were}} obtained using the melt-quenching method, where the initial random silica glass structures {\rred{were}} generated using the {\textsc{Packmol}} package~{\cite{2009Martinez}} with a density of 2.20~g/cm$^3$ and then melted at 3500~K for 2.5~ns. The numbers of Si and O atoms in the simulation cell are 5,000 and 10,000, respectively. The {\rred{melted silica was}} cooled from 3500 K to 300 K at a rate of 1~K/ps, followed by equilibration at 300~K for 500~ps. 
Thereafter, they {\rred{were}} heated to 6100~K at a rate of 10~K/ps and then melted at 6100~K for 1.0~ns. 
Finally, the structures {\rred{were}} cooled from 6100~K to 300~K at a rate of 1~K/ps, where they {\rred{were}} equilibrated for 300 ps every 100~K \red{(Fig.~\ref{SI-temperature})}. 
The {\rred{structures were}} recorded every 1~ps, and the thermodynamic properties, such as density, {\rred{were}} recorded every 0.01 ps. Only the data from the last 150~ps {\rred{was}} used to average the properties. PDs {\rred{were}} computed using only the last configuration for each temperature, {\rred{while}} all {\rred{the}} other properties {\rred{were}} averaged over the recorded values.
In all cases, the equations of motion {\rred{were}} integrated with a time step of 1.0~fs.

\subsection{Structure analysis and Ring statistics}
\begin{figure*}[htbp]
  \centering
  \includegraphics[width=\columnwidth]{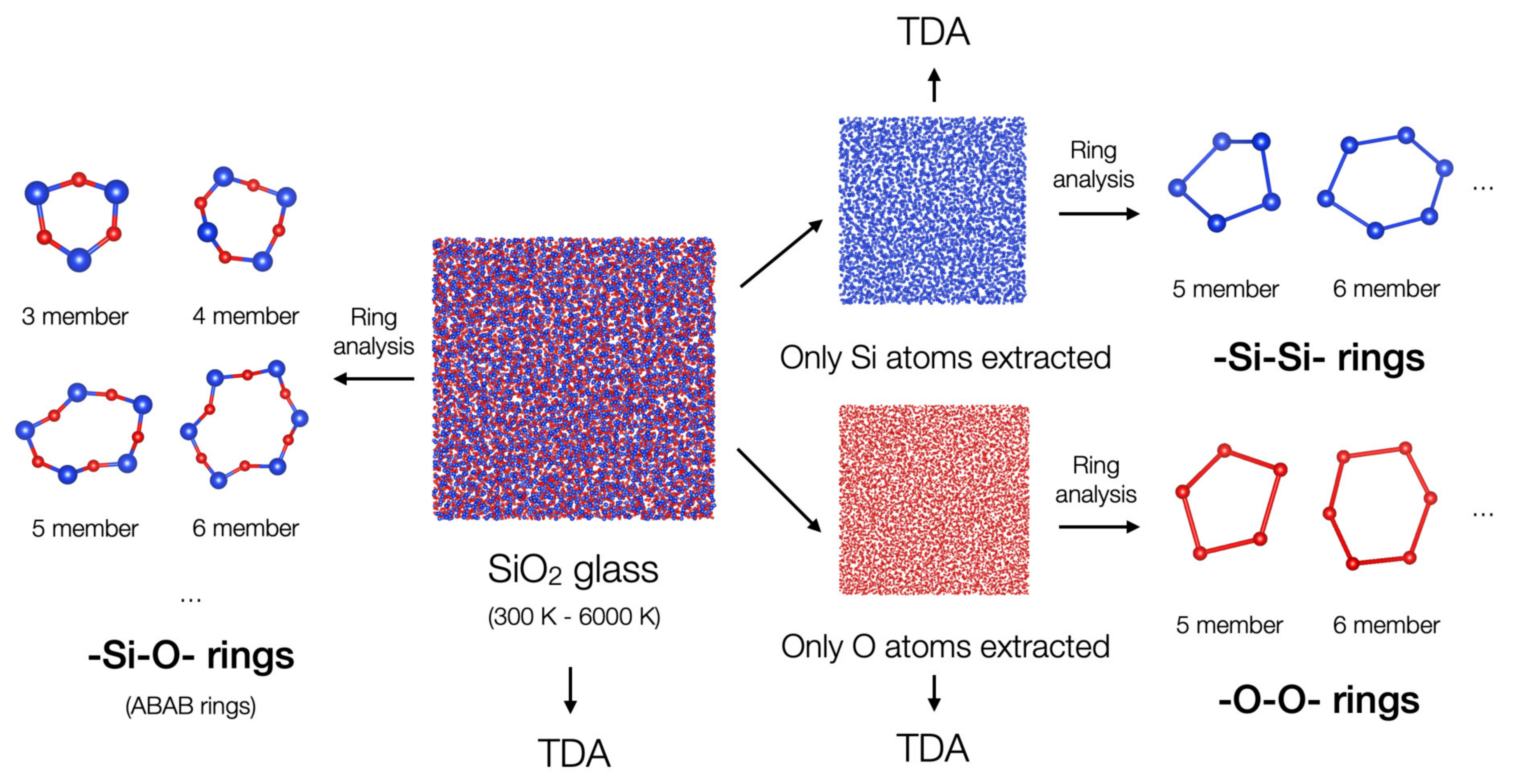}
  \caption{\red{A schematic figure of the TDA and ring statistics analysis procedures, and the obtained rings. Note that the -Si-O- ring size refers to the number of either Si or O atoms in a ring.}
  } 
  \label{schematic}
\end{figure*}

{\rred{The obtained silica structure was analyzed by}} the radial distribution functions (Fig.~\ref{SI-gr}), the first minimums of the radial distribution functions (Fig.~\ref{SI-gr-min}), coordination numbers (Figs.~\ref{SI-population-sio2},~\ref{SI-coord-num}), structure factors (Fig.~\ref{SI-sq}), bond-angle distributions (Figs.~\ref{SI-bad-O-Si-O}, \ref{SI-bad-Si-O-Si}, \ref{SI-bad-Si-Si-Si}, \ref{SI-bad-O-O-O}), and ring statistics. {{\rred They were}} computed using the R.I.N.G.S. code (ver.1.3.4)~{\cite{2010ROU}}. The cutoff radius for each atomic pair {\rred{was}} set as the first minimum of the radial distribution function at each temperature (Fig.~\ref{SI-gr-min}), except for the coordination number analysis and tetrahedrality analysis. The first (second) minimum $r$ in the $g(r)$ plots at 300~K were employed for the cutoffs in 1st nearest-neighbor (up to 2nd nearest-neighbor) coordination number analysis. Instead, all the 4 nearest neighbor O (Si) atoms were considered in the SiO$_4$ (SiSi$_4$) in the tetrahedrality analysis (Figs.~\ref{SI-tetra-SiO4},~\ref{SI-tetra-SiSi4}). We {\rred{computed}} the ring statistics according to King’s~{\cite{1967KIN}}, Guttman’s~{\cite{1990GUT}}, and the primitive~{\cite{1991GOE,2002YUA,2002WOO}} definitions. The differences between these definitions are discussed later.
{\rred{Ring analysis was performed on the entire silica network (-Si-O-) as well as the partial Si (-Si-Si-) and O (-O-O-) networks. As shown in Fig.~\ref{schematic}, the -Si-Si- or -O-O- rings were computed from the structures whereas only Si or O atoms were extracted from MD trajectories. The -Si-O- ring statistics was obtained by counting only ABAB rings~{\cite{2010ROU}}}}.
The population of ring sizes {\rred{was}} computed from the number of rings obtained for each size.

\subsection{Topological Data Analysis}\label{sec:tda}
In this section, we give an account on the computational technique used throughout this work, which combines techniques from topological data analysis (TDA) to {\rred{investigate}} the structures of {\rred{silica obtained by the MD simulations described in Sec. ~{\ref{sec:modeling}}}}.
The approach outlined here was originally devised in the works of the first author and collaborators in Refs.~\onlinecite{tirelli2021a, tirelli2021b}, to which we refer for a detailed account on the topics mentioned in this section. Computational algebraic topology was first applied to machine learning problems in the seminal work of Carlsson~\cite{Carlsson2009}. Since then, it has undergone significant development with widespread applications (e.g., time series analysis~\cite{umeda2017, gidea2018} and computer vision \cite{bernstein2020}). TDA has only recently been applied in material science. In Refs.~\onlinecite{tirelli2021a, tirelli2021b} the authors have used the topological techniques presented here to study the phase transitions of classical and quantum lattice models. 

\vspace{2mm}
The working hypothesis of TDA is that the elements of a point cloud are finite samples from an underlying manifold, whose geometric and topological properties are reflected by the structure of the point cloud. From this starting assumption, one could argue that the qualitative information about such manifolds may help in obtaining knowledge about data and reaching a precise and quantitative understanding on the overall organization of such data at multiple scales. Persistent homology, which is one of the main techniques in TDA, is a mathematical tool whose main purpose is to infer topological information of a data manifold from a finite set of discrete points sampled from it.

\subsubsection{Persistent Homology}\label{subsec:pers}
In what follows, $X$ will denote a point cloud belonging to a given metric space, \footnote{A metric space $X$ is a set endowed with distance function $d$, \textit{i.e.} a function $d:X\times X \rightarrow \mathbb{R}_{>0}$ satisfying certain properties}. Starting from $X$, for any given positive number $\varepsilon$, we can construct a covering of $X$, 
\[
C_{\varepsilon}(X) = \bigcup_{p\in X} B(p, \varepsilon),
\]
given by the union over all points, $p$, belonging to the point cloud of balls $B(p, \varepsilon):=\{y\ |\ d(y, x)\leq \varepsilon\}$. First, note that for $\varepsilon\leq \varepsilon'$, $C_{\varepsilon}(X) \subset C_{\varepsilon'}(X)$. Moreover, variations in the value of $\varepsilon$ imply modifications in  the topology of space $C_{\varepsilon}(X)$. For example, for a sufficiently small $\varepsilon$, the number of connected components (typically referred to as the 0-\textit{th} Betti number in topology) of $C_{\varepsilon}(X)$ is equal to the number of points in $X$. In contrast, for a sufficiently large $\varepsilon$, such a number is equal to 1 (each $B(p, \varepsilon)$ in $C_{\varepsilon}(X)$ has a non-empty intersection with $B(p', \varepsilon)$ for some $p' \in X$). Topological invariants can be summarised by the so-called {\textit Betti numbers}, one for each homological dimension. The 0-$th$ Betti number of topological space $Y$, which is denoted by $b_0(Y)$, is the first of a sequence of topological invariants associated with $Y$ (one for each positive integer, $i\in \mathbb{Z}$), where $b_i(Y)$ denotes the number of \textit{i-th dimensional holes} in $Y$. For example, $b_1(Y)$ is the number of (nontrivial) closed loops of $Y$. Hence, if $Y$ is a circle, then $b_1(Y)=1$.

\vspace{2mm}
The key point is to analyse modifications in topological invariants (such as connected components, loops, and $b_i(C_\varepsilon(X))$) as $\varepsilon$ varies. Specifically, we assign birth value $b$ and death value $d$ to each invariant so that the invariant appears for the first time in $C_b(X)$ and disappears in $C_d(X)$. Therefore, we can associate a pair of positive numbers, $(b, d)$, with each invariant, which is referred to as the \textit{persistence pair} of the invariant. In general, we can associate a \textit{persistence diagram}, $\mathcal{D}(\mathcal{I})$, with a set of topological invariants, $\mathcal{I}$,  by combining all the persistence pairs arising from the elements of $\mathcal{I}$. 
\begin{equation}
	\mathcal{D}(\mathcal{I}) = \{(b_i, d_i) \in \mathbb{R}^2\ |\ i \in \mathcal{I} \},
\end{equation}
where $b_i$ and $d_i$ denote the birth and death of invariant $i$, respectively, for $i\in \mathcal{I}$. $\mathcal{D}(\mathcal{I})$ is simply denoted by $\mathcal{D}$ when $\mathcal{I}$ is fixed and clear from the context (this is always the case in this study, where we compare different point clouds, $X_T$, using persistence diagrams with a fixed topological invariant set, $\mathcal{I}$).

\vspace{2mm}
Therefore, $\mathcal{I}$ is used to build a map,
\begin{equation}\label{top_emb}
	X \longrightarrow \mathcal{D}_X,    
\end{equation}
by associating point cloud $X$ with its persistence diagram, $\mathcal{D}_X$. We refer to equation \,\eqref{top_emb} as a \textit{persistence embedding}. From a topological point of view, such a mapping is used to compare two different point clouds, $X_1$ and $X_2$. Let $\mathcal{PD}$ be the set of all persistence diagrams arising from $\mathcal{I}$. Assuming that we can define distance function $d$ on $\mathcal{PD}$, we can define a distance measure between $X_1$ and $X_2$ as follows: 
\begin{equation}\label{dist_point_cloud}
	\tilde{d}(X_1, X_2) = d(D_{X_1}, D_{X_2}).
\end{equation}
It is clear from equation\eqref{dist_point_cloud} that the comparison of $X_1$ and $X_2$ strongly depends on the selection of $d$ between persistence diagrams. \blue{Among the various possible definitions for $d$}, we mention \blue{the Wasserstein, Bottleneck and Betti distances (notice that, in the formulas below, $\Delta$ indicates the diagional in $\mathbb{R}^2$, \title{i.e.} $\Delta = \{(x, y)\in \mathbb{R}^2\ |\ x=y\}$):}
\begin{itemize}
	\item the \textit{p-Wasserstein distance}~\cite{kerber2017}: Given persistence diagrams $\mathcal{D}_1$ and $\mathcal{D}_2$, it is defined as the infimum over all bijections, $\gamma: \mathcal{D}_1 \cup \Delta \rightarrow \mathcal{D}_2 \cup \Delta$, of 
	\[
	\left( \sum_{x \in \mathcal{D}_1\cup \Delta} || x - \gamma(x)||_{\infty}^p \right)^{1/p},
	\]
	where $||\cdot ||_{\infty}$ is the standard $\infty$-norm on $\mathbb{R}^2$.
	\item the \textit{Bottleneck distance}: This can be obtained from the $p$-Wasserstein distance by taking the limit $p\rightarrow\infty$. It is the infimum over the same set of bijections of the value given by
	\[
	\sup_{x \in \mathcal{D}_1\cup \Delta} || x - \gamma(x)||_{\infty}
	\]
	\item the \textit{Betti distance}: Given persistence diagram $\mathcal{D}$, its \textit{Betti curve} is defined as a function, $\beta_{\mathcal{D}}: \mathbb{R}\rightarrow \mathbb{N}$. The value of the function at $s\in\mathbb{R}$ is the number (counted with multiplicity) of points, $(b_i, d_i)$, in $\mathcal{D}$ such that $b_i\leq s \leq d_i$. The Betti distance between two persistence diagrams is defined as the $L^p$ distance between the Betti curves, $\beta_{\mathcal{D}_1}$ and $\beta_{\mathcal{D}_2}$.
\end{itemize} The set, $\mathcal{PD}$, with any of the distances defined above is a metric space. \red{ The Betti distance was employed in this study.}

\subsubsection{Fuzzy Spectral Clustering}
Given $n$ point clouds, $X_1, \dots, X_n$ (in this work, $X_i$  is an instantaneous configuration of the {\rred{silica}} structure at a given temperature obtained by performing a molecular dynamics (MD) simulation, \red{(Fig.~\ref{schematic})}), one can construct square matrix $M_X$ with dimension $n$, $M_X=(m_{ij})$, by letting
\begin{equation}\label{eq:dist_matrix}
	m_{ij} = \tilde{d}(X_i, X_j).
\end{equation}
We call $M_X$ the \textit{distance matrix} of the set of point clouds $X_i$, for $i=1, \dots, n$. The way in which we interpret the information content of $M_X$ is the following: $\tilde{d}(X_i, X_j)$ decreases as the similarity --in terms of the topological structure-- between $X_i$ and $X_j$ increases. As a consequence, one possibility is to employ \textit{clustering algorithms} that work on similarity matrices such as $M_X$ in order to group the point clouds that share the same topological features and separate those with an inherently different topology. In order to do so, we associate to $M_X$ ad edge labelled graph $G=(V,E)$ as follows: $V=\{1, \dots, n\}$, $E=V\times V$, where $m_{ij}$ is the label on edge $(i, j)$. In this manner, the problem of clustering point clouds on the basis of their distance is converted into that of identifying the communities of nodes in a graph according to their connecting edges.

\vspace{2mm}
A standard way to achieve this through the use of \textit{Spectral clustering} algorithms. These procedures exploit the information obtained from the eigenvalues (spectrum) of special matrices --the laplacian-- built from a graph or dataset. Spectral clustering algorithms, in both their theoretical and practical aspects, are reviewed in Refs.~\onlinecite{von2007, liu2018}. The important point is that the input of spectral clustering is a \textit{similarity matrix}, $S=(s_{ij})$, where $s_{ij}\in \left[0,1\right]$ with $s_{ii}=1$, i.e., the self-similarity of a point is the maximum value of $s_{ij}$. The distance matrix $M$ (e.g., generated using the above mentioned method) can be transformed into a similarity matrix by applying the following Gaussian kernel transformation to each entry of the matrix: 
\[
k(x) = e^{-\frac{x^2}{2\sigma^2}},
\]
where $\sigma$ is a hyperparameter that governs the spread of the Gaussian distribution. In this study, we use a relaxed version of spectral clustering, referred to as \textit{fuzzy spectral clustering}, which is a combination of the standard spectral clustering algorithm with the fuzzy k-means algorithm~\cite{jimenez2008fuzzy, bezdek1984fcm}\footnote{This is achieved as follows: $k$-means is used as an intermediate step in spectral clustering; therefore, we can obtain fuzzy spectral clustering by using fuzzy $k$-means in place of the $k$-means procedure in the original formulation of spectral clustering}. \blue{The choice of employing the fuzzy version of spectral clsutering is motivated by the specific use case at hand: indeed, not only do we want to locate where the transition point but we also intend to study the nature of the transition, \title{i.e.} investigating whether the anomaly point is crossed abruptly or gradually. This information cannot be obtained through standard clustering methods. }

\vspace{2mm}
The application of fuzzy spectral clustering to the kernel of matrix $M_X$ produces a membership degree function,
\[
l=(l_0, l_1): \{X_1, \dots, X_n\} \rightarrow [0, 1]^2, 
\]
such that $l_0(X_i)$ is the membership degree of the first cluster of point cloud $X_i$, $l_1(X_i)$ is the membership degree of the second cluster of point cloud $X_i$, and $l_0(X_i)+l_1(X_i)=1$ for all $i$. We identify the critical point by analysing the following sequence: 
\begin{equation}\label{eq:member_list} 
	\bar{l} = (l_0(X_1), \dots, l_0(X_n)). 
\end{equation}
For a data point, the obtained membership degrees indicate the distribution of the total membership over different clusters.  For instance, if $l_0(X_1)=1$ and  $l_1(X_1)=0$, point cloud $X_1$ perfectly belongs to the first cluster. In contrast, if $l_0(X_2)=0.5$ and  $l_1(X_2)=0.5$, $X_2$ shows the topological properties that characterize clusters 1 and 2 with equal weights. Given this, we can conclude that if two clusters of point clouds have extremely different topologies, then this difference is reflected in considerably sharp changes in the membership functions around the points of such topological modifications. In this case, the corresponding topological invariants are significantly different. Therefore, at large distances, the fuzzy spectral clustering algorithm applied to the obtained distance matrix provides membership values that are either extremely close to one or zero. 

\begin{figure}[htbp]
    \centering
    \includegraphics[scale=1]{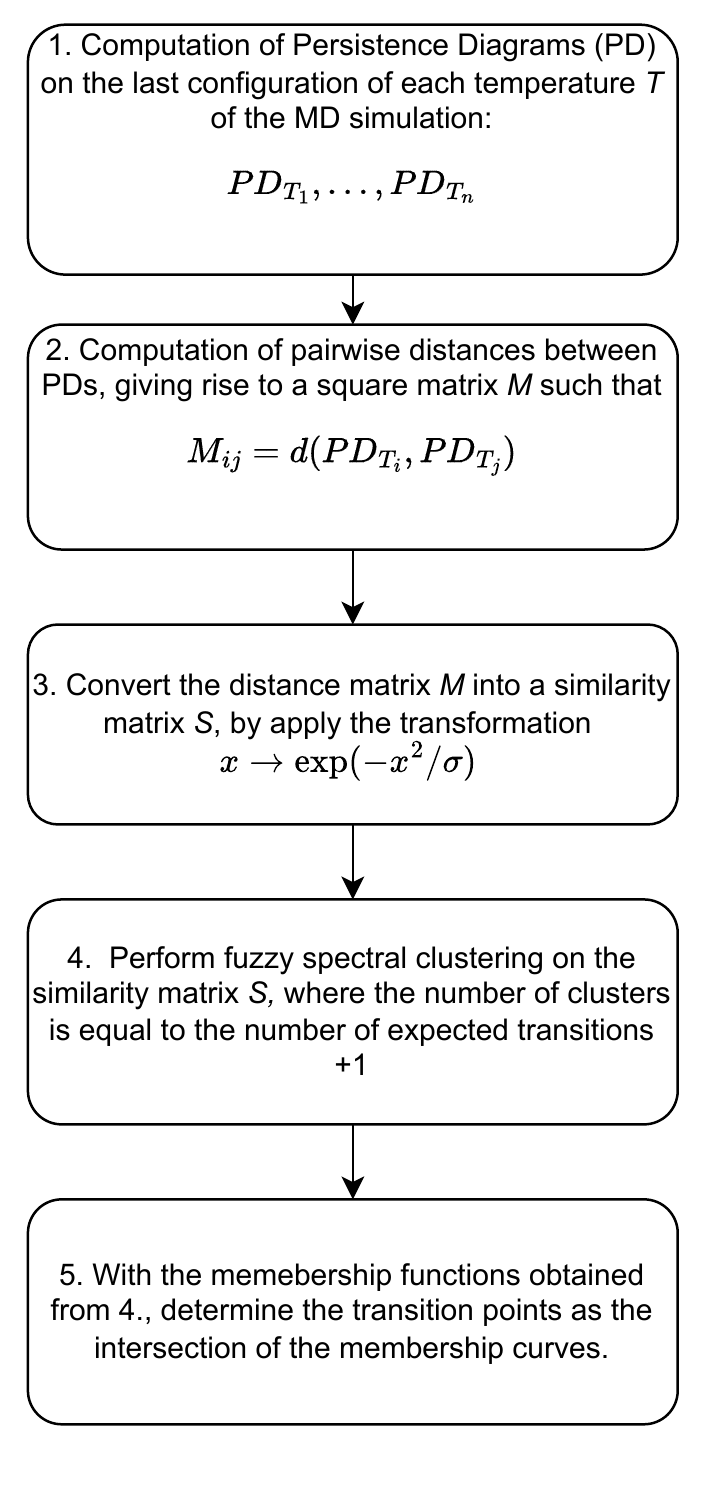}
    \caption{Diagram outlining the procedural steps of our computational workflow.}
    \label{workflow}
\end{figure}

\vspace{2mm}
{\rred{Based on the above theoretical approach, the computational workflow that used in this study is outlined in Fig.~\ref{workflow}. At the first step of the workflow, PDs were computed not only for all the point clouds containing Si and O, but also for the partial point clouds containing either Si or O, as shown in Fig.{\ref{schematic}}. The obtained PDs of the -Si-Si-, -O-O-, and -Si-O- networks at each temperature are shown in the supplementary information (Figs.~{\ref{SI-PD-Si}}--{\ref{SI-PD-SiO}}). 
}}

\begin{figure}[htbp]
  \centering
  \includegraphics[width=\columnwidth]{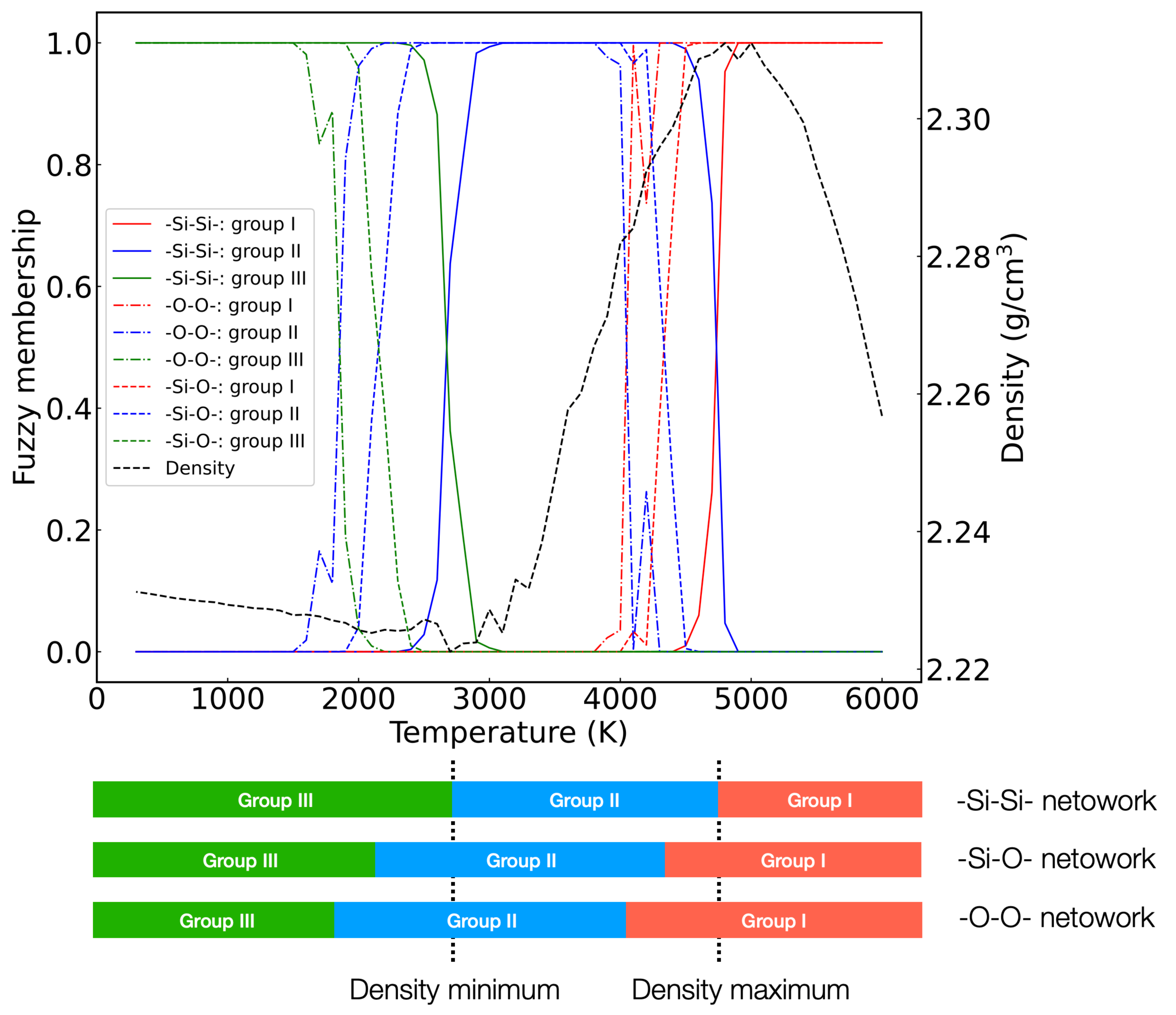}
  \caption{Upper panel: Fuzzy memberships of -Si-Si-, -O-O-, and -Si-O- networks of {\rred{silica}}. The PDs are divided into three different clusters (I, II, and III) using the fuzzy spectral clustering algorithm. The intersection points for groups I (red) and II (blue) are located at 4735~K, 4048~K, and 4334~K for -Si-Si-, -O-O-, and -Si-O- networks, respectively. The intersection points for groups II (blue) and III (green) are located at 2674~K, 1856~K, and 2153~K for -Si-Si-, -O-O-, and -Si-O- networks, respectively. The density of the {\rred{silica}} (15,000 atoms) is plotted on the right $y$ axis as a function of temperature. Lower panel: Visualization of the major groups of each network as a function of temperature, obtained from the TDA.}
  \label{fuzzy}
\end{figure}

\begin{figure*}[htbp]
  \centering
  \includegraphics[width=\columnwidth]{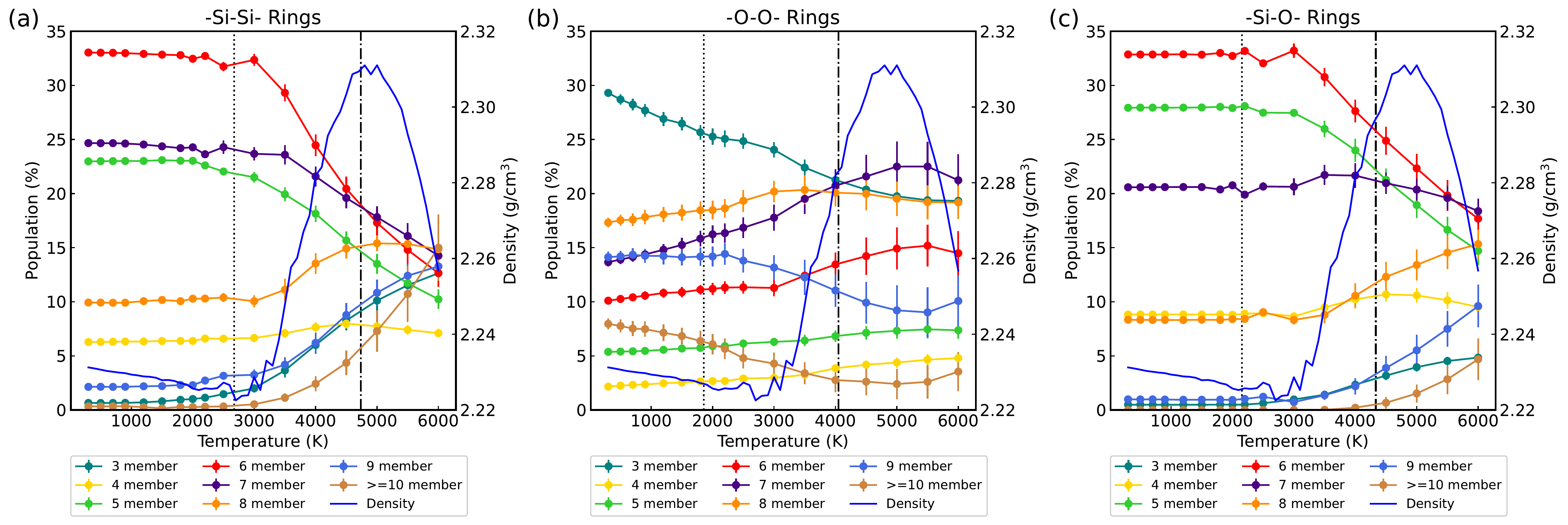}
  \caption{Ring statistics of (a) -Si-Si-, (b) -O-O-, and (c) -Si-O- networks. The vertical broken lines represent the temperature at which TDA indicates a change in topological features. King's~{\cite{1967KIN}} definition is employed for the rings.
  }
  \label{rings}
\end{figure*}


\section{Results}\label{sec:results}
Fig.~{\ref{fuzzy}} shows the change in the density of the {\rred{silica}} (15,000 atoms) as a function of temperature obtained from the MD simulations. The results reproduce the density anomalies reported in MD simulations at approximately 3000~K and 5000~K~{\cite{1996VOL,1996VOL2}}.
The overestimation of the anomaly temperatures compared with the experimental values is a well-known behavior of the BKS potential~{\cite{1955VAN}} employed in this work.
%
Fig.~{\ref{fuzzy}} also shows the graphs of the fuzzy membership functions obtained as the output of the simulations for the -Si-Si-, -O-O-, and -Si-O- networks. \blue{Through our computational workflow,} the PDs are divided into three different clusters (groups I, II, and III). Three clusters are selected because there are two transition points in the density plot. 
{\rred{The intersection points for groups I (red) and II (blue) are located at 4735, 4048, and 4334~K for -Si-Si-, -O-O-, and -Si-O- networks, respectively~{\footnote{{\rred{These values were obtained when the last configuration at each temperature was picked up to compute the PDs, as written in Sec.~{\ref{sec:methods}}. They were estimated to be 4635~$\pm$~55, 4034~$\pm$~74, and 4310~$\pm$~39~K for -Si-Si-, -O-O-, and -Si-O- networks, respectively, by randomly choosing a configuration at each temperature.}}}}, while those for groups II (blue) and III (green) are located at 2674, 1856, and 2153~K for -Si-Si-, -O-O-, and -Si-O- networks, respectively~{\footnote{{\rred{These values were obtained when the last configuration at each temperature was picked up to compute the PDs, as written in Sec.~{\ref{sec:methods}}. They were estimated to be 2670~$\pm$~91, 1850~$\pm$~55, and 2080~$\pm$~38~K for -Si-Si-, -O-O-, and -Si-O- networks, respectively, by randomly choosing a configuration at each temperature.}}}}}}.
{\rred{Thus}}, the following conclusions are obtained from our analysis: (1) All the -Si-Si-, -O-O-, and -Si-O- PDs are clearly divided into three clusters. (2) The change in the -Si-Si- topology agrees with the transition points of density. (3) The changes in the -Si-O- and -O-O- topologies occur consecutively at lower temperatures. Therefore, they do not coincide with the density anomalies.
The first observation implies that topological features of the -Si-Si-, -O-O-, and -Si-O- networks dynamically evolve during the simulations. When a dynamical topological change is not significant, the separation between clusters is considerably less sharp~(see the method section for details). This suggests that there are significant changes in the {\it ring-related} topological features of the -Si-Si-, -O-O-, and -Si-O- networks because the one-dimensional topology is strongly related to the presence of ring structures in the networks. From a purely topological point of view, rings can be considered as nontrivial \blue{(non-contractible)} loops, which constitute the generators of the first persistent homology groups, i.e., the {\it one-dimensional topology}.
The second and third observations indicate that in the cooling process of {\rred{liquid silica}}, the one-dimensional topology of the -Si-Si- network changes at the temperatures of the density maximum and minimum. In contrast, the one-dimensional topologies of the -Si-O- and -O-O- networks change at lower temperatures.
This is quite surprising because the structural origin of the density maximum and minimum has been historically considered to be the stronger vibrations of O atoms located at the corner of silica tetrahedra~{\cite{2011SOU}}, which is completely opposite to our \blue{findings}. 

\vspace{2mm}
{\rred{Note that}} the fuzzy clustering algorithm depends on a hyperparameter $\lambda$ that controls the {\it fuzziness degree} allowed in the partitioning between clusters. Sharp clustering can be recovered from the fuzzy version by setting $\lambda \sim 1$ and, on the other hand, complete fuzziness -- {\it i.e.} such that there is no point with clear membership to a specific cluster -- is reached by letting $\lambda \rightarrow +\infty$. In our study, we found that the transition points detected by the algorithm are stabile (with differences around only $\pm 10$~K), and the figures displaying the fuzzy membership functions are obtained with a value of $1.2$.  This shows that the values detected for the transitions points are robust and do not depend on the choice of $\lambda$ and that, on the other hand, since we choose a value of $\lambda$ that allows some degree of fuzziness, we can study how gradual the transition between phases is.

\vspace{2mm}
\red{To interpret the changes in topological features obtained by TDA, we computed other standard local and global descriptors, such as radial distribution functions \red{(Fig.~\ref{SI-gr},~\ref{SI-gr-min})}, coordination numbers \red{(Fig.~\ref{SI-population-sio2},~\ref{SI-coord-num})}, structure factors \red{(Fig.~\ref{SI-sq})}, bond-angle distributions \red{(Figs.~\ref{SI-bad-O-Si-O}, \ref{SI-bad-Si-O-Si}, \ref{SI-bad-Si-Si-Si}, \ref{SI-bad-O-O-O})}, and tetrahedral ordering parameters \red{(Figs.~\ref{SI-tetra-SiO4},~\ref{SI-tetra-SiSi4})}. These results imply that TDA captures the topological features that are more difficult to identify using \blue{such} local and global descriptors.
For instance, there are {\it kinks} in the averages of the O-Si-O \red{(Fig.~\ref{SI-bad-O-Si-O})} and O-O-O \red{(Fig.~\ref{SI-bad-O-O-O})} bond-angle distributions at the temperatures of the density maximum (5000K) and minimum (3000~K), which are not consistent with the temperatures at which the TDA detects the change in the topological features for the -Si-O- and -O-O- networks. This is a resonable outcome because the local and global (i.e. averaged over the three-dimensional space) \blue{descriptors} are obviously not suitable for capturing medium-range structural orders that the TDA is able to capture. Although we also tried to interpret the TDA results using the variances, skewnesses, and kurtosises of the local and global descriptors \red{(Figs.~\ref{SI-bad-O-Si-O}, \ref{SI-bad-Si-O-Si}, \ref{SI-bad-Si-Si-Si}, \ref{SI-bad-O-O-O}, \ref{SI-tetra-SiO4}, \ref{SI-tetra-SiSi4})}, we could not find reasonable interpretation consistent with the \blue{findings}.
}

\begin{figure*}[htbp]
  \centering
  \includegraphics[width=\columnwidth]{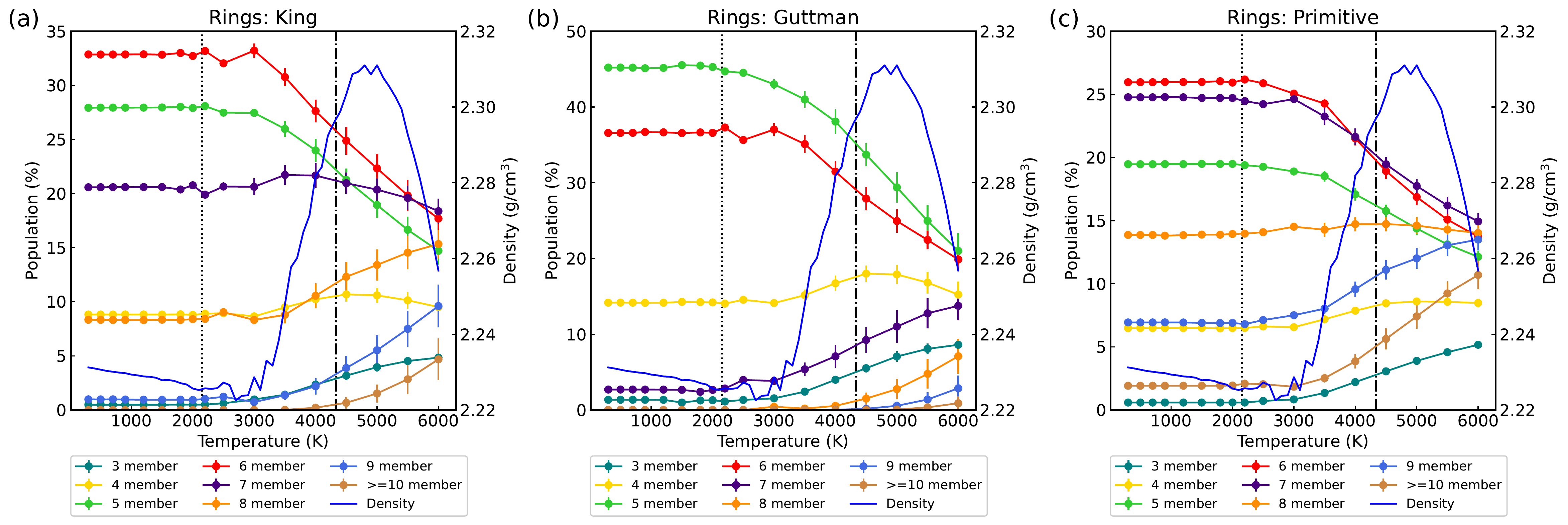}
  \caption{Ring statistics of -Si-O- networks computed using (a) King’s~{\cite{1967KIN}}, (b) Guttman’s~{\cite{1990GUT}}, and (c) primitive~{\cite{1991GOE,2002YUA,2002WOO}} definitions. The vertical broken lines represent the temperature at which TDA indicates a change in topological features.
  }
  \label{rings-SiO-comparison}
\end{figure*}

\vspace{2mm}
\red{\blue{On the other hand,} we found that conventional ring statistics shows a transition behaviour that is consistent with the \blue{our} results.} Such an agreement is reasonable because the one-dimensional topology is strongly related to the presence of nontrivial rings in the system. Fig.~{\ref{rings}} shows the populations of the -Si-Si-, -O-O-, and -Si-O- ring sizes as a function of temperature. The -Si-Si- or -O-O- ring sizes were computed from the structures where only Si or O atoms were extracted from MD trajectories \red{(Fig.~\ref{schematic})}, respectively. The -Si-O- ring size was computed from MD trajectories by counting only ABAB rings~{\cite{2010ROU}}. Note that the -Si-O- ring size refers to the number of Si atoms in a ring. The vertical broken lines denote the temperatures at which topological features change according to TDA. The densities are plotted in Fig.~{\ref{rings}}.
Fig.~{\ref{rings}} shows that at high temperatures, the transition points obtained using \blue{our TDA approach} agree with the temperatures at which {\it dominant ring sizes are swapped}. For instance, in the -Si-Si- ring statistics, the maximum ring size changes from 7 to 6 at approximately 4700~K. For the -O-O- ring statistics, the maximum ring size changes from 7 to 3 at approximately 4000~K. Shih et al. have reported that the structural origin of the anomalous density maximum in {\rred{liquid silica}} is the change in the dominance of -Si-O- rings~{\cite{2019SHI}}. We also observe the same change (the population of the 6-member rings becomes the most dominant instead of 7-member rings) in the -Si-O- ring statistics~{\footnote{ Shih et al. have used the primitive definition of rings, whereas we have used King's definition (Fig.~{\ref{rings}}). As shown in Fig.~{\ref{rings-SiO-comparison}}, the same swap is observed in the analysis performed using the primitive definition. However, it is at a lower temperature compared to the analysis performed using King's definition.}}. 
However, TDA implies that the change in topological features is more significant at lower temperatures (approximately 4300~K). At such temperatures, we can observe the swaps between 7-member and 5-member rings (second majority) and between 8-member and 4-member rings (third majority). 
Fig.~{\ref{rings}} also reveals that at low temperatures, the transition points obtained via TDA agree with the temperatures at which the populations of rings become stable. For instance, in the -Si-O- and -Si-Si- ring statistics, the populations of the ring sizes become completely flat at the temperature threshold obtained in the TDA. In the -O-O- ring statistics, we cannot draw a conclusion as clearly as that for the -Si-O- and -Si-Si- ring statistics. This is probably because the sensitivity of TDA to topological features is higher than that of ring statistics.

\section{Discussion}\label{sec:discussion}

\vspace{2mm}
\red{
Our TDA results reveal that the topology of the partial -Si-Si- network changes at the density maximum and minimum rather than the entire -Si-O- networks, which are also supported by the above conventional ring analysis.
}
However, the ring analysis for silica glasses has been historically performed considering only ABAB rings (-Si-O- rings)~{\cite{1993RIN, 2004HUA, 2008TAK2, 2015KOZ, 2019ATI, 2019ONO, 2020YAN, 2020SHI, 2021SHI}}\red{, though there are several seminal works focusing on the -Si-Si- networks in compressed silica glasses~{\cite{1990STI, 2002TRA}}}. The reasons for this are \blue{the following}:
Silica glass is typically categorized as a ``tetrahedral liquid” or ``water-type liquid,” which includes water, silicon, germanium, and beryllium fluoride~{\cite{2012SHA, 2019TAN}}. Among these, water has been studied the most intensively because of its ubiquity. 
On one hand, in the glass community, the tetrahedral unit network of silica is generally considered as the unit of SiO$_4$~{\cite{1993RIN, 2004HUA, 2015KOZ, 2019ONO, 2020YAN, 2020SHI}}. Thus, ring statistics are historically performed for the -Si-O- network, and the tetrahedral ordering parameter is computed for the tetrahedrons composed of Si and O$_4$ atoms~{\cite{2021TAN}}.
%
On the other hand, in the water community, the tetrahedral unit network of water is always considered to be composed of only O atoms. Thus, ring statistics and the tetrahedral ordering parameter are computed for tetrahedrons composed of O atoms~{\cite{2019SHI,2021FOF}}. 
Other descriptors for water also focus on the O tetrahedral network~{\cite{2011MEG,2013GAL,2014RUS,2014JON,2016YIC,2016YIC2,2016PAT,2021FOF}}. 
Thus, the studies that consider silica as ``a tetrahedral liquid" focus on SiSi$_4$~{\cite{2019SHI}}, not on SiO$_4$, and calculate ordering parameters for tetrahedrons composed of Si and Si$_4$ atoms~{\cite{2019SHI}}.
The \blue{findings} obtained \blue{via TDA} in our work imply that silica glass should be considered as ``a tetrahedral liquid," and one should focus on SiSi$_4$ tetrahedrons to understand its structural transitions in more detail.
In fact, the tetrahedral ordering parameters of SiO$_4$ (Fig.~\ref{SI-tetra-SiO4}) and SiSi$_4$ (Fig.~\ref{SI-tetra-SiSi4}) show quantitative different behaviors as temperature increases. The distribution of the SiO$_4$ tetrahedral ordering parameter (Fig.~\ref{SI-tetra-SiO4}) becomes broader as the temperature increases, while that of SiSi$_4$ (Fig.~\ref{SI-tetra-SiSi4}) shows a bimodal distribution at high temperature. The SiSi$_4$ tetrahedron could be a key descriptor to understand the structural origin of the density anomalies of {\rred{silica}}.

\vspace{2mm}
Three ring definitions (King~{\cite{1967KIN}}, Guttman~{\cite{1990GUT}}, and primitive (most common)~{\cite{1991GOE,2002YUA,2002WOO}}) have been adopted for silica glasses~{\cite{1993RIN, 2004HUA, 2015KOZ, 2019ATI, 2019ONO, 2020YAN, 2020SHI, 2021SHI}}. While we interpreted the TDA results based on the ring statistics computed using King's definition (Fig.~{\ref{rings}}), we also investigated the other definitions. We found that the ring statistics computed according to Guttman's definition~{\cite{1990GUT}} do not show any swap at the temperatures obtained using TDA, unlike the ring statistics computed using King's and the primitive definitions. Fig.~{\ref{rings-SiO-comparison}} shows the comparison of the -Si-O- ring statistics computed using the three different definitions. Apparently, the ring statistics computed using Guttman's definition do not show any swap. The ring statistics computed using King's criterion show swaps between 5-member and 7-member rings and between 4-member and 8-member rings at approximately 4300~K. The ring statistics computed using the primitive definition show swaps between 6-member and 7-member rings. The results indicate that the topological features that can be captured using ring statistics strongly depend on the ring definition. This implies that Guttman’s definition is inferior to the other two definition in terms of its power of capturing topological features. This is because, as recently reported in Ref.~{\onlinecite{2021ZHO}}, Guttman’s definition provides the {\it narrowest} distribution of ring sizes, whereas King's definition provides more varieties of ring sizes. The distribution obtained using the primitive definition is between those obtained using King's and Guttman's definitions.{\footnote{The ring analysis (Fig.~{\ref{rings-SiO-comparison}}) performed in this work shows the same tendency. 6-member rings are the most dominant for King's and the primitive definitions, whereas 5-member rings are the most dominant for Guttman's definition.}} Indeed, a descriptor is more powerful in describing topological features if it includes more varieties of rings and more redundant rings. This is probably the reason why Guttman's definition does not capture the topological features that TDA does.
However, it should be noted that Zhou et al.~{\cite{2021ZHO}} reported Guttman’s definition as the most suitable for describing the ring distribution derived from the FSDP of experimental scattering patterns. They explained the reason for this as follows: ``{\it the Guttman definition is suitable for capturing medium-range order patterns matching those can be captured by the FSDP, while the King’s and primitive definitions are associated with larger diameters and, hence, would be only very weakly captured by the FSDP}." This indicates that the ring definition should be based on the topological features to be captured.

\vspace{2mm}
As mentioned in the introduction, new descriptors are being developed to understand structural ordering of glassy materials in more detail~{\cite{2019TAN}}. One of the most recent and relevant studies is Ref.~\onlinecite{2021TAN}, where the authors have used the D-measure~\cite{2017SCH} to compare the graphs generated by the configurations of silica glass obtained through MD simulations. One of the main differences between our approach and that reported in Ref.~\onlinecite{2021TAN} is that our approach directly extracts topological invariants from the configurations, whereas the other approach must first construct a graph. This additional step requires hyperparameters, thus leading to potential approximations and loss of information. Moreover, in Ref.~\onlinecite{2021TAN}, the (dis)similarity between different glass configurations is not quantitatively assessed through the D-measure, and the main arguments concern only the {\it qualitative} differences in the distributions of the D-measure. In contrast, we use the rigorous and quantitative approach given by fuzzy spectral clustering. Finally, the D-measure has no clear topological meaning, unlike PDs. This makes the geometric and topological interpretation of the differences between two glass configurations more difficult. FInally, we mention that an interesting direction for future research would be to understand the structural ordering of glassy materials more comprehensively.

\vspace{2mm}
{\rred{
In this study, we employed the classical BKS potential and demonstrated the application of the TDA workflow to investigate the topological change in glassy materials, even though other sophisticated classical~{\cite{1988TSU, 1990SOU, 2004TAK, 2008CAR, 2011SOU, 2018SUN}} or machine-learning~{\cite{2020BAL,2021KOB,2021URAa,2021URAb,2022ERH}} potentials have been proposed.
Recently, Erhard et al.~{\cite{2022ERH}} reported that the FSDP of the structure factor simulated using the BKS potential is slightly biased compared with the experimental observation~{\cite{2007MEI, 2008MEI}}. It indicates that the BKS classical potential does not perfectly reproduce the realistic medium-range ordering of the glass structure~{\cite{2021ZHO}}. This problem can be solved by using atomistic glass structures obtained from experiments. For instance, the so-called force-enhanced atomic refinement (FEAR) modeling approach~{\cite{2015PAN}}, which relies on an iterative combination of reverse Monte Carlo (RMC) refinement and energy minimization cycle, is a state-of-the-art method to obtain realistic glass structures from experimental results~{\cite{2018LIM, 2020ZHO, 2021ZHO}}. Obviously, our TDA works not only with structures obtained from MD simulations, but also with those obtained from experiments. Applying our TDA computational approach on realistic glass structures is an intriguing future work.
}}


\section{Conclusion}\label{sec:conclusion}

\vspace{2mm}
This work revealed the importance of the partial -Si-Si- network in {\rred{silica}} and demonstrated the usefulness of TDA for understanding the transition behaviour of materials. TDA detects complex topological features, which are quantitatively compared via a rigorous mathematical approach and then clustered using state-of-the-art unsupervised ML methods. In principle, TDA can detect the topological invariants of material configurations with arbitrary dimensions. This makes it essential in the study of materials that show transitions with unidentified origins. As shown in our case study on {\rred{silica}}, the topological analysis can be specialized to a specific homological dimension, which has a well-defined physical interpretation, to understand the type of invariants responsible for the transition. Furthermore, our workflow computes general topological properties. We compute standard topological features (such as \red{ring statistics}) and show that the results obtained through these features can be remapped to the results of our analysis. We expect that our work will promote the application of topological techniques from TDA in the investigation of the key structural origins of certain transitions in liquid and glassy materials. 

\section{Acknowledgments} \label{sec:ack}
K.N. is grateful for the computational resources provided by the Research Center for Advanced Computing Infrastructure at Japan Advanced Institute of Science and Technology (JAIST).
A.T. acknowledges  the financial  support  from the  MIUR  Progetti  di  Ricerca  di  Rilevante  Interesse Nazionale  (PRIN)  Bando  2017  (grant number  2017BZPKSZ).
K.N. acknowledges the support from the JSPS Overseas Research Fellowships.

\section{Data availability} \label{sec:data}
{\rred{Our GitHub repository [\url{https://github.com/kousuke-nakano/TDA_examples}] contains the codes with the implementation of the TDA computational workflow.}} {\rred{Other}} data that support the findings of this study are available from the corresponding author upon reasonable request.

\section{Code availability} \label{sec:code}
The codes that were used to perform MD simulations (LAMMPS~{\cite{1995Plimpton}}) and analyse trajectories (R.I.N.G.S.~{\cite{2010ROU}}) are available from their websites. The codes that were used to generate and analyze PDs are available from the corresponding author upon reasonable request.

\section{Author contributions} \label{sec:contribution}
K.N. conceived the study. K.N. carried out the MD simulations using LAMMPS and computed the local and global descriptors using the R.I.N.G.S. code. A.T. implemented the TDA workflow in \verb|Python| and analysed the topological features of MD trajectories. A.T. and K.N. contributed to writing the manuscript and approved the final version for submission. 

\section{Conflict of interest} \label{sec:interests}
The authors declare no conflict of interest.

\bibliography{references}

\begin{thebibliography}{112}%
\makeatletter
\providecommand \@ifxundefined [1]{%
 \@ifx{#1\undefined}
}%
\providecommand \@ifnum [1]{%
 \ifnum #1\expandafter \@firstoftwo
 \else \expandafter \@secondoftwo
 \fi
}%
\providecommand \@ifx [1]{%
 \ifx #1\expandafter \@firstoftwo
 \else \expandafter \@secondoftwo
 \fi
}%
\providecommand \natexlab [1]{#1}%
\providecommand \enquote  [1]{``#1''}%
\providecommand \bibnamefont  [1]{#1}%
\providecommand \bibfnamefont [1]{#1}%
\providecommand \citenamefont [1]{#1}%
\providecommand \href@noop [0]{\@secondoftwo}%
\providecommand \href [0]{\begingroup \@sanitize@url \@href}%
\providecommand \@href[1]{\@@startlink{#1}\@@href}%
\providecommand \@@href[1]{\endgroup#1\@@endlink}%
\providecommand \@sanitize@url [0]{\catcode `\\12\catcode `\$12\catcode
  `\&12\catcode `\#12\catcode `\^12\catcode `\_12\catcode `\%12\relax}%
\providecommand \@@startlink[1]{}%
\providecommand \@@endlink[0]{}%
\providecommand \url  [0]{\begingroup\@sanitize@url \@url }%
\providecommand \@url [1]{\endgroup\@href {#1}{\urlprefix }}%
\providecommand \urlprefix  [0]{URL }%
\providecommand \Eprint [0]{\href }%
\providecommand \doibase [0]{https://doi.org/}%
\providecommand \selectlanguage [0]{\@gobble}%
\providecommand \bibinfo  [0]{\@secondoftwo}%
\providecommand \bibfield  [0]{\@secondoftwo}%
\providecommand \translation [1]{[#1]}%
\providecommand \BibitemOpen [0]{}%
\providecommand \bibitemStop [0]{}%
\providecommand \bibitemNoStop [0]{.\EOS\space}%
\providecommand \EOS [0]{\spacefactor3000\relax}%
\providecommand \BibitemShut  [1]{\csname bibitem#1\endcsname}%
\let\auto@bib@innerbib\@empty
\bibitem [{\citenamefont {Anderson}(1995)}]{1995AND}%
  \BibitemOpen
  \bibfield  {author} {\bibinfo {author} {\bibfnamefont {P.~W.}\ \bibnamefont
  {Anderson}},\ }\href {https://doi.org/10.1126/science.267.5204.1610.a}
  {\bibfield  {journal} {\bibinfo  {journal} {Science}\ }\textbf {\bibinfo
  {volume} {267}},\ \bibinfo {pages} {1615} (\bibinfo {year}
  {1995})}\BibitemShut {NoStop}%
\bibitem [{\citenamefont {Tanaka}\ \emph {et~al.}(2019)\citenamefont {Tanaka},
  \citenamefont {Tong}, \citenamefont {Shi},\ and\ \citenamefont
  {Russo}}]{2019TAN}%
  \BibitemOpen
  \bibfield  {author} {\bibinfo {author} {\bibfnamefont {H.}~\bibnamefont
  {Tanaka}}, \bibinfo {author} {\bibfnamefont {H.}~\bibnamefont {Tong}},
  \bibinfo {author} {\bibfnamefont {R.}~\bibnamefont {Shi}},\ and\ \bibinfo
  {author} {\bibfnamefont {J.}~\bibnamefont {Russo}},\ }\href
  {https://doi.org/10.1038/s42254-019-0053-3} {\bibfield  {journal} {\bibinfo
  {journal} {Nat. Rev. Phys.}\ }\textbf {\bibinfo {volume} {1}},\ \bibinfo
  {pages} {333} (\bibinfo {year} {2019})}\BibitemShut {NoStop}%
\bibitem [{\citenamefont {Bacon}\ \emph {et~al.}(1960)\citenamefont {Bacon},
  \citenamefont {Hasapis},\ and\ \citenamefont {Wholley~Jr}}]{1960BAC}%
  \BibitemOpen
  \bibfield  {author} {\bibinfo {author} {\bibfnamefont {J.}~\bibnamefont
  {Bacon}}, \bibinfo {author} {\bibfnamefont {A.}~\bibnamefont {Hasapis}},\
  and\ \bibinfo {author} {\bibfnamefont {J.}~\bibnamefont {Wholley~Jr}},\
  }\href@noop {} {\bibfield  {journal} {\bibinfo  {journal} {Phys. Chem.
  Glasses}\ }\textbf {\bibinfo {volume} {1}},\ \bibinfo {pages} {90} (\bibinfo
  {year} {1960})}\BibitemShut {NoStop}%
\bibitem [{\citenamefont {Brueckner}(1970)}]{1970BRU}%
  \BibitemOpen
  \bibfield  {author} {\bibinfo {author} {\bibfnamefont {R.}~\bibnamefont
  {Brueckner}},\ }\href {https://doi.org/10.1016/0022-3093(70)90190-0}
  {\bibfield  {journal} {\bibinfo  {journal} {J. Non-Cryst. Solids}\ }\textbf
  {\bibinfo {volume} {5}},\ \bibinfo {pages} {123} (\bibinfo {year}
  {1970})}\BibitemShut {NoStop}%
\bibitem [{\citenamefont {Kakiuchida}\ \emph {et~al.}(2003)\citenamefont
  {Kakiuchida}, \citenamefont {Saito},\ and\ \citenamefont
  {Ikushima}}]{2003KAK}%
  \BibitemOpen
  \bibfield  {author} {\bibinfo {author} {\bibfnamefont {H.}~\bibnamefont
  {Kakiuchida}}, \bibinfo {author} {\bibfnamefont {K.}~\bibnamefont {Saito}},\
  and\ \bibinfo {author} {\bibfnamefont {A.~J.}\ \bibnamefont {Ikushima}},\
  }\href {https://doi.org/10.1063/1.1527206} {\bibfield  {journal} {\bibinfo
  {journal} {J. Appl. Phys.}\ }\textbf {\bibinfo {volume} {93}},\ \bibinfo
  {pages} {777} (\bibinfo {year} {2003})}\BibitemShut {NoStop}%
\bibitem [{\citenamefont {Skinner}\ \emph {et~al.}(2013)\citenamefont
  {Skinner}, \citenamefont {Benmore}, \citenamefont {Weber}, \citenamefont
  {Wilding}, \citenamefont {Tumber},\ and\ \citenamefont {Parise}}]{2013SKI}%
  \BibitemOpen
  \bibfield  {author} {\bibinfo {author} {\bibfnamefont {L.}~\bibnamefont
  {Skinner}}, \bibinfo {author} {\bibfnamefont {C.}~\bibnamefont {Benmore}},
  \bibinfo {author} {\bibfnamefont {J.}~\bibnamefont {Weber}}, \bibinfo
  {author} {\bibfnamefont {M.}~\bibnamefont {Wilding}}, \bibinfo {author}
  {\bibfnamefont {S.}~\bibnamefont {Tumber}},\ and\ \bibinfo {author}
  {\bibfnamefont {J.}~\bibnamefont {Parise}},\ }\href
  {https://doi.org/10.1039/C3CP44347G} {\bibfield  {journal} {\bibinfo
  {journal} {Phys. Chem. Chem. Phys.}\ }\textbf {\bibinfo {volume} {15}},\
  \bibinfo {pages} {8566} (\bibinfo {year} {2013})}\BibitemShut {NoStop}%
\bibitem [{\citenamefont {Yamahara}\ \emph {et~al.}(2001)\citenamefont
  {Yamahara}, \citenamefont {Okazaki},\ and\ \citenamefont
  {Kawamura}}]{2001YAM}%
  \BibitemOpen
  \bibfield  {author} {\bibinfo {author} {\bibfnamefont {K.}~\bibnamefont
  {Yamahara}}, \bibinfo {author} {\bibfnamefont {K.}~\bibnamefont {Okazaki}},\
  and\ \bibinfo {author} {\bibfnamefont {K.}~\bibnamefont {Kawamura}},\ }\href
  {https://doi.org/10.1016/S0022-3093(01)00795-5} {\bibfield  {journal}
  {\bibinfo  {journal} {J. Non. Cryst. Solids}\ }\textbf {\bibinfo {volume}
  {291}},\ \bibinfo {pages} {32} (\bibinfo {year} {2001})}\BibitemShut
  {NoStop}%
\bibitem [{\citenamefont {Soules}\ \emph {et~al.}(2011)\citenamefont {Soules},
  \citenamefont {Gilmer}, \citenamefont {Matthews}, \citenamefont {Stolken},\
  and\ \citenamefont {Feit}}]{2011SOU}%
  \BibitemOpen
  \bibfield  {author} {\bibinfo {author} {\bibfnamefont {T.~F.}\ \bibnamefont
  {Soules}}, \bibinfo {author} {\bibfnamefont {G.~H.}\ \bibnamefont {Gilmer}},
  \bibinfo {author} {\bibfnamefont {M.~J.}\ \bibnamefont {Matthews}}, \bibinfo
  {author} {\bibfnamefont {J.~S.}\ \bibnamefont {Stolken}},\ and\ \bibinfo
  {author} {\bibfnamefont {M.~D.}\ \bibnamefont {Feit}},\ }\href
  {https://doi.org/10.1016/j.jnoncrysol.2011.01.009} {\bibfield  {journal}
  {\bibinfo  {journal} {J. Non. Cryst. Solids}\ }\textbf {\bibinfo {volume}
  {357}},\ \bibinfo {pages} {1564} (\bibinfo {year} {2011})}\BibitemShut
  {NoStop}%
\bibitem [{\citenamefont {Shi}\ and\ \citenamefont {Tanaka}(2019)}]{2019SHI}%
  \BibitemOpen
  \bibfield  {author} {\bibinfo {author} {\bibfnamefont {R.}~\bibnamefont
  {Shi}}\ and\ \bibinfo {author} {\bibfnamefont {H.}~\bibnamefont {Tanaka}},\
  }\href {https://doi.org/10.1126/sciadv.aav3194} {\bibfield  {journal}
  {\bibinfo  {journal} {Sci. Adv.}\ }\textbf {\bibinfo {volume} {5}},\ \bibinfo
  {pages} {eaav3194} (\bibinfo {year} {2019})}\BibitemShut {NoStop}%
\bibitem [{\citenamefont {Rui~Tan}\ \emph {et~al.}(2021)\citenamefont
  {Rui~Tan}, \citenamefont {Urata}, \citenamefont {Yamada},\ and\ \citenamefont
  {G{\'o}mez-Bombarelli}}]{2021TAN}%
  \BibitemOpen
  \bibfield  {author} {\bibinfo {author} {\bibfnamefont {A.}~\bibnamefont
  {Rui~Tan}}, \bibinfo {author} {\bibfnamefont {S.}~\bibnamefont {Urata}},
  \bibinfo {author} {\bibfnamefont {M.}~\bibnamefont {Yamada}},\ and\ \bibinfo
  {author} {\bibfnamefont {R.}~\bibnamefont {G{\'o}mez-Bombarelli}},\ }\href
  {https://doi.org/10.48550/arXiv.2111.07452} {\bibfield  {journal} {\bibinfo
  {journal} {arXiv e-prints}\ ,\ \bibinfo {pages} {arXiv}} (\bibinfo {year}
  {2021})}\BibitemShut {NoStop}%
\bibitem [{\citenamefont {Rebelo}\ \emph {et~al.}(1998)\citenamefont {Rebelo},
  \citenamefont {Debenedetti},\ and\ \citenamefont {Sastry}}]{1998LUI}%
  \BibitemOpen
  \bibfield  {author} {\bibinfo {author} {\bibfnamefont {L.~P.~N.}\
  \bibnamefont {Rebelo}}, \bibinfo {author} {\bibfnamefont {P.~G.}\
  \bibnamefont {Debenedetti}},\ and\ \bibinfo {author} {\bibfnamefont
  {S.}~\bibnamefont {Sastry}},\ }\href {https://doi.org/10.1063/1.476600}
  {\bibfield  {journal} {\bibinfo  {journal} {J. Chem. Phys.}\ }\textbf
  {\bibinfo {volume} {109}},\ \bibinfo {pages} {626} (\bibinfo {year}
  {1998})}\BibitemShut {NoStop}%
\bibitem [{\citenamefont {Tanaka}(2000)}]{2000TAN}%
  \BibitemOpen
  \bibfield  {author} {\bibinfo {author} {\bibfnamefont {H.}~\bibnamefont
  {Tanaka}},\ }\href {https://doi.org/10.1209/epl/i2000-00276-4} {\bibfield
  {journal} {\bibinfo  {journal} {EPL (Europhysics Letters)}\ }\textbf
  {\bibinfo {volume} {50}},\ \bibinfo {pages} {340} (\bibinfo {year}
  {2000})}\BibitemShut {NoStop}%
\bibitem [{\citenamefont {Errington}\ and\ \citenamefont
  {Debenedetti}(2001)}]{2001JEF}%
  \BibitemOpen
  \bibfield  {author} {\bibinfo {author} {\bibfnamefont {J.~R.}\ \bibnamefont
  {Errington}}\ and\ \bibinfo {author} {\bibfnamefont {P.~G.}\ \bibnamefont
  {Debenedetti}},\ }\href {https://doi.org/10.1038/35053024} {\bibfield
  {journal} {\bibinfo  {journal} {Nature}\ }\textbf {\bibinfo {volume} {409}},\
  \bibinfo {pages} {318} (\bibinfo {year} {2001})}\BibitemShut {NoStop}%
\bibitem [{\citenamefont {Tanaka}(2002)}]{2002TAN}%
  \BibitemOpen
  \bibfield  {author} {\bibinfo {author} {\bibfnamefont {H.}~\bibnamefont
  {Tanaka}},\ }\href {https://doi.org/10.1103/PhysRevB.66.064202} {\bibfield
  {journal} {\bibinfo  {journal} {Phys. Rev. B}\ }\textbf {\bibinfo {volume}
  {66}},\ \bibinfo {pages} {064202} (\bibinfo {year} {2002})}\BibitemShut
  {NoStop}%
\bibitem [{\citenamefont {Errington}\ \emph {et~al.}(2002)\citenamefont
  {Errington}, \citenamefont {Debenedetti},\ and\ \citenamefont
  {Torquato}}]{2002JEF}%
  \BibitemOpen
  \bibfield  {author} {\bibinfo {author} {\bibfnamefont {J.~R.}\ \bibnamefont
  {Errington}}, \bibinfo {author} {\bibfnamefont {P.~G.}\ \bibnamefont
  {Debenedetti}},\ and\ \bibinfo {author} {\bibfnamefont {S.}~\bibnamefont
  {Torquato}},\ }\href {https://doi.org/10.1103/PhysRevLett.89.215503}
  {\bibfield  {journal} {\bibinfo  {journal} {Phys. Rev. Lett.}\ }\textbf
  {\bibinfo {volume} {89}},\ \bibinfo {pages} {215503} (\bibinfo {year}
  {2002})}\BibitemShut {NoStop}%
\bibitem [{\citenamefont {Errington}\ \emph {et~al.}(2006)\citenamefont
  {Errington}, \citenamefont {Truskett},\ and\ \citenamefont
  {Mittal}}]{2006JEF}%
  \BibitemOpen
  \bibfield  {author} {\bibinfo {author} {\bibfnamefont {J.~R.}\ \bibnamefont
  {Errington}}, \bibinfo {author} {\bibfnamefont {T.~M.}\ \bibnamefont
  {Truskett}},\ and\ \bibinfo {author} {\bibfnamefont {J.}~\bibnamefont
  {Mittal}},\ }\href {https://doi.org/10.1063/1.2409932} {\bibfield  {journal}
  {\bibinfo  {journal} {J. Chem. Phys.}\ }\textbf {\bibinfo {volume} {125}},\
  \bibinfo {pages} {244502} (\bibinfo {year} {2006})}\BibitemShut {NoStop}%
\bibitem [{\citenamefont {Molinero}\ \emph {et~al.}(2006)\citenamefont
  {Molinero}, \citenamefont {Sastry},\ and\ \citenamefont {Angell}}]{2006VAL}%
  \BibitemOpen
  \bibfield  {author} {\bibinfo {author} {\bibfnamefont {V.}~\bibnamefont
  {Molinero}}, \bibinfo {author} {\bibfnamefont {S.}~\bibnamefont {Sastry}},\
  and\ \bibinfo {author} {\bibfnamefont {C.~A.}\ \bibnamefont {Angell}},\
  }\href {https://doi.org/10.1103/PhysRevLett.97.075701} {\bibfield  {journal}
  {\bibinfo  {journal} {Phys. Rev. Lett.}\ }\textbf {\bibinfo {volume} {97}},\
  \bibinfo {pages} {075701} (\bibinfo {year} {2006})}\BibitemShut {NoStop}%
\bibitem [{\citenamefont {Sharma}\ \emph {et~al.}(2006)\citenamefont {Sharma},
  \citenamefont {Chakraborty},\ and\ \citenamefont {Chakravarty}}]{2006RUC}%
  \BibitemOpen
  \bibfield  {author} {\bibinfo {author} {\bibfnamefont {R.}~\bibnamefont
  {Sharma}}, \bibinfo {author} {\bibfnamefont {S.~N.}\ \bibnamefont
  {Chakraborty}},\ and\ \bibinfo {author} {\bibfnamefont {C.}~\bibnamefont
  {Chakravarty}},\ }\href {https://doi.org/10.1063/1.2390710} {\bibfield
  {journal} {\bibinfo  {journal} {J. Chem. Phys.}\ }\textbf {\bibinfo {volume}
  {125}},\ \bibinfo {pages} {204501} (\bibinfo {year} {2006})}\BibitemShut
  {NoStop}%
\bibitem [{\citenamefont {Agarwal}\ \emph {et~al.}(2007)\citenamefont
  {Agarwal}, \citenamefont {Sharma},\ and\ \citenamefont
  {Chakravarty}}]{2007AGA}%
  \BibitemOpen
  \bibfield  {author} {\bibinfo {author} {\bibfnamefont {M.}~\bibnamefont
  {Agarwal}}, \bibinfo {author} {\bibfnamefont {R.}~\bibnamefont {Sharma}},\
  and\ \bibinfo {author} {\bibfnamefont {C.}~\bibnamefont {Chakravarty}},\
  }\href {https://doi.org/10.1063/1.2794766} {\bibfield  {journal} {\bibinfo
  {journal} {J. Chem. Phys.}\ }\textbf {\bibinfo {volume} {127}},\ \bibinfo
  {pages} {164502} (\bibinfo {year} {2007})}\BibitemShut {NoStop}%
\bibitem [{\citenamefont {Agarwal}\ and\ \citenamefont
  {Chakravarty}(2007)}]{2007AGA2}%
  \BibitemOpen
  \bibfield  {author} {\bibinfo {author} {\bibfnamefont {M.}~\bibnamefont
  {Agarwal}}\ and\ \bibinfo {author} {\bibfnamefont {C.}~\bibnamefont
  {Chakravarty}},\ }\href {https://doi.org/10.1021/jp0753272} {\bibfield
  {journal} {\bibinfo  {journal} {J. Phys. Chem. B}\ }\textbf {\bibinfo
  {volume} {111}},\ \bibinfo {pages} {13294} (\bibinfo {year}
  {2007})}\BibitemShut {NoStop}%
\bibitem [{\citenamefont {Sharma}\ \emph {et~al.}(2008)\citenamefont {Sharma},
  \citenamefont {Agarwal},\ and\ \citenamefont {Chakravarty}}]{2008RUC}%
  \BibitemOpen
  \bibfield  {author} {\bibinfo {author} {\bibfnamefont {R.}~\bibnamefont
  {Sharma}}, \bibinfo {author} {\bibfnamefont {M.}~\bibnamefont {Agarwal}},\
  and\ \bibinfo {author} {\bibfnamefont {C.}~\bibnamefont {Chakravarty}},\
  }\href {https://doi.org/10.1080/00268970802378662} {\bibfield  {journal}
  {\bibinfo  {journal} {Mol. Phys.}\ }\textbf {\bibinfo {volume} {106}},\
  \bibinfo {pages} {1925} (\bibinfo {year} {2008})}\BibitemShut {NoStop}%
\bibitem [{\citenamefont {Agarwal}\ and\ \citenamefont
  {Chakravarty}(2009)}]{2009MAN}%
  \BibitemOpen
  \bibfield  {author} {\bibinfo {author} {\bibfnamefont {M.}~\bibnamefont
  {Agarwal}}\ and\ \bibinfo {author} {\bibfnamefont {C.}~\bibnamefont
  {Chakravarty}},\ }\href {https://doi.org/10.1103/PhysRevE.79.030202}
  {\bibfield  {journal} {\bibinfo  {journal} {Phys. Rev. E}\ }\textbf {\bibinfo
  {volume} {79}},\ \bibinfo {pages} {030202} (\bibinfo {year}
  {2009})}\BibitemShut {NoStop}%
\bibitem [{\citenamefont {Agarwal}\ \emph {et~al.}(2009)\citenamefont
  {Agarwal}, \citenamefont {Ganguly},\ and\ \citenamefont
  {Chakravarty}}]{2009AGA}%
  \BibitemOpen
  \bibfield  {author} {\bibinfo {author} {\bibfnamefont {M.}~\bibnamefont
  {Agarwal}}, \bibinfo {author} {\bibfnamefont {A.}~\bibnamefont {Ganguly}},\
  and\ \bibinfo {author} {\bibfnamefont {C.}~\bibnamefont {Chakravarty}},\
  }\href {https://doi.org/10.1021/jp903694b} {\bibfield  {journal} {\bibinfo
  {journal} {J. Phys. Chem. B}\ }\textbf {\bibinfo {volume} {113}},\ \bibinfo
  {pages} {15284} (\bibinfo {year} {2009})}\BibitemShut {NoStop}%
\bibitem [{\citenamefont {Jabes}\ \emph {et~al.}(2010)\citenamefont {Jabes},
  \citenamefont {Agarwal},\ and\ \citenamefont {Chakravarty}}]{2010SHA}%
  \BibitemOpen
  \bibfield  {author} {\bibinfo {author} {\bibfnamefont {B.~S.}\ \bibnamefont
  {Jabes}}, \bibinfo {author} {\bibfnamefont {M.}~\bibnamefont {Agarwal}},\
  and\ \bibinfo {author} {\bibfnamefont {C.}~\bibnamefont {Chakravarty}},\
  }\href {https://doi.org/10.1063/1.3439593} {\bibfield  {journal} {\bibinfo
  {journal} {J. Chem. Phys.}\ }\textbf {\bibinfo {volume} {132}},\ \bibinfo
  {pages} {234507} (\bibinfo {year} {2010})}\BibitemShut {NoStop}%
\bibitem [{\citenamefont {Hujo}\ \emph {et~al.}(2011)\citenamefont {Hujo},
  \citenamefont {Shadrack~Jabes}, \citenamefont {Rana}, \citenamefont
  {Chakravarty},\ and\ \citenamefont {Molinero}}]{2011HUJ}%
  \BibitemOpen
  \bibfield  {author} {\bibinfo {author} {\bibfnamefont {W.}~\bibnamefont
  {Hujo}}, \bibinfo {author} {\bibfnamefont {B.}~\bibnamefont
  {Shadrack~Jabes}}, \bibinfo {author} {\bibfnamefont {V.~K.}\ \bibnamefont
  {Rana}}, \bibinfo {author} {\bibfnamefont {C.}~\bibnamefont {Chakravarty}},\
  and\ \bibinfo {author} {\bibfnamefont {V.}~\bibnamefont {Molinero}},\ }\href
  {https://doi.org/10.1007/s10955-011-0293-9} {\bibfield  {journal} {\bibinfo
  {journal} {J. Stat. Phys.}\ }\textbf {\bibinfo {volume} {145}},\ \bibinfo
  {pages} {293} (\bibinfo {year} {2011})}\BibitemShut {NoStop}%
\bibitem [{\citenamefont {Agarwal}\ \emph {et~al.}(2011)\citenamefont
  {Agarwal}, \citenamefont {Alam},\ and\ \citenamefont
  {Chakravarty}}]{2011AGA}%
  \BibitemOpen
  \bibfield  {author} {\bibinfo {author} {\bibfnamefont {M.}~\bibnamefont
  {Agarwal}}, \bibinfo {author} {\bibfnamefont {M.~P.}\ \bibnamefont {Alam}},\
  and\ \bibinfo {author} {\bibfnamefont {C.}~\bibnamefont {Chakravarty}},\
  }\href {https://doi.org/10.1021/jp110695t} {\bibfield  {journal} {\bibinfo
  {journal} {J. Phys. Chem. B}\ }\textbf {\bibinfo {volume} {115}},\ \bibinfo
  {pages} {6935} (\bibinfo {year} {2011})}\BibitemShut {NoStop}%
\bibitem [{\citenamefont {Shih}\ \emph {et~al.}(2020)\citenamefont {Shih},
  \citenamefont {Sundararaman},\ and\ \citenamefont {Huang}}]{2020SHI}%
  \BibitemOpen
  \bibfield  {author} {\bibinfo {author} {\bibfnamefont {Y.-T.}\ \bibnamefont
  {Shih}}, \bibinfo {author} {\bibfnamefont {S.}~\bibnamefont {Sundararaman}},\
  and\ \bibinfo {author} {\bibfnamefont {L.}~\bibnamefont {Huang}},\ }\href
  {https://doi.org/10.1111/jace.16850} {\bibfield  {journal} {\bibinfo
  {journal} {J. Am. Ceram. Soc.}\ }\textbf {\bibinfo {volume} {103}},\ \bibinfo
  {pages} {3942} (\bibinfo {year} {2020})}\BibitemShut {NoStop}%
\bibitem [{\citenamefont {Sen}\ \emph {et~al.}(2004)\citenamefont {Sen},
  \citenamefont {Andrus}, \citenamefont {Baker},\ and\ \citenamefont
  {Murtagh}}]{2004SEN}%
  \BibitemOpen
  \bibfield  {author} {\bibinfo {author} {\bibfnamefont {S.}~\bibnamefont
  {Sen}}, \bibinfo {author} {\bibfnamefont {R.~L.}\ \bibnamefont {Andrus}},
  \bibinfo {author} {\bibfnamefont {D.~E.}\ \bibnamefont {Baker}},\ and\
  \bibinfo {author} {\bibfnamefont {M.~T.}\ \bibnamefont {Murtagh}},\ }\href
  {https://doi.org/10.1103/PhysRevLett.93.125902} {\bibfield  {journal}
  {\bibinfo  {journal} {Phys. Rev. Lett.}\ }\textbf {\bibinfo {volume} {93}},\
  \bibinfo {pages} {125902} (\bibinfo {year} {2004})}\BibitemShut {NoStop}%
\bibitem [{\citenamefont {Saika-Voivod}\ \emph {et~al.}(2001)\citenamefont
  {Saika-Voivod}, \citenamefont {Poole},\ and\ \citenamefont
  {Sciortino}}]{2001VOL}%
  \BibitemOpen
  \bibfield  {author} {\bibinfo {author} {\bibfnamefont {I.}~\bibnamefont
  {Saika-Voivod}}, \bibinfo {author} {\bibfnamefont {P.~H.}\ \bibnamefont
  {Poole}},\ and\ \bibinfo {author} {\bibfnamefont {F.}~\bibnamefont
  {Sciortino}},\ }\href {https://doi.org/10.1038/35087524} {\bibfield
  {journal} {\bibinfo  {journal} {Nature}\ }\textbf {\bibinfo {volume} {412}},\
  \bibinfo {pages} {514} (\bibinfo {year} {2001})}\BibitemShut {NoStop}%
\bibitem [{\citenamefont {Takada}\ \emph {et~al.}(2009)\citenamefont {Takada},
  \citenamefont {Richet},\ and\ \citenamefont {Atake}}]{2009TAK}%
  \BibitemOpen
  \bibfield  {author} {\bibinfo {author} {\bibfnamefont {A.}~\bibnamefont
  {Takada}}, \bibinfo {author} {\bibfnamefont {P.}~\bibnamefont {Richet}},\
  and\ \bibinfo {author} {\bibfnamefont {T.}~\bibnamefont {Atake}},\ }\href
  {https://doi.org/10.1016/j.jnoncrysol.2008.11.024} {\bibfield  {journal}
  {\bibinfo  {journal} {J. Non-Cryst. Solids}\ }\textbf {\bibinfo {volume}
  {355}},\ \bibinfo {pages} {694} (\bibinfo {year} {2009})}\BibitemShut
  {NoStop}%
\bibitem [{\citenamefont {Cuthbertson}\ and\ \citenamefont
  {Poole}(2011)}]{2011MEG}%
  \BibitemOpen
  \bibfield  {author} {\bibinfo {author} {\bibfnamefont {M.~J.}\ \bibnamefont
  {Cuthbertson}}\ and\ \bibinfo {author} {\bibfnamefont {P.~H.}\ \bibnamefont
  {Poole}},\ }\href {https://doi.org/10.1103/PhysRevLett.106.115706} {\bibfield
   {journal} {\bibinfo  {journal} {Phys. Rev. Lett.}\ }\textbf {\bibinfo
  {volume} {106}},\ \bibinfo {pages} {115706} (\bibinfo {year}
  {2011})}\BibitemShut {NoStop}%
\bibitem [{\citenamefont {Gallet}\ and\ \citenamefont
  {Pietrucci}(2013)}]{2013GAL}%
  \BibitemOpen
  \bibfield  {author} {\bibinfo {author} {\bibfnamefont {G.~A.}\ \bibnamefont
  {Gallet}}\ and\ \bibinfo {author} {\bibfnamefont {F.}~\bibnamefont
  {Pietrucci}},\ }\href {https://doi.org/10.1063/1.4818005} {\bibfield
  {journal} {\bibinfo  {journal} {J. Chem. Phys.}\ }\textbf {\bibinfo {volume}
  {139}},\ \bibinfo {pages} {074101} (\bibinfo {year} {2013})}\BibitemShut
  {NoStop}%
\bibitem [{\citenamefont {Russo}\ and\ \citenamefont {Tanaka}(2014)}]{2014RUS}%
  \BibitemOpen
  \bibfield  {author} {\bibinfo {author} {\bibfnamefont {J.}~\bibnamefont
  {Russo}}\ and\ \bibinfo {author} {\bibfnamefont {H.}~\bibnamefont {Tanaka}},\
  }\href {https://doi.org/10.1038/ncomms4556} {\bibfield  {journal} {\bibinfo
  {journal} {Nat. Commun.}\ }\textbf {\bibinfo {volume} {5}},\ \bibinfo {pages}
  {1} (\bibinfo {year} {2014})}\BibitemShut {NoStop}%
\bibitem [{\citenamefont {Sellberg}\ \emph {et~al.}(2014)\citenamefont
  {Sellberg}, \citenamefont {Huang}, \citenamefont {McQueen}, \citenamefont
  {Loh}, \citenamefont {Laksmono}, \citenamefont {Schlesinger}, \citenamefont
  {Sierra}, \citenamefont {Nordlund}, \citenamefont {Hampton}, \citenamefont
  {Starodub}, \citenamefont {DePonte}, \citenamefont {Beye}, \citenamefont
  {Chen}, \citenamefont {Martin}, \citenamefont {Barty}, \citenamefont
  {Wikfeldt}, \citenamefont {Weiss}, \citenamefont {Caronna}, \citenamefont
  {Feldkamp}, \citenamefont {Skinner}, \citenamefont {Seibert}, \citenamefont
  {Messerschmidt}, \citenamefont {Williams}, \citenamefont {Boutet},
  \citenamefont {Pettersson}, \citenamefont {Bogan},\ and\ \citenamefont
  {Nilsson}}]{2014JON}%
  \BibitemOpen
  \bibfield  {author} {\bibinfo {author} {\bibfnamefont {J.}~\bibnamefont
  {Sellberg}}, \bibinfo {author} {\bibfnamefont {C.}~\bibnamefont {Huang}},
  \bibinfo {author} {\bibfnamefont {T.}~\bibnamefont {McQueen}}, \bibinfo
  {author} {\bibfnamefont {N.}~\bibnamefont {Loh}}, \bibinfo {author}
  {\bibfnamefont {H.}~\bibnamefont {Laksmono}}, \bibinfo {author}
  {\bibfnamefont {D.}~\bibnamefont {Schlesinger}}, \bibinfo {author}
  {\bibfnamefont {R.}~\bibnamefont {Sierra}}, \bibinfo {author} {\bibfnamefont
  {D.}~\bibnamefont {Nordlund}}, \bibinfo {author} {\bibfnamefont
  {C.}~\bibnamefont {Hampton}}, \bibinfo {author} {\bibfnamefont
  {D.}~\bibnamefont {Starodub}}, \bibinfo {author} {\bibfnamefont
  {D.}~\bibnamefont {DePonte}}, \bibinfo {author} {\bibfnamefont
  {M.}~\bibnamefont {Beye}}, \bibinfo {author} {\bibfnamefont {C.}~\bibnamefont
  {Chen}}, \bibinfo {author} {\bibfnamefont {A.}~\bibnamefont {Martin}},
  \bibinfo {author} {\bibfnamefont {A.}~\bibnamefont {Barty}}, \bibinfo
  {author} {\bibfnamefont {K.}~\bibnamefont {Wikfeldt}}, \bibinfo {author}
  {\bibfnamefont {T.}~\bibnamefont {Weiss}}, \bibinfo {author} {\bibfnamefont
  {C.}~\bibnamefont {Caronna}}, \bibinfo {author} {\bibfnamefont
  {J.}~\bibnamefont {Feldkamp}}, \bibinfo {author} {\bibfnamefont
  {L.}~\bibnamefont {Skinner}}, \bibinfo {author} {\bibfnamefont
  {M.}~\bibnamefont {Seibert}}, \bibinfo {author} {\bibfnamefont
  {M.}~\bibnamefont {Messerschmidt}}, \bibinfo {author} {\bibfnamefont
  {G.}~\bibnamefont {Williams}}, \bibinfo {author} {\bibfnamefont
  {S.}~\bibnamefont {Boutet}}, \bibinfo {author} {\bibfnamefont
  {L.}~\bibnamefont {Pettersson}}, \bibinfo {author} {\bibfnamefont
  {M.}~\bibnamefont {Bogan}},\ and\ \bibinfo {author} {\bibfnamefont
  {A.}~\bibnamefont {Nilsson}},\ }\href {https://doi.org/10.1038/nature13266}
  {\bibfield  {journal} {\bibinfo  {journal} {Nature}\ }\textbf {\bibinfo
  {volume} {510}},\ \bibinfo {pages} {381} (\bibinfo {year}
  {2014})}\BibitemShut {NoStop}%
\bibitem [{\citenamefont {Hiraoka}\ \emph {et~al.}(2016)\citenamefont
  {Hiraoka}, \citenamefont {Nakamura}, \citenamefont {Hirata}, \citenamefont
  {Escolar}, \citenamefont {Matsue},\ and\ \citenamefont {Nishiura}}]{2016HIR}%
  \BibitemOpen
  \bibfield  {author} {\bibinfo {author} {\bibfnamefont {Y.}~\bibnamefont
  {Hiraoka}}, \bibinfo {author} {\bibfnamefont {T.}~\bibnamefont {Nakamura}},
  \bibinfo {author} {\bibfnamefont {A.}~\bibnamefont {Hirata}}, \bibinfo
  {author} {\bibfnamefont {E.~G.}\ \bibnamefont {Escolar}}, \bibinfo {author}
  {\bibfnamefont {K.}~\bibnamefont {Matsue}},\ and\ \bibinfo {author}
  {\bibfnamefont {Y.}~\bibnamefont {Nishiura}},\ }\href
  {https://doi.org/10.1073/pnas.1520877113} {\bibfield  {journal} {\bibinfo
  {journal} {Proc. Natl. Acad. Sci. U.S.A.}\ }\textbf {\bibinfo {volume}
  {113}},\ \bibinfo {pages} {7035} (\bibinfo {year} {2016})}\BibitemShut
  {NoStop}%
\bibitem [{\citenamefont {Ni}\ and\ \citenamefont
  {Skinner}(2016{\natexlab{a}})}]{2016YIC}%
  \BibitemOpen
  \bibfield  {author} {\bibinfo {author} {\bibfnamefont {Y.}~\bibnamefont
  {Ni}}\ and\ \bibinfo {author} {\bibfnamefont {J.}~\bibnamefont {Skinner}},\
  }\href {https://doi.org/10.1063/1.4952991} {\bibfield  {journal} {\bibinfo
  {journal} {J. Chem. Phys.}\ }\textbf {\bibinfo {volume} {144}},\ \bibinfo
  {pages} {214501} (\bibinfo {year} {2016}{\natexlab{a}})}\BibitemShut
  {NoStop}%
\bibitem [{\citenamefont {Ni}\ and\ \citenamefont
  {Skinner}(2016{\natexlab{b}})}]{2016YIC2}%
  \BibitemOpen
  \bibfield  {author} {\bibinfo {author} {\bibfnamefont {Y.}~\bibnamefont
  {Ni}}\ and\ \bibinfo {author} {\bibfnamefont {J.}~\bibnamefont {Skinner}},\
  }\href {https://doi.org/10.1063/1.4963736} {\bibfield  {journal} {\bibinfo
  {journal} {J. Chem. Phys.}\ }\textbf {\bibinfo {volume} {145}},\ \bibinfo
  {pages} {124509} (\bibinfo {year} {2016}{\natexlab{b}})}\BibitemShut
  {NoStop}%
\bibitem [{\citenamefont {Pathak}\ \emph {et~al.}(2016)\citenamefont {Pathak},
  \citenamefont {Palmer}, \citenamefont {Schlesinger}, \citenamefont
  {Wikfeldt}, \citenamefont {Sellberg}, \citenamefont {Pettersson},\ and\
  \citenamefont {Nilsson}}]{2016PAT}%
  \BibitemOpen
  \bibfield  {author} {\bibinfo {author} {\bibfnamefont {H.}~\bibnamefont
  {Pathak}}, \bibinfo {author} {\bibfnamefont {J.}~\bibnamefont {Palmer}},
  \bibinfo {author} {\bibfnamefont {D.}~\bibnamefont {Schlesinger}}, \bibinfo
  {author} {\bibfnamefont {K.~T.}\ \bibnamefont {Wikfeldt}}, \bibinfo {author}
  {\bibfnamefont {J.~A.}\ \bibnamefont {Sellberg}}, \bibinfo {author}
  {\bibfnamefont {L.~G.}\ \bibnamefont {Pettersson}},\ and\ \bibinfo {author}
  {\bibfnamefont {A.}~\bibnamefont {Nilsson}},\ }\href
  {https://doi.org/10.1063/1.4963913} {\bibfield  {journal} {\bibinfo
  {journal} {J. Chem. Phys.}\ }\textbf {\bibinfo {volume} {145}},\ \bibinfo
  {pages} {134507} (\bibinfo {year} {2016})}\BibitemShut {NoStop}%
\bibitem [{\citenamefont {Takada}(2018)}]{2018TAK}%
  \BibitemOpen
  \bibfield  {author} {\bibinfo {author} {\bibfnamefont {A.}~\bibnamefont
  {Takada}},\ }\href {https://doi.org/10.1016/j.jnoncrysol.2018.07.037}
  {\bibfield  {journal} {\bibinfo  {journal} {J. Non-Cryst. Solids}\ }\textbf
  {\bibinfo {volume} {499}},\ \bibinfo {pages} {309} (\bibinfo {year}
  {2018})}\BibitemShut {NoStop}%
\bibitem [{\citenamefont {Bapst}\ \emph {et~al.}(2020)\citenamefont {Bapst},
  \citenamefont {Keck}, \citenamefont {Grabska-Barwi{\'n}ska}, \citenamefont
  {Donner}, \citenamefont {Cubuk}, \citenamefont {Schoenholz}, \citenamefont
  {Obika}, \citenamefont {Nelson}, \citenamefont {Back}, \citenamefont
  {Hassabis},\ and\ \citenamefont {Kohli}}]{2020VIC}%
  \BibitemOpen
  \bibfield  {author} {\bibinfo {author} {\bibfnamefont {V.}~\bibnamefont
  {Bapst}}, \bibinfo {author} {\bibfnamefont {T.}~\bibnamefont {Keck}},
  \bibinfo {author} {\bibfnamefont {A.}~\bibnamefont {Grabska-Barwi{\'n}ska}},
  \bibinfo {author} {\bibfnamefont {C.}~\bibnamefont {Donner}}, \bibinfo
  {author} {\bibfnamefont {E.~D.}\ \bibnamefont {Cubuk}}, \bibinfo {author}
  {\bibfnamefont {S.~S.}\ \bibnamefont {Schoenholz}}, \bibinfo {author}
  {\bibfnamefont {A.}~\bibnamefont {Obika}}, \bibinfo {author} {\bibfnamefont
  {A.~W.}\ \bibnamefont {Nelson}}, \bibinfo {author} {\bibfnamefont
  {T.}~\bibnamefont {Back}}, \bibinfo {author} {\bibfnamefont {D.}~\bibnamefont
  {Hassabis}},\ and\ \bibinfo {author} {\bibfnamefont {P.}~\bibnamefont
  {Kohli}},\ }\href {https://doi.org/10.1038/s41567-020-0842-8} {\bibfield
  {journal} {\bibinfo  {journal} {Nat. Phys.}\ }\textbf {\bibinfo {volume}
  {16}},\ \bibinfo {pages} {448} (\bibinfo {year} {2020})}\BibitemShut
  {NoStop}%
\bibitem [{\citenamefont {Foffi}\ \emph {et~al.}(2021)\citenamefont {Foffi},
  \citenamefont {Russo},\ and\ \citenamefont {Sciortino}}]{2021FOF}%
  \BibitemOpen
  \bibfield  {author} {\bibinfo {author} {\bibfnamefont {R.}~\bibnamefont
  {Foffi}}, \bibinfo {author} {\bibfnamefont {J.}~\bibnamefont {Russo}},\ and\
  \bibinfo {author} {\bibfnamefont {F.}~\bibnamefont {Sciortino}},\ }\href
  {https://doi.org/10.1063/5.0049299} {\bibfield  {journal} {\bibinfo
  {journal} {J. Chem. Phys.}\ }\textbf {\bibinfo {volume} {154}},\ \bibinfo
  {pages} {184506} (\bibinfo {year} {2021})}\BibitemShut {NoStop}%
\bibitem [{\citenamefont {Wasserman}(2018)}]{2018LAR}%
  \BibitemOpen
  \bibfield  {author} {\bibinfo {author} {\bibfnamefont {L.}~\bibnamefont
  {Wasserman}},\ }\href
  {https://doi.org/10.1146/annurev-statistics-031017-100045} {\bibfield
  {journal} {\bibinfo  {journal} {Annu. Rev. Stat. Appl.}\ }\textbf {\bibinfo
  {volume} {5}},\ \bibinfo {pages} {501} (\bibinfo {year} {2018})}\BibitemShut
  {NoStop}%
\bibitem [{\citenamefont {Chazal}\ and\ \citenamefont
  {Michel}(2021)}]{2021CHA}%
  \BibitemOpen
  \bibfield  {author} {\bibinfo {author} {\bibfnamefont {F.}~\bibnamefont
  {Chazal}}\ and\ \bibinfo {author} {\bibfnamefont {B.}~\bibnamefont
  {Michel}},\ }\href {https://doi.org/10.3389/frai.2021.667963} {\bibfield
  {journal} {\bibinfo  {journal} {Front. Artif. Intell. Appl.}\ }\textbf
  {\bibinfo {volume} {4}},\ \bibinfo {pages} {667963} (\bibinfo {year}
  {2021})}\BibitemShut {NoStop}%
\bibitem [{\citenamefont {Chen}\ \emph {et~al.}(2015)\citenamefont {Chen},
  \citenamefont {Ho}, \citenamefont {Freeman}, \citenamefont {Genovese},\ and\
  \citenamefont {Wasserman}}]{2015CHE}%
  \BibitemOpen
  \bibfield  {author} {\bibinfo {author} {\bibfnamefont {Y.-C.}\ \bibnamefont
  {Chen}}, \bibinfo {author} {\bibfnamefont {S.}~\bibnamefont {Ho}}, \bibinfo
  {author} {\bibfnamefont {P.~E.}\ \bibnamefont {Freeman}}, \bibinfo {author}
  {\bibfnamefont {C.~R.}\ \bibnamefont {Genovese}},\ and\ \bibinfo {author}
  {\bibfnamefont {L.}~\bibnamefont {Wasserman}},\ }\href
  {https://doi.org/10.1093/mnras/stv1996} {\bibfield  {journal} {\bibinfo
  {journal} {Mon. Notices Royal Astron. Soc.}\ }\textbf {\bibinfo {volume}
  {454}},\ \bibinfo {pages} {1140} (\bibinfo {year} {2015})}\BibitemShut
  {NoStop}%
\bibitem [{\citenamefont {Tirelli}\ and\ \citenamefont
  {Costa}(2021)}]{tirelli2021a}%
  \BibitemOpen
  \bibfield  {author} {\bibinfo {author} {\bibfnamefont {A.}~\bibnamefont
  {Tirelli}}\ and\ \bibinfo {author} {\bibfnamefont {N.~C.}\ \bibnamefont
  {Costa}},\ }\href {https://doi.org/10.1103/PhysRevB.104.235146} {\bibfield
  {journal} {\bibinfo  {journal} {Phys. Rev. B}\ }\textbf {\bibinfo {volume}
  {104}},\ \bibinfo {pages} {235146} (\bibinfo {year} {2021})}\BibitemShut
  {NoStop}%
\bibitem [{\citenamefont {Park}\ \emph {et~al.}(2022)\citenamefont {Park},
  \citenamefont {Hwang},\ and\ \citenamefont {Yang}}]{2022SUN}%
  \BibitemOpen
  \bibfield  {author} {\bibinfo {author} {\bibfnamefont {S.}~\bibnamefont
  {Park}}, \bibinfo {author} {\bibfnamefont {Y.}~\bibnamefont {Hwang}},\ and\
  \bibinfo {author} {\bibfnamefont {B.-J.}\ \bibnamefont {Yang}},\ }\href
  {https://doi.org/10.1103/PhysRevB.105.195115} {\bibfield  {journal} {\bibinfo
   {journal} {Phys. Rev. B}\ }\textbf {\bibinfo {volume} {105}},\ \bibinfo
  {pages} {195115} (\bibinfo {year} {2022})}\BibitemShut {NoStop}%
\bibitem [{\citenamefont {Torshin}\ and\ \citenamefont
  {Rudakov}(2020)}]{2020TOR}%
  \BibitemOpen
  \bibfield  {author} {\bibinfo {author} {\bibfnamefont {I.~Y.}\ \bibnamefont
  {Torshin}}\ and\ \bibinfo {author} {\bibfnamefont {K.}~\bibnamefont
  {Rudakov}},\ }\href {https://doi.org/10.1134/S1054661820020157} {\bibfield
  {journal} {\bibinfo  {journal} {Pattern Recognit. Image Anal.}\ }\textbf
  {\bibinfo {volume} {30}},\ \bibinfo {pages} {264} (\bibinfo {year}
  {2020})}\BibitemShut {NoStop}%
\bibitem [{\citenamefont {Broderick}\ \emph {et~al.}(2021)\citenamefont
  {Broderick}, \citenamefont {Dongol}, \citenamefont {Zhang},\ and\
  \citenamefont {Rajan}}]{2021SCO}%
  \BibitemOpen
  \bibfield  {author} {\bibinfo {author} {\bibfnamefont {S.}~\bibnamefont
  {Broderick}}, \bibinfo {author} {\bibfnamefont {R.}~\bibnamefont {Dongol}},
  \bibinfo {author} {\bibfnamefont {T.}~\bibnamefont {Zhang}},\ and\ \bibinfo
  {author} {\bibfnamefont {K.}~\bibnamefont {Rajan}},\ }\href
  {https://doi.org/10.1038/s41598-021-90070-4} {\bibfield  {journal} {\bibinfo
  {journal} {Sci. Rep.}\ }\textbf {\bibinfo {volume} {11}},\ \bibinfo {pages}
  {1} (\bibinfo {year} {2021})}\BibitemShut {NoStop}%
\bibitem [{\citenamefont {Anand}\ \emph {et~al.}(2022)\citenamefont {Anand},
  \citenamefont {Xu}, \citenamefont {Wee}, \citenamefont {Xia},\ and\
  \citenamefont {Sum}}]{2022ANA}%
  \BibitemOpen
  \bibfield  {author} {\bibinfo {author} {\bibfnamefont {D.~V.}\ \bibnamefont
  {Anand}}, \bibinfo {author} {\bibfnamefont {Q.}~\bibnamefont {Xu}}, \bibinfo
  {author} {\bibfnamefont {J.}~\bibnamefont {Wee}}, \bibinfo {author}
  {\bibfnamefont {K.}~\bibnamefont {Xia}},\ and\ \bibinfo {author}
  {\bibfnamefont {T.~C.}\ \bibnamefont {Sum}},\ }\href
  {https://doi.org/10.1038/s41524-022-00883-8} {\bibfield  {journal} {\bibinfo
  {journal} {Npj Comput. Mater.}\ }\textbf {\bibinfo {volume} {8}},\ \bibinfo
  {pages} {1} (\bibinfo {year} {2022})}\BibitemShut {NoStop}%
\bibitem [{\citenamefont {Smith}\ \emph {et~al.}(2021)\citenamefont {Smith},
  \citenamefont {D{\l}otko},\ and\ \citenamefont {Zavala}}]{2021ALE}%
  \BibitemOpen
  \bibfield  {author} {\bibinfo {author} {\bibfnamefont {A.~D.}\ \bibnamefont
  {Smith}}, \bibinfo {author} {\bibfnamefont {P.}~\bibnamefont {D{\l}otko}},\
  and\ \bibinfo {author} {\bibfnamefont {V.~M.}\ \bibnamefont {Zavala}},\
  }\href {https://doi.org/10.1016/j.compchemeng.2020.107202} {\bibfield
  {journal} {\bibinfo  {journal} {Comput. Chem. Eng.}\ }\textbf {\bibinfo
  {volume} {146}},\ \bibinfo {pages} {107202} (\bibinfo {year}
  {2021})}\BibitemShut {NoStop}%
\bibitem [{\citenamefont {Onodera}\ \emph {et~al.}(2019)\citenamefont
  {Onodera}, \citenamefont {Takimoto}, \citenamefont {Hijiya}, \citenamefont
  {Taniguchi}, \citenamefont {Urata}, \citenamefont {Inaba}, \citenamefont
  {Fujita}, \citenamefont {Obayashi}, \citenamefont {Hiraoka},\ and\
  \citenamefont {Kohara}}]{2019ONO}%
  \BibitemOpen
  \bibfield  {author} {\bibinfo {author} {\bibfnamefont {Y.}~\bibnamefont
  {Onodera}}, \bibinfo {author} {\bibfnamefont {Y.}~\bibnamefont {Takimoto}},
  \bibinfo {author} {\bibfnamefont {H.}~\bibnamefont {Hijiya}}, \bibinfo
  {author} {\bibfnamefont {T.}~\bibnamefont {Taniguchi}}, \bibinfo {author}
  {\bibfnamefont {S.}~\bibnamefont {Urata}}, \bibinfo {author} {\bibfnamefont
  {S.}~\bibnamefont {Inaba}}, \bibinfo {author} {\bibfnamefont
  {S.}~\bibnamefont {Fujita}}, \bibinfo {author} {\bibfnamefont
  {I.}~\bibnamefont {Obayashi}}, \bibinfo {author} {\bibfnamefont
  {Y.}~\bibnamefont {Hiraoka}},\ and\ \bibinfo {author} {\bibfnamefont
  {S.}~\bibnamefont {Kohara}},\ }\href
  {https://doi.org/10.1038/s41427-019-0180-4} {\bibfield  {journal} {\bibinfo
  {journal} {NPG Asia Mater.}\ }\textbf {\bibinfo {volume} {11}},\ \bibinfo
  {pages} {1} (\bibinfo {year} {2019})}\BibitemShut {NoStop}%
\bibitem [{\citenamefont {Sørensen}\ \emph {et~al.}(2020)\citenamefont
  {Sørensen}, \citenamefont {Biscio}, \citenamefont {Bauchy}, \citenamefont
  {Fajstrup},\ and\ \citenamefont {Smedskjaer}}]{2020SOR}%
  \BibitemOpen
  \bibfield  {author} {\bibinfo {author} {\bibfnamefont {S.~S.}\ \bibnamefont
  {Sørensen}}, \bibinfo {author} {\bibfnamefont {C.~A.~N.}\ \bibnamefont
  {Biscio}}, \bibinfo {author} {\bibfnamefont {M.}~\bibnamefont {Bauchy}},
  \bibinfo {author} {\bibfnamefont {L.}~\bibnamefont {Fajstrup}},\ and\
  \bibinfo {author} {\bibfnamefont {M.~M.}\ \bibnamefont {Smedskjaer}},\ }\href
  {https://doi.org/10.1126/sciadv.abc2320} {\bibfield  {journal} {\bibinfo
  {journal} {Science Advances}\ }\textbf {\bibinfo {volume} {6}},\ \bibinfo
  {pages} {eabc2320} (\bibinfo {year} {2020})}\BibitemShut {NoStop}%
\bibitem [{\citenamefont {Kusano}\ \emph {et~al.}(2016)\citenamefont {Kusano},
  \citenamefont {Hiraoka},\ and\ \citenamefont {Fukumizu}}]{2016KUS}%
  \BibitemOpen
  \bibfield  {author} {\bibinfo {author} {\bibfnamefont {G.}~\bibnamefont
  {Kusano}}, \bibinfo {author} {\bibfnamefont {Y.}~\bibnamefont {Hiraoka}},\
  and\ \bibinfo {author} {\bibfnamefont {K.}~\bibnamefont {Fukumizu}},\ }in\
  \href@noop {} {\emph {\bibinfo {booktitle} {International conference on
  machine learning}}}\ (\bibinfo {organization} {PMLR},\ \bibinfo {year}
  {2016})\ pp.\ \bibinfo {pages} {2004--2013}\BibitemShut {NoStop}%
\bibitem [{\citenamefont {Hirata}\ \emph {et~al.}(2020)\citenamefont {Hirata},
  \citenamefont {Wada}, \citenamefont {Obayashi},\ and\ \citenamefont
  {Hiraoka}}]{2020HIR}%
  \BibitemOpen
  \bibfield  {author} {\bibinfo {author} {\bibfnamefont {A.}~\bibnamefont
  {Hirata}}, \bibinfo {author} {\bibfnamefont {T.}~\bibnamefont {Wada}},
  \bibinfo {author} {\bibfnamefont {I.}~\bibnamefont {Obayashi}},\ and\
  \bibinfo {author} {\bibfnamefont {Y.}~\bibnamefont {Hiraoka}},\ }\href
  {https://doi.org/10.1038/s43246-020-00100-3} {\bibfield  {journal} {\bibinfo
  {journal} {Commun. Mater.}\ }\textbf {\bibinfo {volume} {1}},\ \bibinfo
  {pages} {1} (\bibinfo {year} {2020})}\BibitemShut {NoStop}%
\bibitem [{\citenamefont {Jimenez}(2008)}]{jimenez2008fuzzy}%
  \BibitemOpen
  \bibfield  {author} {\bibinfo {author} {\bibfnamefont {R.}~\bibnamefont
  {Jimenez}},\ }\href {https://doi.org/10.1007/s00603-007-0155-6} {\bibfield
  {journal} {\bibinfo  {journal} {Rock Mech. Rock. Eng.}\ }\textbf {\bibinfo
  {volume} {41}},\ \bibinfo {pages} {929} (\bibinfo {year} {2008})}\BibitemShut
  {NoStop}%
\bibitem [{\citenamefont {Bezdek}\ \emph {et~al.}(1984)\citenamefont {Bezdek},
  \citenamefont {Ehrlich},\ and\ \citenamefont {Full}}]{bezdek1984fcm}%
  \BibitemOpen
  \bibfield  {author} {\bibinfo {author} {\bibfnamefont {J.~C.}\ \bibnamefont
  {Bezdek}}, \bibinfo {author} {\bibfnamefont {R.}~\bibnamefont {Ehrlich}},\
  and\ \bibinfo {author} {\bibfnamefont {W.}~\bibnamefont {Full}},\ }\href
  {https://doi.org/10.1016/0098-3004(84)90020-7} {\bibfield  {journal}
  {\bibinfo  {journal} {Comput. Geosci.}\ }\textbf {\bibinfo {volume} {10}},\
  \bibinfo {pages} {191} (\bibinfo {year} {1984})}\BibitemShut {NoStop}%
\bibitem [{Note1()}]{Note1}%
  \BibitemOpen
  \bibinfo {note} {{\protect \it One-dimensional topology} refers to all the
  topological invariants that generate the first persistent homology group of
  the point clouds on which we perform TDA. Nontrivial (noncontractible) loops
  are an example of topological invariants at the first homology group
  level.}\BibitemShut {Stop}%
\bibitem [{\citenamefont {Plimpton}(1995)}]{1995Plimpton}%
  \BibitemOpen
  \bibfield  {author} {\bibinfo {author} {\bibfnamefont {S.}~\bibnamefont
  {Plimpton}},\ }\href {https://doi.org/10.1006/jcph.1995.1039} {\bibfield
  {journal} {\bibinfo  {journal} {J. Comput. Phys.}\ }\textbf {\bibinfo
  {volume} {117}},\ \bibinfo {pages} {1} (\bibinfo {year} {1995})}\BibitemShut
  {NoStop}%
\bibitem [{\citenamefont {van Beest}\ \emph {et~al.}(1990)\citenamefont {van
  Beest}, \citenamefont {Kramer},\ and\ \citenamefont {van Santen}}]{1955VAN}%
  \BibitemOpen
  \bibfield  {author} {\bibinfo {author} {\bibfnamefont {B.~W.~H.}\
  \bibnamefont {van Beest}}, \bibinfo {author} {\bibfnamefont {G.~J.}\
  \bibnamefont {Kramer}},\ and\ \bibinfo {author} {\bibfnamefont {R.~A.}\
  \bibnamefont {van Santen}},\ }\href
  {https://doi.org/10.1103/PhysRevLett.64.1955} {\bibfield  {journal} {\bibinfo
   {journal} {Phys. Rev. Lett.}\ }\textbf {\bibinfo {volume} {64}},\ \bibinfo
  {pages} {1955} (\bibinfo {year} {1990})}\BibitemShut {NoStop}%
\bibitem [{\citenamefont {Vollmayr}\ \emph {et~al.}(1996)\citenamefont
  {Vollmayr}, \citenamefont {Kob},\ and\ \citenamefont {Binder}}]{1996VOL}%
  \BibitemOpen
  \bibfield  {author} {\bibinfo {author} {\bibfnamefont {K.}~\bibnamefont
  {Vollmayr}}, \bibinfo {author} {\bibfnamefont {W.}~\bibnamefont {Kob}},\ and\
  \bibinfo {author} {\bibfnamefont {K.}~\bibnamefont {Binder}},\ }\href
  {https://doi.org/10.1103/PhysRevB.54.15808} {\bibfield  {journal} {\bibinfo
  {journal} {Phys. Rev. B}\ }\textbf {\bibinfo {volume} {54}},\ \bibinfo
  {pages} {15808} (\bibinfo {year} {1996})}\BibitemShut {NoStop}%
\bibitem [{\citenamefont {Lane}(2015)}]{2015LAN}%
  \BibitemOpen
  \bibfield  {author} {\bibinfo {author} {\bibfnamefont {J.~M.~D.}\
  \bibnamefont {Lane}},\ }\href {https://doi.org/10.1103/PhysRevE.92.012320}
  {\bibfield  {journal} {\bibinfo  {journal} {Phys. Rev. E}\ }\textbf {\bibinfo
  {volume} {92}},\ \bibinfo {pages} {012320} (\bibinfo {year}
  {2015})}\BibitemShut {NoStop}%
\bibitem [{\citenamefont {Nos{\'e}}(1984)}]{1989NOSE}%
  \BibitemOpen
  \bibfield  {author} {\bibinfo {author} {\bibfnamefont {S.}~\bibnamefont
  {Nos{\'e}}},\ }\href {https://doi.org/10.1063/1.447334} {\bibfield  {journal}
  {\bibinfo  {journal} {J. Chem. Phys.}\ }\textbf {\bibinfo {volume} {81}},\
  \bibinfo {pages} {511} (\bibinfo {year} {1984})}\BibitemShut {NoStop}%
\bibitem [{\citenamefont {Tuckerman}\ \emph {et~al.}(2006)\citenamefont
  {Tuckerman}, \citenamefont {Alejandre}, \citenamefont {L{\'o}pez-Rend{\'o}n},
  \citenamefont {Jochim},\ and\ \citenamefont {Martyna}}]{2006Tuckerman}%
  \BibitemOpen
  \bibfield  {author} {\bibinfo {author} {\bibfnamefont {M.~E.}\ \bibnamefont
  {Tuckerman}}, \bibinfo {author} {\bibfnamefont {J.}~\bibnamefont
  {Alejandre}}, \bibinfo {author} {\bibfnamefont {R.}~\bibnamefont
  {L{\'o}pez-Rend{\'o}n}}, \bibinfo {author} {\bibfnamefont {A.~L.}\
  \bibnamefont {Jochim}},\ and\ \bibinfo {author} {\bibfnamefont {G.~J.}\
  \bibnamefont {Martyna}},\ }\href
  {https://doi.org/10.1088/0305-4470/39/19/S18} {\bibfield  {journal} {\bibinfo
   {journal} {J. Phys. A: Math. Gen.}\ }\textbf {\bibinfo {volume} {39}},\
  \bibinfo {pages} {5629} (\bibinfo {year} {2006})}\BibitemShut {NoStop}%
\bibitem [{\citenamefont {Mart{\'\i}nez}\ \emph {et~al.}(2009)\citenamefont
  {Mart{\'\i}nez}, \citenamefont {Andrade}, \citenamefont {Birgin},\ and\
  \citenamefont {Mart{\'\i}nez}}]{2009Martinez}%
  \BibitemOpen
  \bibfield  {author} {\bibinfo {author} {\bibfnamefont {L.}~\bibnamefont
  {Mart{\'\i}nez}}, \bibinfo {author} {\bibfnamefont {R.}~\bibnamefont
  {Andrade}}, \bibinfo {author} {\bibfnamefont {E.~G.}\ \bibnamefont
  {Birgin}},\ and\ \bibinfo {author} {\bibfnamefont {J.~M.}\ \bibnamefont
  {Mart{\'\i}nez}},\ }\href {https://doi.org/10.1002/jcc.21224} {\bibfield
  {journal} {\bibinfo  {journal} {J. Comput. Chem.}\ }\textbf {\bibinfo
  {volume} {30}},\ \bibinfo {pages} {2157} (\bibinfo {year}
  {2009})}\BibitemShut {NoStop}%
\bibitem [{\citenamefont {Le~Roux}\ and\ \citenamefont {Jund}(2010)}]{2010ROU}%
  \BibitemOpen
  \bibfield  {author} {\bibinfo {author} {\bibfnamefont {S.}~\bibnamefont
  {Le~Roux}}\ and\ \bibinfo {author} {\bibfnamefont {P.}~\bibnamefont {Jund}},\
  }\href {https://doi.org/10.1016/j.commatsci.2010.04.023} {\bibfield
  {journal} {\bibinfo  {journal} {Comput. Mater. Sci.}\ }\textbf {\bibinfo
  {volume} {49}},\ \bibinfo {pages} {70} (\bibinfo {year} {2010})}\BibitemShut
  {NoStop}%
\bibitem [{\citenamefont {King}(1967)}]{1967KIN}%
  \BibitemOpen
  \bibfield  {author} {\bibinfo {author} {\bibfnamefont {S.~V.}\ \bibnamefont
  {King}},\ }\href {https://doi.org/10.1038/2131112a0} {\bibfield  {journal}
  {\bibinfo  {journal} {Nature}\ }\textbf {\bibinfo {volume} {213}},\ \bibinfo
  {pages} {1112} (\bibinfo {year} {1967})}\BibitemShut {NoStop}%
\bibitem [{\citenamefont {Guttman}(1990)}]{1990GUT}%
  \BibitemOpen
  \bibfield  {author} {\bibinfo {author} {\bibfnamefont {L.}~\bibnamefont
  {Guttman}},\ }\href {https://doi.org/10.1016/0022-3093(90)90686-G} {\bibfield
   {journal} {\bibinfo  {journal} {J. Non-Cryst. Solids}\ }\textbf {\bibinfo
  {volume} {116}},\ \bibinfo {pages} {145} (\bibinfo {year}
  {1990})}\BibitemShut {NoStop}%
\bibitem [{\citenamefont {Goetzke}\ and\ \citenamefont
  {Klein}(1991)}]{1991GOE}%
  \BibitemOpen
  \bibfield  {author} {\bibinfo {author} {\bibfnamefont {K.}~\bibnamefont
  {Goetzke}}\ and\ \bibinfo {author} {\bibfnamefont {H.-J.}\ \bibnamefont
  {Klein}},\ }\href {https://doi.org/10.1016/0022-3093(91)90145-V} {\bibfield
  {journal} {\bibinfo  {journal} {J. Non-Cryst. Solids}\ }\textbf {\bibinfo
  {volume} {127}},\ \bibinfo {pages} {215} (\bibinfo {year}
  {1991})}\BibitemShut {NoStop}%
\bibitem [{\citenamefont {Yuan}\ and\ \citenamefont {Cormack}(2002)}]{2002YUA}%
  \BibitemOpen
  \bibfield  {author} {\bibinfo {author} {\bibfnamefont {X.}~\bibnamefont
  {Yuan}}\ and\ \bibinfo {author} {\bibfnamefont {A.}~\bibnamefont {Cormack}},\
  }\href {https://doi.org/10.1016/S0927-0256(01)00256-7} {\bibfield  {journal}
  {\bibinfo  {journal} {Comput. Mater. Sci.}\ }\textbf {\bibinfo {volume}
  {24}},\ \bibinfo {pages} {343} (\bibinfo {year} {2002})}\BibitemShut
  {NoStop}%
\bibitem [{\citenamefont {Wooten}(2002)}]{2002WOO}%
  \BibitemOpen
  \bibfield  {author} {\bibinfo {author} {\bibfnamefont {F.}~\bibnamefont
  {Wooten}},\ }\href {https://doi.org/10.1107/s0108767302006669} {\bibfield
  {journal} {\bibinfo  {journal} {Acta Crystallogr. A}\ }\textbf {\bibinfo
  {volume} {58}},\ \bibinfo {pages} {346} (\bibinfo {year} {2002})}\BibitemShut
  {NoStop}%
\bibitem [{\citenamefont {Tirelli}\ \emph {et~al.}(2022)\citenamefont
  {Tirelli}, \citenamefont {Carvalho}, \citenamefont {Oliveira}, \citenamefont
  {de~Lima}, \citenamefont {Costa},\ and\ \citenamefont {dos
  Santos}}]{tirelli2021b}%
  \BibitemOpen
  \bibfield  {author} {\bibinfo {author} {\bibfnamefont {A.}~\bibnamefont
  {Tirelli}}, \bibinfo {author} {\bibfnamefont {D.~O.}\ \bibnamefont
  {Carvalho}}, \bibinfo {author} {\bibfnamefont {L.~A.}\ \bibnamefont
  {Oliveira}}, \bibinfo {author} {\bibfnamefont {J.~P.}\ \bibnamefont
  {de~Lima}}, \bibinfo {author} {\bibfnamefont {N.~C.}\ \bibnamefont {Costa}},\
  and\ \bibinfo {author} {\bibfnamefont {R.~R.}\ \bibnamefont {dos Santos}},\
  }\href {https://doi.org/10.1140/epjb/s10051-022-00453-3} {\bibfield
  {journal} {\bibinfo  {journal} {Eur. Phys. J. B}\ }\textbf {\bibinfo {volume}
  {95}},\ \bibinfo {pages} {1} (\bibinfo {year} {2022})}\BibitemShut {NoStop}%
\bibitem [{\citenamefont {Carlsson}(2009)}]{Carlsson2009}%
  \BibitemOpen
  \bibfield  {author} {\bibinfo {author} {\bibfnamefont {G.}~\bibnamefont
  {Carlsson}},\ }\href {https://doi.org/10.1090/S0273-0979-09-01249-X}
  {\bibfield  {journal} {\bibinfo  {journal} {Bull. Am. Math. Soc.}\ }\textbf
  {\bibinfo {volume} {46}},\ \bibinfo {pages} {255} (\bibinfo {year}
  {2009})}\BibitemShut {NoStop}%
\bibitem [{\citenamefont {Umeda}(2017)}]{umeda2017}%
  \BibitemOpen
  \bibfield  {author} {\bibinfo {author} {\bibfnamefont {Y.}~\bibnamefont
  {Umeda}},\ }\href {https://doi.org/10.11185/imt.12.228} {\bibfield  {journal}
  {\bibinfo  {journal} {Information and Media Technologies}\ }\textbf {\bibinfo
  {volume} {12}},\ \bibinfo {pages} {228} (\bibinfo {year} {2017})}\BibitemShut
  {NoStop}%
\bibitem [{\citenamefont {Gidea}\ and\ \citenamefont {Katz}(2018)}]{gidea2018}%
  \BibitemOpen
  \bibfield  {author} {\bibinfo {author} {\bibfnamefont {M.}~\bibnamefont
  {Gidea}}\ and\ \bibinfo {author} {\bibfnamefont {Y.}~\bibnamefont {Katz}},\
  }\href {https://doi.org/10.1016/j.physa.2017.09.028} {\bibfield  {journal}
  {\bibinfo  {journal} {Phys. A: Stat. Mech. Appl.}\ }\textbf {\bibinfo
  {volume} {491}},\ \bibinfo {pages} {820} (\bibinfo {year}
  {2018})}\BibitemShut {NoStop}%
\bibitem [{\citenamefont {Bernstein}\ \emph {et~al.}(2020)\citenamefont
  {Bernstein}, \citenamefont {Burnaev}, \citenamefont {Sharaev}, \citenamefont
  {Kondrateva},\ and\ \citenamefont {Kachan}}]{bernstein2020}%
  \BibitemOpen
  \bibfield  {author} {\bibinfo {author} {\bibfnamefont {A.}~\bibnamefont
  {Bernstein}}, \bibinfo {author} {\bibfnamefont {E.}~\bibnamefont {Burnaev}},
  \bibinfo {author} {\bibfnamefont {M.}~\bibnamefont {Sharaev}}, \bibinfo
  {author} {\bibfnamefont {E.}~\bibnamefont {Kondrateva}},\ and\ \bibinfo
  {author} {\bibfnamefont {O.}~\bibnamefont {Kachan}},\ }in\ \href
  {https://doi.org/10.1117/12.2562501} {\emph {\bibinfo {booktitle} {Twelfth
  International Conference on Machine Vision (ICMV 2019)}}},\ Vol.\ \bibinfo
  {volume} {11433}\ (\bibinfo {organization} {International Society for Optics
  and Photonics},\ \bibinfo {year} {2020})\ p.\ \bibinfo {pages}
  {114332H}\BibitemShut {NoStop}%
\bibitem [{Note2()}]{Note2}%
  \BibitemOpen
  \bibinfo {note} {A metric space $X$ is a set endowed with distance function
  $d$, \protect \textit {i.e.} a function $d:X\times X \rightarrow \protect
  \mathbb {R}_{>0}$ satisfying certain properties}\BibitemShut {NoStop}%
\bibitem [{\citenamefont {Kerber}\ \emph {et~al.}(2017)\citenamefont {Kerber},
  \citenamefont {Morozov},\ and\ \citenamefont {Nigmetov}}]{kerber2017}%
  \BibitemOpen
  \bibfield  {author} {\bibinfo {author} {\bibfnamefont {M.}~\bibnamefont
  {Kerber}}, \bibinfo {author} {\bibfnamefont {D.}~\bibnamefont {Morozov}},\
  and\ \bibinfo {author} {\bibfnamefont {A.}~\bibnamefont {Nigmetov}},\
  }\href@noop {} {\bibinfo {title} {Geometry helps to compare persistence
  diagrams}} (\bibinfo {year} {2017})\BibitemShut {NoStop}%
\bibitem [{\citenamefont {Von~Luxburg}(2007)}]{von2007}%
  \BibitemOpen
  \bibfield  {author} {\bibinfo {author} {\bibfnamefont {U.}~\bibnamefont
  {Von~Luxburg}},\ }\href {https://doi.org/10.1007/s11222-007-9033-z}
  {\bibfield  {journal} {\bibinfo  {journal} {Stat. Comput.}\ }\textbf
  {\bibinfo {volume} {17}},\ \bibinfo {pages} {395} (\bibinfo {year}
  {2007})}\BibitemShut {NoStop}%
\bibitem [{\citenamefont {Liu}\ and\ \citenamefont {Han}(2018)}]{liu2018}%
  \BibitemOpen
  \bibfield  {author} {\bibinfo {author} {\bibfnamefont {J.}~\bibnamefont
  {Liu}}\ and\ \bibinfo {author} {\bibfnamefont {J.}~\bibnamefont {Han}},\ }in\
  \href@noop {} {\emph {\bibinfo {booktitle} {Data Clustering}}}\ (\bibinfo
  {publisher} {Chapman and Hall/CRC},\ \bibinfo {year} {2018})\ pp.\ \bibinfo
  {pages} {177--200}\BibitemShut {NoStop}%
\bibitem [{Note3()}]{Note3}%
  \BibitemOpen
  \bibinfo {note} {This is achieved as follows: $k$-means is used as an
  intermediate step in spectral clustering; therefore, we can obtain fuzzy
  spectral clustering by using fuzzy $k$-means in place of the $k$-means
  procedure in the original formulation of spectral clustering}\BibitemShut
  {NoStop}%
\bibitem [{\citenamefont {Vollmayr}\ and\ \citenamefont
  {Kob}(1996)}]{1996VOL2}%
  \BibitemOpen
  \bibfield  {author} {\bibinfo {author} {\bibfnamefont {K.}~\bibnamefont
  {Vollmayr}}\ and\ \bibinfo {author} {\bibfnamefont {W.}~\bibnamefont {Kob}},\
  }\href {https://doi.org/10.1002/bbpc.19961000906} {\bibfield  {journal}
  {\bibinfo  {journal} {Ber. Bunsenges. Phys. Chem.}\ }\textbf {\bibinfo
  {volume} {100}},\ \bibinfo {pages} {1399} (\bibinfo {year}
  {1996})}\BibitemShut {NoStop}%
\bibitem [{Note4()}]{Note4}%
  \BibitemOpen
  \bibinfo {note} {{\protect \leavevmode {\protect \color {black}\protect \rm
  {These values were obtained when the last configuration at each temperature
  was picked up to compute the PDs, as written in Sec.~{\ref {sec:methods}}.
  They were estimated to be 4635~$\pm $~55, 4034~$\pm $~74, and 4310~$\pm
  $~39~K for -Si-Si-, -O-O-, and -Si-O- networks, respectively, by randomly
  choosing a configuration at each temperature.}}}}\BibitemShut {Stop}%
\bibitem [{Note5()}]{Note5}%
  \BibitemOpen
  \bibinfo {note} {{\protect \leavevmode {\protect \color {black}\protect \rm
  {These values were obtained when the last configuration at each temperature
  was picked up to compute the PDs, as written in Sec.~{\ref {sec:methods}}.
  They were estimated to be 2670~$\pm $~91, 1850~$\pm $~55, and 2080~$\pm
  $~38~K for -Si-Si-, -O-O-, and -Si-O- networks, respectively, by randomly
  choosing a configuration at each temperature.}}}}\BibitemShut {Stop}%
\bibitem [{Note6()}]{Note6}%
  \BibitemOpen
  \bibinfo {note} {Shih et al. have used the primitive definition of rings,
  whereas we have used King's definition (Fig.~{\ref {rings}}). As shown in
  Fig.~{\ref {rings-SiO-comparison}}, the same swap is observed in the analysis
  performed using the primitive definition. However, it is at a lower
  temperature compared to the analysis performed using King's
  definition.}\BibitemShut {Stop}%
\bibitem [{\citenamefont {Rino}\ \emph {et~al.}(1993)\citenamefont {Rino},
  \citenamefont {Ebbsj{\"o}}, \citenamefont {Kalia}, \citenamefont {Nakano},\
  and\ \citenamefont {Vashishta}}]{1993RIN}%
  \BibitemOpen
  \bibfield  {author} {\bibinfo {author} {\bibfnamefont {J.~P.}\ \bibnamefont
  {Rino}}, \bibinfo {author} {\bibfnamefont {I.}~\bibnamefont {Ebbsj{\"o}}},
  \bibinfo {author} {\bibfnamefont {R.~K.}\ \bibnamefont {Kalia}}, \bibinfo
  {author} {\bibfnamefont {A.}~\bibnamefont {Nakano}},\ and\ \bibinfo {author}
  {\bibfnamefont {P.}~\bibnamefont {Vashishta}},\ }\href
  {https://doi.org/10.1103/PhysRevB.47.3053} {\bibfield  {journal} {\bibinfo
  {journal} {Phys. Rev. B}\ }\textbf {\bibinfo {volume} {47}},\ \bibinfo
  {pages} {3053} (\bibinfo {year} {1993})}\BibitemShut {NoStop}%
\bibitem [{\citenamefont {Huang}\ and\ \citenamefont
  {Kieffer}(2004)}]{2004HUA}%
  \BibitemOpen
  \bibfield  {author} {\bibinfo {author} {\bibfnamefont {L.}~\bibnamefont
  {Huang}}\ and\ \bibinfo {author} {\bibfnamefont {J.}~\bibnamefont
  {Kieffer}},\ }\href {https://doi.org/10.1103/PhysRevB.69.224203} {\bibfield
  {journal} {\bibinfo  {journal} {Phys. Rev. B}\ }\textbf {\bibinfo {volume}
  {69}},\ \bibinfo {pages} {224203} (\bibinfo {year} {2004})}\BibitemShut
  {NoStop}%
\bibitem [{\citenamefont {Takada}\ \emph {et~al.}(2008)\citenamefont {Takada},
  \citenamefont {Richet}, \citenamefont {Catlow},\ and\ \citenamefont
  {Price}}]{2008TAK2}%
  \BibitemOpen
  \bibfield  {author} {\bibinfo {author} {\bibfnamefont {A.}~\bibnamefont
  {Takada}}, \bibinfo {author} {\bibfnamefont {P.}~\bibnamefont {Richet}},
  \bibinfo {author} {\bibfnamefont {C.}~\bibnamefont {Catlow}},\ and\ \bibinfo
  {author} {\bibfnamefont {G.}~\bibnamefont {Price}},\ }\href
  {https://doi.org/10.1016/j.jnoncrysol.2007.07.062} {\bibfield  {journal}
  {\bibinfo  {journal} {J. Non-Cryst. Solids}\ }\textbf {\bibinfo {volume}
  {354}},\ \bibinfo {pages} {181} (\bibinfo {year} {2008})}\BibitemShut
  {NoStop}%
\bibitem [{\citenamefont {Koziatek}\ \emph {et~al.}(2015)\citenamefont
  {Koziatek}, \citenamefont {Barrat},\ and\ \citenamefont {Rodney}}]{2015KOZ}%
  \BibitemOpen
  \bibfield  {author} {\bibinfo {author} {\bibfnamefont {P.}~\bibnamefont
  {Koziatek}}, \bibinfo {author} {\bibfnamefont {J.}~\bibnamefont {Barrat}},\
  and\ \bibinfo {author} {\bibfnamefont {D.}~\bibnamefont {Rodney}},\ }\href
  {https://doi.org/10.1016/j.jnoncrysol.2015.01.009} {\bibfield  {journal}
  {\bibinfo  {journal} {J. Non-Cryst. Solids}\ }\textbf {\bibinfo {volume}
  {414}},\ \bibinfo {pages} {7} (\bibinfo {year} {2015})}\BibitemShut {NoStop}%
\bibitem [{\citenamefont {Atila}\ \emph {et~al.}(2019)\citenamefont {Atila},
  \citenamefont {Ghardi}, \citenamefont {Hasnaoui},\ and\ \citenamefont
  {Ouaskit}}]{2019ATI}%
  \BibitemOpen
  \bibfield  {author} {\bibinfo {author} {\bibfnamefont {A.}~\bibnamefont
  {Atila}}, \bibinfo {author} {\bibfnamefont {E.~M.}\ \bibnamefont {Ghardi}},
  \bibinfo {author} {\bibfnamefont {A.}~\bibnamefont {Hasnaoui}},\ and\
  \bibinfo {author} {\bibfnamefont {S.}~\bibnamefont {Ouaskit}},\ }\href
  {https://doi.org/10.1016/j.jnoncrysol.2019.119470} {\bibfield  {journal}
  {\bibinfo  {journal} {J. Non-Cryst. Solids}\ }\textbf {\bibinfo {volume}
  {525}},\ \bibinfo {pages} {119470} (\bibinfo {year} {2019})}\BibitemShut
  {NoStop}%
\bibitem [{\citenamefont {Yang}\ \emph {et~al.}(2020)\citenamefont {Yang},
  \citenamefont {Tokunaga}, \citenamefont {Ono}, \citenamefont {Hayashi},\ and\
  \citenamefont {Mauro}}]{2020YAN}%
  \BibitemOpen
  \bibfield  {author} {\bibinfo {author} {\bibfnamefont {Y.}~\bibnamefont
  {Yang}}, \bibinfo {author} {\bibfnamefont {H.}~\bibnamefont {Tokunaga}},
  \bibinfo {author} {\bibfnamefont {M.}~\bibnamefont {Ono}}, \bibinfo {author}
  {\bibfnamefont {K.}~\bibnamefont {Hayashi}},\ and\ \bibinfo {author}
  {\bibfnamefont {J.~C.}\ \bibnamefont {Mauro}},\ }\href
  {https://doi.org/10.1111/jace.17126} {\bibfield  {journal} {\bibinfo
  {journal} {J. Am. Ceram. Soc.}\ }\textbf {\bibinfo {volume} {103}},\ \bibinfo
  {pages} {4256} (\bibinfo {year} {2020})}\BibitemShut {NoStop}%
\bibitem [{\citenamefont {Kohara}\ \emph {et~al.}(2021)\citenamefont {Kohara},
  \citenamefont {Shiga}, \citenamefont {Onodera}, \citenamefont {Masai},
  \citenamefont {Hirata}, \citenamefont {Murakami}, \citenamefont {Morishita},
  \citenamefont {Kimura},\ and\ \citenamefont {Hayashi}}]{2021SHI}%
  \BibitemOpen
  \bibfield  {author} {\bibinfo {author} {\bibfnamefont {S.}~\bibnamefont
  {Kohara}}, \bibinfo {author} {\bibfnamefont {M.}~\bibnamefont {Shiga}},
  \bibinfo {author} {\bibfnamefont {Y.}~\bibnamefont {Onodera}}, \bibinfo
  {author} {\bibfnamefont {H.}~\bibnamefont {Masai}}, \bibinfo {author}
  {\bibfnamefont {A.}~\bibnamefont {Hirata}}, \bibinfo {author} {\bibfnamefont
  {M.}~\bibnamefont {Murakami}}, \bibinfo {author} {\bibfnamefont
  {T.}~\bibnamefont {Morishita}}, \bibinfo {author} {\bibfnamefont
  {K.}~\bibnamefont {Kimura}},\ and\ \bibinfo {author} {\bibfnamefont
  {K.}~\bibnamefont {Hayashi}},\ }\href
  {https://doi.org/10.1038/s41598-021-00965-5} {\bibfield  {journal} {\bibinfo
  {journal} {Sci. Rep.}\ }\textbf {\bibinfo {volume} {11}},\ \bibinfo {pages}
  {1} (\bibinfo {year} {2021})}\BibitemShut {NoStop}%
\bibitem [{\citenamefont {Stixrude}\ and\ \citenamefont
  {Bukowinski}(1990)}]{1990STI}%
  \BibitemOpen
  \bibfield  {author} {\bibinfo {author} {\bibfnamefont {L.}~\bibnamefont
  {Stixrude}}\ and\ \bibinfo {author} {\bibfnamefont {M.}~\bibnamefont
  {Bukowinski}},\ }\href {https://doi.org/10.1126/science.250.4980.541}
  {\bibfield  {journal} {\bibinfo  {journal} {Science}\ }\textbf {\bibinfo
  {volume} {250}},\ \bibinfo {pages} {541} (\bibinfo {year}
  {1990})}\BibitemShut {NoStop}%
\bibitem [{\citenamefont {Trave}\ \emph {et~al.}(2002)\citenamefont {Trave},
  \citenamefont {Tangney}, \citenamefont {Scandolo}, \citenamefont
  {Pasquarello},\ and\ \citenamefont {Car}}]{2002TRA}%
  \BibitemOpen
  \bibfield  {author} {\bibinfo {author} {\bibfnamefont {A.}~\bibnamefont
  {Trave}}, \bibinfo {author} {\bibfnamefont {P.}~\bibnamefont {Tangney}},
  \bibinfo {author} {\bibfnamefont {S.}~\bibnamefont {Scandolo}}, \bibinfo
  {author} {\bibfnamefont {A.}~\bibnamefont {Pasquarello}},\ and\ \bibinfo
  {author} {\bibfnamefont {R.}~\bibnamefont {Car}},\ }\href
  {https://doi.org/10.1103/PhysRevLett.89.245504} {\bibfield  {journal}
  {\bibinfo  {journal} {Phys. Rev. Lett.}\ }\textbf {\bibinfo {volume} {89}},\
  \bibinfo {pages} {245504} (\bibinfo {year} {2002})}\BibitemShut {NoStop}%
\bibitem [{\citenamefont {Jabes}\ \emph {et~al.}(2012)\citenamefont {Jabes},
  \citenamefont {Nayar}, \citenamefont {Dhabal}, \citenamefont {Molinero},\
  and\ \citenamefont {Chakravarty}}]{2012SHA}%
  \BibitemOpen
  \bibfield  {author} {\bibinfo {author} {\bibfnamefont {B.~S.}\ \bibnamefont
  {Jabes}}, \bibinfo {author} {\bibfnamefont {D.}~\bibnamefont {Nayar}},
  \bibinfo {author} {\bibfnamefont {D.}~\bibnamefont {Dhabal}}, \bibinfo
  {author} {\bibfnamefont {V.}~\bibnamefont {Molinero}},\ and\ \bibinfo
  {author} {\bibfnamefont {C.}~\bibnamefont {Chakravarty}},\ }\href
  {https://doi.org/10.1088/0953-8984/24/28/284116} {\bibfield  {journal}
  {\bibinfo  {journal} {J. Phys. Condens. Matter}\ }\textbf {\bibinfo {volume}
  {24}},\ \bibinfo {pages} {284116} (\bibinfo {year} {2012})}\BibitemShut
  {NoStop}%
\bibitem [{\citenamefont {Zhou}\ \emph {et~al.}(2021)\citenamefont {Zhou},
  \citenamefont {Shi}, \citenamefont {Deng}, \citenamefont {Neuefeind},\ and\
  \citenamefont {Bauchy}}]{2021ZHO}%
  \BibitemOpen
  \bibfield  {author} {\bibinfo {author} {\bibfnamefont {Q.}~\bibnamefont
  {Zhou}}, \bibinfo {author} {\bibfnamefont {Y.}~\bibnamefont {Shi}}, \bibinfo
  {author} {\bibfnamefont {B.}~\bibnamefont {Deng}}, \bibinfo {author}
  {\bibfnamefont {J.}~\bibnamefont {Neuefeind}},\ and\ \bibinfo {author}
  {\bibfnamefont {M.}~\bibnamefont {Bauchy}},\ }\href
  {https://doi.org/10.1126/sciadv.abh1761} {\bibfield  {journal} {\bibinfo
  {journal} {Sci. Adv.}\ }\textbf {\bibinfo {volume} {7}},\ \bibinfo {pages}
  {eabh1761} (\bibinfo {year} {2021})}\BibitemShut {NoStop}%
\bibitem [{Note7()}]{Note7}%
  \BibitemOpen
  \bibinfo {note} {The ring analysis (Fig.~{\ref {rings-SiO-comparison}})
  performed in this work shows the same tendency. 6-member rings are the most
  dominant for King's and the primitive definitions, whereas 5-member rings are
  the most dominant for Guttman's definition.}\BibitemShut {Stop}%
\bibitem [{\citenamefont {Schieber}\ \emph {et~al.}(2017)\citenamefont
  {Schieber}, \citenamefont {Carpi}, \citenamefont {D{\'\i}az-Guilera},
  \citenamefont {Pardalos}, \citenamefont {Masoller},\ and\ \citenamefont
  {Ravetti}}]{2017SCH}%
  \BibitemOpen
  \bibfield  {author} {\bibinfo {author} {\bibfnamefont {T.~A.}\ \bibnamefont
  {Schieber}}, \bibinfo {author} {\bibfnamefont {L.}~\bibnamefont {Carpi}},
  \bibinfo {author} {\bibfnamefont {A.}~\bibnamefont {D{\'\i}az-Guilera}},
  \bibinfo {author} {\bibfnamefont {P.~M.}\ \bibnamefont {Pardalos}}, \bibinfo
  {author} {\bibfnamefont {C.}~\bibnamefont {Masoller}},\ and\ \bibinfo
  {author} {\bibfnamefont {M.~G.}\ \bibnamefont {Ravetti}},\ }\href
  {https://doi.org/10.1038/ncomms13928} {\bibfield  {journal} {\bibinfo
  {journal} {Nat. Commun.}\ }\textbf {\bibinfo {volume} {8}},\ \bibinfo {pages}
  {1} (\bibinfo {year} {2017})}\BibitemShut {NoStop}%
\bibitem [{\citenamefont {Tsuneyuki}\ \emph {et~al.}(1988)\citenamefont
  {Tsuneyuki}, \citenamefont {Tsukada}, \citenamefont {Aoki},\ and\
  \citenamefont {Matsui}}]{1988TSU}%
  \BibitemOpen
  \bibfield  {author} {\bibinfo {author} {\bibfnamefont {S.}~\bibnamefont
  {Tsuneyuki}}, \bibinfo {author} {\bibfnamefont {M.}~\bibnamefont {Tsukada}},
  \bibinfo {author} {\bibfnamefont {H.}~\bibnamefont {Aoki}},\ and\ \bibinfo
  {author} {\bibfnamefont {Y.}~\bibnamefont {Matsui}},\ }\href
  {https://doi.org/10.1103/PhysRevLett.61.869} {\bibfield  {journal} {\bibinfo
  {journal} {Phys. Rev. Lett.}\ }\textbf {\bibinfo {volume} {61}},\ \bibinfo
  {pages} {869} (\bibinfo {year} {1988})}\BibitemShut {NoStop}%
\bibitem [{\citenamefont {Soules}(1990)}]{1990SOU}%
  \BibitemOpen
  \bibfield  {author} {\bibinfo {author} {\bibfnamefont {T.~F.}\ \bibnamefont
  {Soules}},\ }\href {https://doi.org/10.1016/0022-3093(90)90773-F} {\bibfield
  {journal} {\bibinfo  {journal} {J. Non. Cryst. Solids}\ }\textbf {\bibinfo
  {volume} {123}},\ \bibinfo {pages} {48} (\bibinfo {year} {1990})}\BibitemShut
  {NoStop}%
\bibitem [{\citenamefont {Takada}\ \emph {et~al.}(2004)\citenamefont {Takada},
  \citenamefont {Richet}, \citenamefont {Catlow},\ and\ \citenamefont
  {Price}}]{2004TAK}%
  \BibitemOpen
  \bibfield  {author} {\bibinfo {author} {\bibfnamefont {A.}~\bibnamefont
  {Takada}}, \bibinfo {author} {\bibfnamefont {P.}~\bibnamefont {Richet}},
  \bibinfo {author} {\bibfnamefont {C.}~\bibnamefont {Catlow}},\ and\ \bibinfo
  {author} {\bibfnamefont {G.}~\bibnamefont {Price}},\ }\href
  {https://doi.org/10.1016/j.jnoncrysol.2004.08.247} {\bibfield  {journal}
  {\bibinfo  {journal} {J. Non. Cryst. Solids}\ }\textbf {\bibinfo {volume}
  {345}},\ \bibinfo {pages} {224} (\bibinfo {year} {2004})}\BibitemShut
  {NoStop}%
\bibitem [{\citenamefont {Carre}\ \emph {et~al.}(2008)\citenamefont {Carre},
  \citenamefont {Horbach}, \citenamefont {Ispas},\ and\ \citenamefont
  {Kob}}]{2008CAR}%
  \BibitemOpen
  \bibfield  {author} {\bibinfo {author} {\bibfnamefont {A.}~\bibnamefont
  {Carre}}, \bibinfo {author} {\bibfnamefont {J.}~\bibnamefont {Horbach}},
  \bibinfo {author} {\bibfnamefont {S.}~\bibnamefont {Ispas}},\ and\ \bibinfo
  {author} {\bibfnamefont {W.}~\bibnamefont {Kob}},\ }\href
  {https://doi.org/10.1209/0295-5075/82/17001} {\bibfield  {journal} {\bibinfo
  {journal} {EPL (Europhysics Letters)}\ }\textbf {\bibinfo {volume} {82}},\
  \bibinfo {pages} {17001} (\bibinfo {year} {2008})}\BibitemShut {NoStop}%
\bibitem [{\citenamefont {Sundararaman}\ \emph {et~al.}(2018)\citenamefont
  {Sundararaman}, \citenamefont {Huang}, \citenamefont {Ispas},\ and\
  \citenamefont {Kob}}]{2018SUN}%
  \BibitemOpen
  \bibfield  {author} {\bibinfo {author} {\bibfnamefont {S.}~\bibnamefont
  {Sundararaman}}, \bibinfo {author} {\bibfnamefont {L.}~\bibnamefont {Huang}},
  \bibinfo {author} {\bibfnamefont {S.}~\bibnamefont {Ispas}},\ and\ \bibinfo
  {author} {\bibfnamefont {W.}~\bibnamefont {Kob}},\ }\href
  {https://doi.org/10.1063/1.5023707} {\bibfield  {journal} {\bibinfo
  {journal} {J. Chem. Phys.}\ }\textbf {\bibinfo {volume} {148}},\ \bibinfo
  {pages} {194504} (\bibinfo {year} {2018})}\BibitemShut {NoStop}%
\bibitem [{\citenamefont {Balyakin}\ \emph {et~al.}(2020)\citenamefont
  {Balyakin}, \citenamefont {Rempel}, \citenamefont {Ryltsev},\ and\
  \citenamefont {Rempel}}]{2020BAL}%
  \BibitemOpen
  \bibfield  {author} {\bibinfo {author} {\bibfnamefont {I.~A.}\ \bibnamefont
  {Balyakin}}, \bibinfo {author} {\bibfnamefont {S.~V.}\ \bibnamefont
  {Rempel}}, \bibinfo {author} {\bibfnamefont {R.~E.}\ \bibnamefont
  {Ryltsev}},\ and\ \bibinfo {author} {\bibfnamefont {A.~A.}\ \bibnamefont
  {Rempel}},\ }\href {https://doi.org/10.1103/PhysRevE.102.052125} {\bibfield
  {journal} {\bibinfo  {journal} {Phys. Rev. E}\ }\textbf {\bibinfo {volume}
  {102}},\ \bibinfo {pages} {052125} (\bibinfo {year} {2020})}\BibitemShut
  {NoStop}%
\bibitem [{\citenamefont {Kobayashi}\ \emph {et~al.}(2021)\citenamefont
  {Kobayashi}, \citenamefont {Nagai}, \citenamefont {Itakura},\ and\
  \citenamefont {Shiga}}]{2021KOB}%
  \BibitemOpen
  \bibfield  {author} {\bibinfo {author} {\bibfnamefont {K.}~\bibnamefont
  {Kobayashi}}, \bibinfo {author} {\bibfnamefont {Y.}~\bibnamefont {Nagai}},
  \bibinfo {author} {\bibfnamefont {M.}~\bibnamefont {Itakura}},\ and\ \bibinfo
  {author} {\bibfnamefont {M.}~\bibnamefont {Shiga}},\ }\href
  {https://doi.org/10.1063/5.0055341} {\bibfield  {journal} {\bibinfo
  {journal} {J. Chem. Phys.}\ }\textbf {\bibinfo {volume} {155}},\ \bibinfo
  {pages} {034106} (\bibinfo {year} {2021})}\BibitemShut {NoStop}%
\bibitem [{\citenamefont {Urata}\ \emph
  {et~al.}(2021{\natexlab{a}})\citenamefont {Urata}, \citenamefont {Nakamura},
  \citenamefont {Aiba}, \citenamefont {Tada},\ and\ \citenamefont
  {Hosono}}]{2021URAa}%
  \BibitemOpen
  \bibfield  {author} {\bibinfo {author} {\bibfnamefont {S.}~\bibnamefont
  {Urata}}, \bibinfo {author} {\bibfnamefont {N.}~\bibnamefont {Nakamura}},
  \bibinfo {author} {\bibfnamefont {K.}~\bibnamefont {Aiba}}, \bibinfo {author}
  {\bibfnamefont {T.}~\bibnamefont {Tada}},\ and\ \bibinfo {author}
  {\bibfnamefont {H.}~\bibnamefont {Hosono}},\ }\href
  {https://doi.org/10.1016/j.matdes.2020.109210} {\bibfield  {journal}
  {\bibinfo  {journal} {Mater. Des.}\ }\textbf {\bibinfo {volume} {197}},\
  \bibinfo {pages} {109210} (\bibinfo {year} {2021}{\natexlab{a}})}\BibitemShut
  {NoStop}%
\bibitem [{\citenamefont {Urata}\ \emph
  {et~al.}(2021{\natexlab{b}})\citenamefont {Urata}, \citenamefont {Nakamura},
  \citenamefont {Tada},\ and\ \citenamefont {Hosono}}]{2021URAb}%
  \BibitemOpen
  \bibfield  {author} {\bibinfo {author} {\bibfnamefont {S.}~\bibnamefont
  {Urata}}, \bibinfo {author} {\bibfnamefont {N.}~\bibnamefont {Nakamura}},
  \bibinfo {author} {\bibfnamefont {T.}~\bibnamefont {Tada}},\ and\ \bibinfo
  {author} {\bibfnamefont {H.}~\bibnamefont {Hosono}},\ }\href
  {https://doi.org/10.1111/jace.17774} {\bibfield  {journal} {\bibinfo
  {journal} {J. Am. Ceram. Soc.}\ }\textbf {\bibinfo {volume} {104}},\ \bibinfo
  {pages} {5001} (\bibinfo {year} {2021}{\natexlab{b}})}\BibitemShut {NoStop}%
\bibitem [{\citenamefont {Erhard}\ \emph {et~al.}(2022)\citenamefont {Erhard},
  \citenamefont {Rohrer}, \citenamefont {Albe},\ and\ \citenamefont
  {Deringer}}]{2022ERH}%
  \BibitemOpen
  \bibfield  {author} {\bibinfo {author} {\bibfnamefont {L.~C.}\ \bibnamefont
  {Erhard}}, \bibinfo {author} {\bibfnamefont {J.}~\bibnamefont {Rohrer}},
  \bibinfo {author} {\bibfnamefont {K.}~\bibnamefont {Albe}},\ and\ \bibinfo
  {author} {\bibfnamefont {V.~L.}\ \bibnamefont {Deringer}},\ }\href
  {https://doi.org/10.1038/s41524-022-00768-w} {\bibfield  {journal} {\bibinfo
  {journal} {Npj Comput. Mater.}\ }\textbf {\bibinfo {volume} {8}},\ \bibinfo
  {pages} {1} (\bibinfo {year} {2022})}\BibitemShut {NoStop}%
\bibitem [{\citenamefont {Mei}\ \emph {et~al.}(2007)\citenamefont {Mei},
  \citenamefont {Benmore},\ and\ \citenamefont {Weber}}]{2007MEI}%
  \BibitemOpen
  \bibfield  {author} {\bibinfo {author} {\bibfnamefont {Q.}~\bibnamefont
  {Mei}}, \bibinfo {author} {\bibfnamefont {C.~J.}\ \bibnamefont {Benmore}},\
  and\ \bibinfo {author} {\bibfnamefont {J.~K.~R.}\ \bibnamefont {Weber}},\
  }\href {https://doi.org/10.1103/PhysRevLett.98.057802} {\bibfield  {journal}
  {\bibinfo  {journal} {Phys. Rev. Lett.}\ }\textbf {\bibinfo {volume} {98}},\
  \bibinfo {pages} {057802} (\bibinfo {year} {2007})}\BibitemShut {NoStop}%
\bibitem [{\citenamefont {Mei}\ \emph {et~al.}(2008)\citenamefont {Mei},
  \citenamefont {Benmore}, \citenamefont {Sen}, \citenamefont {Sharma},\ and\
  \citenamefont {Yarger}}]{2008MEI}%
  \BibitemOpen
  \bibfield  {author} {\bibinfo {author} {\bibfnamefont {Q.}~\bibnamefont
  {Mei}}, \bibinfo {author} {\bibfnamefont {C.~J.}\ \bibnamefont {Benmore}},
  \bibinfo {author} {\bibfnamefont {S.}~\bibnamefont {Sen}}, \bibinfo {author}
  {\bibfnamefont {R.}~\bibnamefont {Sharma}},\ and\ \bibinfo {author}
  {\bibfnamefont {J.~L.}\ \bibnamefont {Yarger}},\ }\href
  {https://doi.org/10.1103/PhysRevB.78.144204} {\bibfield  {journal} {\bibinfo
  {journal} {Phys. Rev. B}\ }\textbf {\bibinfo {volume} {78}},\ \bibinfo
  {pages} {144204} (\bibinfo {year} {2008})}\BibitemShut {NoStop}%
\bibitem [{\citenamefont {Pandey}\ \emph {et~al.}(2015)\citenamefont {Pandey},
  \citenamefont {Biswas},\ and\ \citenamefont {Drabold}}]{2015PAN}%
  \BibitemOpen
  \bibfield  {author} {\bibinfo {author} {\bibfnamefont {A.}~\bibnamefont
  {Pandey}}, \bibinfo {author} {\bibfnamefont {P.}~\bibnamefont {Biswas}},\
  and\ \bibinfo {author} {\bibfnamefont {D.~A.}\ \bibnamefont {Drabold}},\
  }\href {https://doi.org/10.1103/PhysRevB.92.155205} {\bibfield  {journal}
  {\bibinfo  {journal} {Phys. Rev. B}\ }\textbf {\bibinfo {volume} {92}},\
  \bibinfo {pages} {155205} (\bibinfo {year} {2015})}\BibitemShut {NoStop}%
\bibitem [{\citenamefont {Limbu}\ \emph {et~al.}(2018)\citenamefont {Limbu},
  \citenamefont {Atta-Fynn}, \citenamefont {Drabold}, \citenamefont {Elliott},\
  and\ \citenamefont {Biswas}}]{2018LIM}%
  \BibitemOpen
  \bibfield  {author} {\bibinfo {author} {\bibfnamefont {D.~K.}\ \bibnamefont
  {Limbu}}, \bibinfo {author} {\bibfnamefont {R.}~\bibnamefont {Atta-Fynn}},
  \bibinfo {author} {\bibfnamefont {D.~A.}\ \bibnamefont {Drabold}}, \bibinfo
  {author} {\bibfnamefont {S.~R.}\ \bibnamefont {Elliott}},\ and\ \bibinfo
  {author} {\bibfnamefont {P.}~\bibnamefont {Biswas}},\ }\href
  {https://doi.org/10.1103/PhysRevMaterials.2.115602} {\bibfield  {journal}
  {\bibinfo  {journal} {Phys. Rev. Materials}\ }\textbf {\bibinfo {volume}
  {2}},\ \bibinfo {pages} {115602} (\bibinfo {year} {2018})}\BibitemShut
  {NoStop}%
\bibitem [{\citenamefont {Zhou}\ \emph {et~al.}(2020)\citenamefont {Zhou},
  \citenamefont {Du}, \citenamefont {Guo}, \citenamefont {Smedskjaer},\ and\
  \citenamefont {Bauchy}}]{2020ZHO}%
  \BibitemOpen
  \bibfield  {author} {\bibinfo {author} {\bibfnamefont {Q.}~\bibnamefont
  {Zhou}}, \bibinfo {author} {\bibfnamefont {T.}~\bibnamefont {Du}}, \bibinfo
  {author} {\bibfnamefont {L.}~\bibnamefont {Guo}}, \bibinfo {author}
  {\bibfnamefont {M.~M.}\ \bibnamefont {Smedskjaer}},\ and\ \bibinfo {author}
  {\bibfnamefont {M.}~\bibnamefont {Bauchy}},\ }\href
  {https://doi.org/10.1016/j.jnoncrysol.2020.120006} {\bibfield  {journal}
  {\bibinfo  {journal} {J. Non. Cryst. Solids}\ }\textbf {\bibinfo {volume}
  {536}},\ \bibinfo {pages} {120006} (\bibinfo {year} {2020})}\BibitemShut
  {NoStop}%
\end{thebibliography}%


\begin{thebibliography}{3}%
\makeatletter
\providecommand \@ifxundefined [1]{%
 \@ifx{#1\undefined}
}%
\providecommand \@ifnum [1]{%
 \ifnum #1\expandafter \@firstoftwo
 \else \expandafter \@secondoftwo
 \fi
}%
\providecommand \@ifx [1]{%
 \ifx #1\expandafter \@firstoftwo
 \else \expandafter \@secondoftwo
 \fi
}%
\providecommand \natexlab [1]{#1}%
\providecommand \enquote  [1]{``#1''}%
\providecommand \bibnamefont  [1]{#1}%
\providecommand \bibfnamefont [1]{#1}%
\providecommand \citenamefont [1]{#1}%
\providecommand \href@noop [0]{\@secondoftwo}%
\providecommand \href [0]{\begingroup \@sanitize@url \@href}%
\providecommand \@href[1]{\@@startlink{#1}\@@href}%
\providecommand \@@href[1]{\endgroup#1\@@endlink}%
\providecommand \@sanitize@url [0]{\catcode `\\12\catcode `\$12\catcode
  `\&12\catcode `\#12\catcode `\^12\catcode `\_12\catcode `\%12\relax}%
\providecommand \@@startlink[1]{}%
\providecommand \@@endlink[0]{}%
\providecommand \url  [0]{\begingroup\@sanitize@url \@url }%
\providecommand \@url [1]{\endgroup\@href {#1}{\urlprefix }}%
\providecommand \urlprefix  [0]{URL }%
\providecommand \Eprint [0]{\href }%
\providecommand \doibase [0]{https://doi.org/}%
\providecommand \selectlanguage [0]{\@gobble}%
\providecommand \bibinfo  [0]{\@secondoftwo}%
\providecommand \bibfield  [0]{\@secondoftwo}%
\providecommand \translation [1]{[#1]}%
\providecommand \BibitemOpen [0]{}%
\providecommand \bibitemStop [0]{}%
\providecommand \bibitemNoStop [0]{.\EOS\space}%
\providecommand \EOS [0]{\spacefactor3000\relax}%
\providecommand \BibitemShut  [1]{\csname bibitem#1\endcsname}%
\let\auto@bib@innerbib\@empty
\bibitem [{\citenamefont {Le~Roux}\ and\ \citenamefont {Jund}(2010)}]{2010ROU}%
  \BibitemOpen
  \bibfield  {author} {\bibinfo {author} {\bibfnamefont {S.}~\bibnamefont
  {Le~Roux}}\ and\ \bibinfo {author} {\bibfnamefont {P.}~\bibnamefont {Jund}},\
  }\href {https://doi.org/10.1016/j.commatsci.2010.04.023} {\bibfield
  {journal} {\bibinfo  {journal} {Comput. Mater. Sci.}\ }\textbf {\bibinfo
  {volume} {49}},\ \bibinfo {pages} {70} (\bibinfo {year} {2010})}\BibitemShut
  {NoStop}%
\bibitem [{\citenamefont {Chau}\ and\ \citenamefont
  {Hardwick}(1998)}]{1998CHA}%
  \BibitemOpen
  \bibfield  {author} {\bibinfo {author} {\bibfnamefont {P.-L.}\ \bibnamefont
  {Chau}}\ and\ \bibinfo {author} {\bibfnamefont {A.}~\bibnamefont
  {Hardwick}},\ }\href {https://doi.org/10.1080/002689798169195} {\bibfield
  {journal} {\bibinfo  {journal} {Mol. Phys.}\ }\textbf {\bibinfo {volume}
  {93}},\ \bibinfo {pages} {511} (\bibinfo {year} {1998})}\BibitemShut
  {NoStop}%
\bibitem [{\citenamefont {Errington}\ and\ \citenamefont
  {Debenedetti}(2001)}]{2001JEF}%
  \BibitemOpen
  \bibfield  {author} {\bibinfo {author} {\bibfnamefont {J.~R.}\ \bibnamefont
  {Errington}}\ and\ \bibinfo {author} {\bibfnamefont {P.~G.}\ \bibnamefont
  {Debenedetti}},\ }\href {https://doi.org/10.1038/35053024} {\bibfield
  {journal} {\bibinfo  {journal} {Nature}\ }\textbf {\bibinfo {volume} {409}},\
  \bibinfo {pages} {318} (\bibinfo {year} {2001})}\BibitemShut {NoStop}%
\end{thebibliography}%
\end{document}


\title{Supplementary Information for Topological data analysis for revealing structural origin of density anomalies in silica glass}

\author{Andrea Tirelli} 
\affiliation{International School for Advanced Studies (SISSA), Via Bonomea 265, 34136 Trieste, Italy}

\author{Kousuke Nakano} 
\email{kousuke\_1123@icloud.com}
\affiliation{International School for Advanced Studies (SISSA), Via Bonomea 265, 34136 Trieste, Italy}
\affiliation{School of Information Science, JAIST, Asahidai 1-1, Nomi, Ishikawa 923-1292, Japan}

\makeatletter
\renewcommand{\refname}{}
\renewcommand*{\citenumfont}[1]{#1}
\renewcommand*{\bibnumfmt}[1]{[#1]}
\makeatother

\setcounter{table}{0}
\setcounter{equation}{0}
\setcounter{figure}{0}
\renewcommand{\thetable}{S-\Roman{table}}
\renewcommand{\thefigure}{S-\arabic{figure}}
\renewcommand{\theequation}{S-\arabic{equation}}

\date{\today}
\maketitle
\begin{figure}[htbp]
  \centering
  \includegraphics[width=0.50\columnwidth]{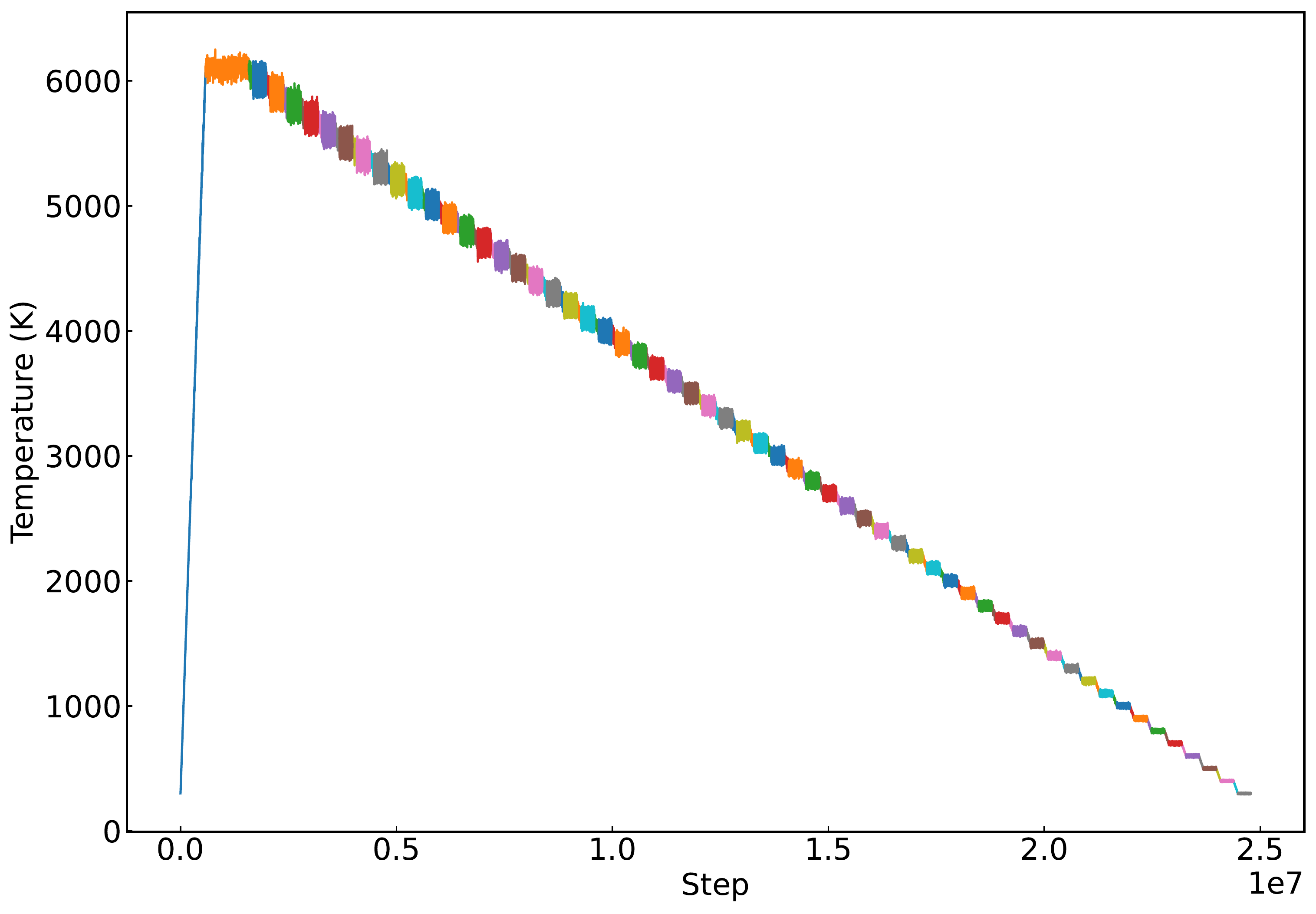}
  \caption{
  Temperature profile during cooling.
  }
  \label{temperature}
\end{figure}
\begin{figure*}[htbp]
  \centering
  \includegraphics[width=\columnwidth]{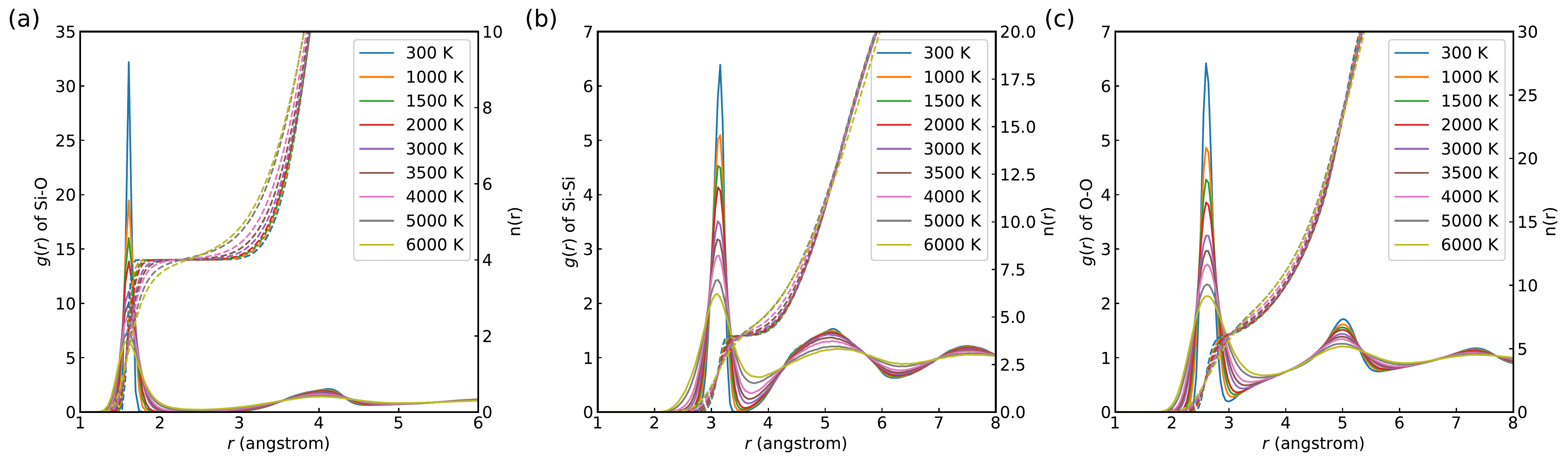}
  \caption{
  Averaged $g(r)$ for (a) Si-O, (b) Si-Si, and (c) O-O.
  }
  \label{gr}
\end{figure*}
\begin{figure*}[htbp]
  \centering
  \includegraphics[width=\columnwidth]{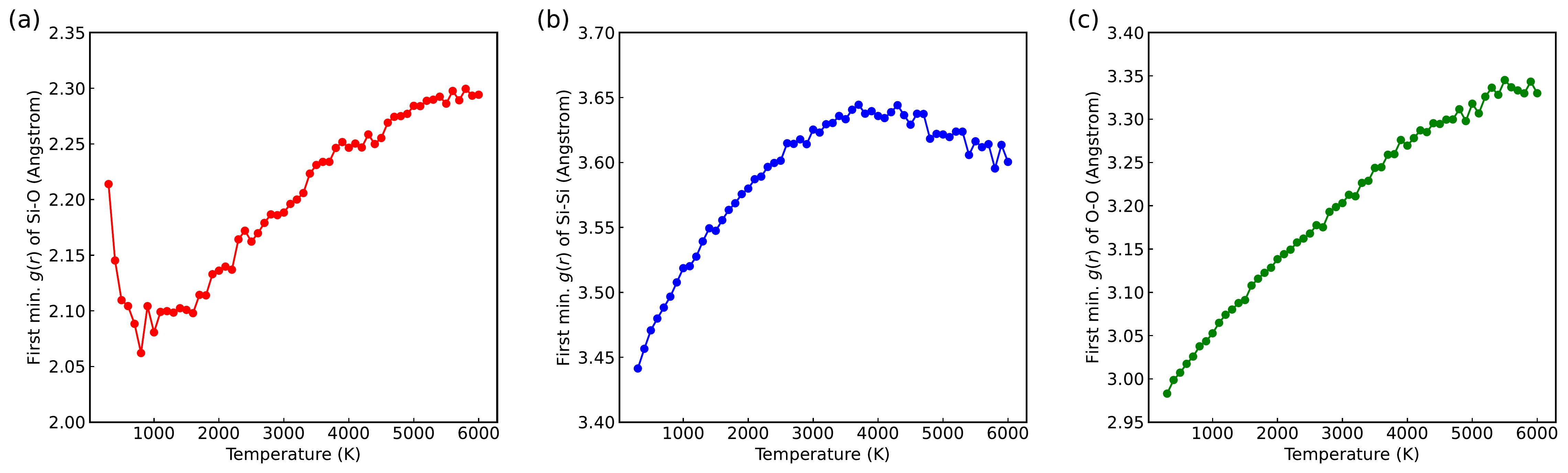}
  \caption{
  Averaged first minimum, $r$, of $g(r)$ for (a) Si-O, (b) Si-Si, and (c) O-O.
  }
  \label{gr-min}
\end{figure*}
\begin{figure*}[htbp]
  \centering
  \includegraphics[width=0.8\columnwidth]{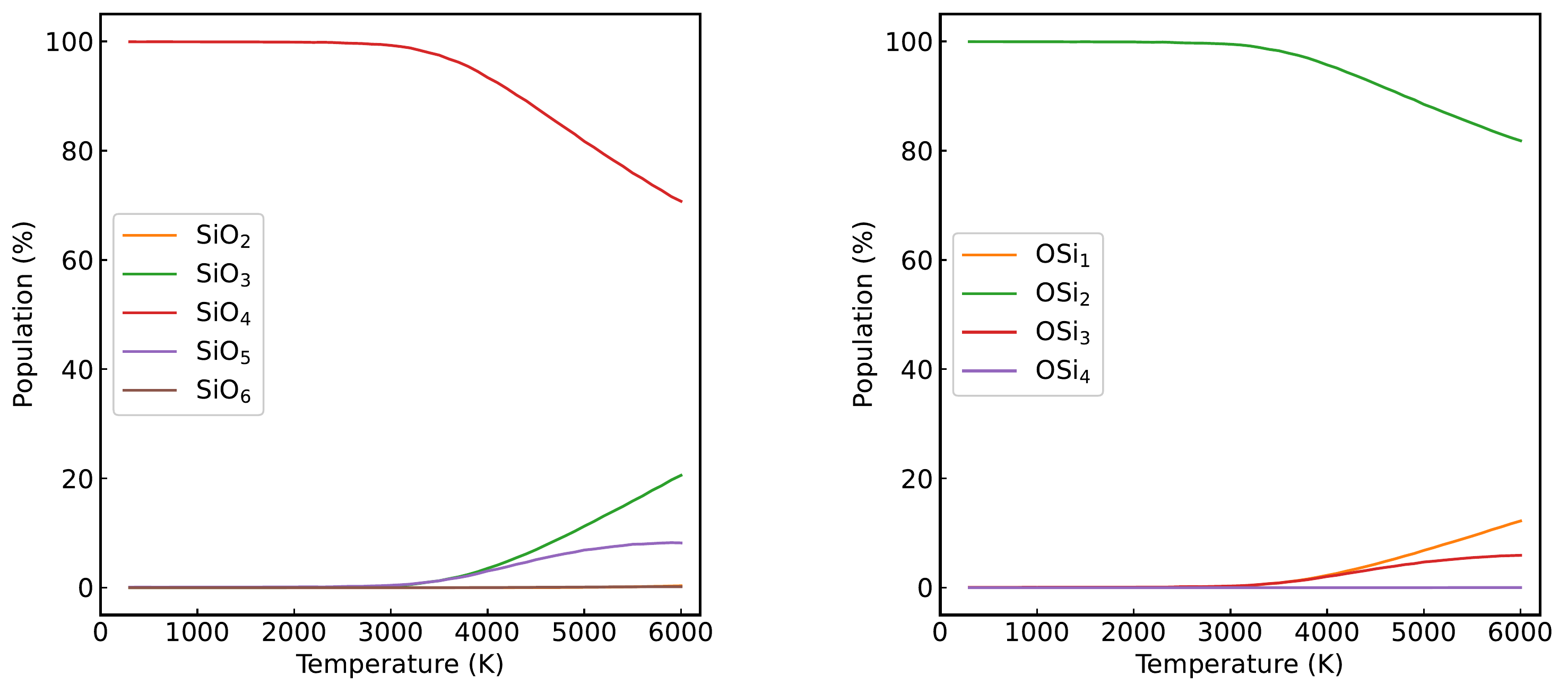}
  \caption{\red{Populations of Si (left) and oxygen (right) atoms with different coordination numbers in the silica as a function of temperature. The cutoff distance is 2.21~$\AA$.}
  }
  \label{population-sio2}
\end{figure*}
\begin{figure*}[htbp]
  \centering
  \includegraphics[width=0.8\columnwidth]{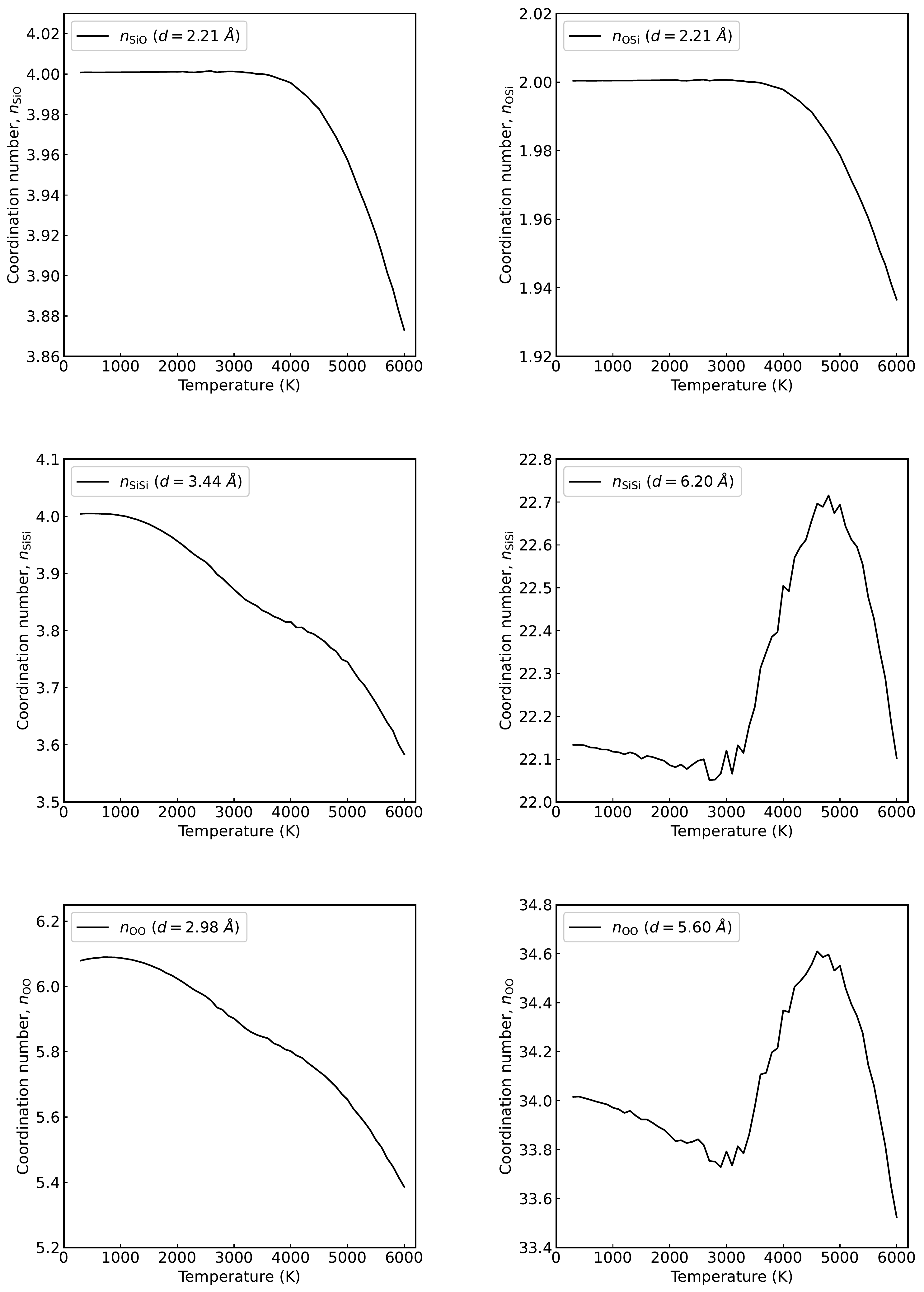}
  \caption{\red{Integrated coordination numbers $n_{i,j}(r)$ for the silica glass as a function of temperature. The cutoff values are written in the legends.}
  }
  \label{coord-num}
\end{figure*}
\begin{figure*}[htbp]
  \centering
  \includegraphics[width=\columnwidth]{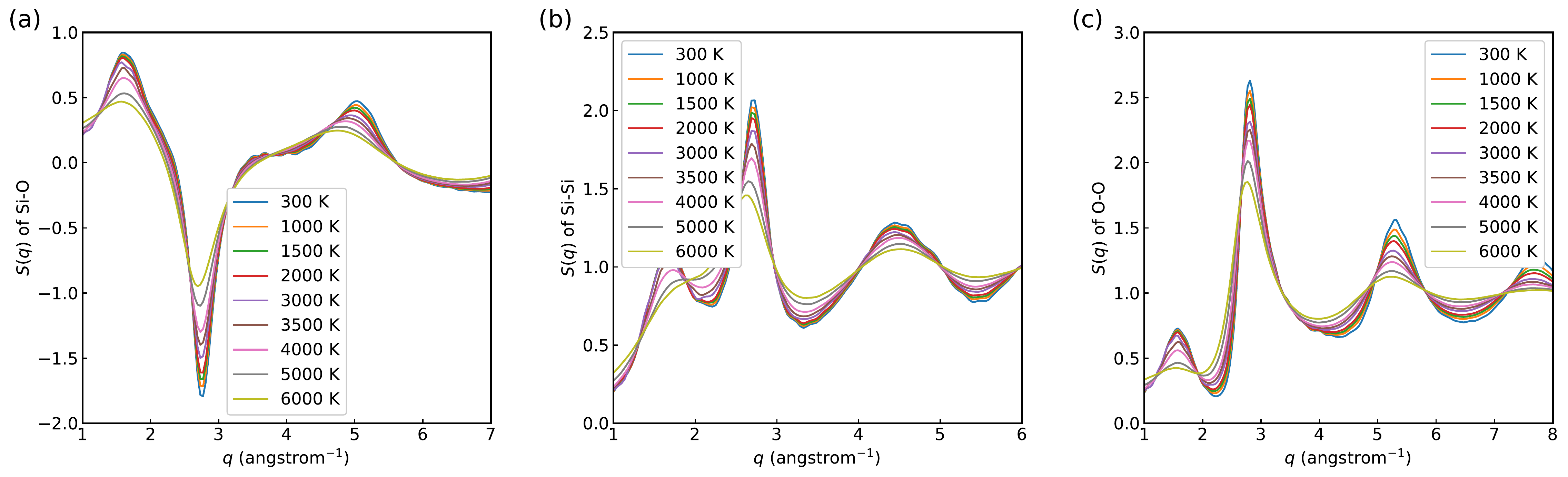}
  \caption{
  Averaged $S(q)$ for (a) Si-O, (b) Si-Si, and (c) O-O.
  }
  \label{sq}
\end{figure*}
\begin{figure*}[htbp]
  \centering
  \includegraphics[width=\columnwidth]{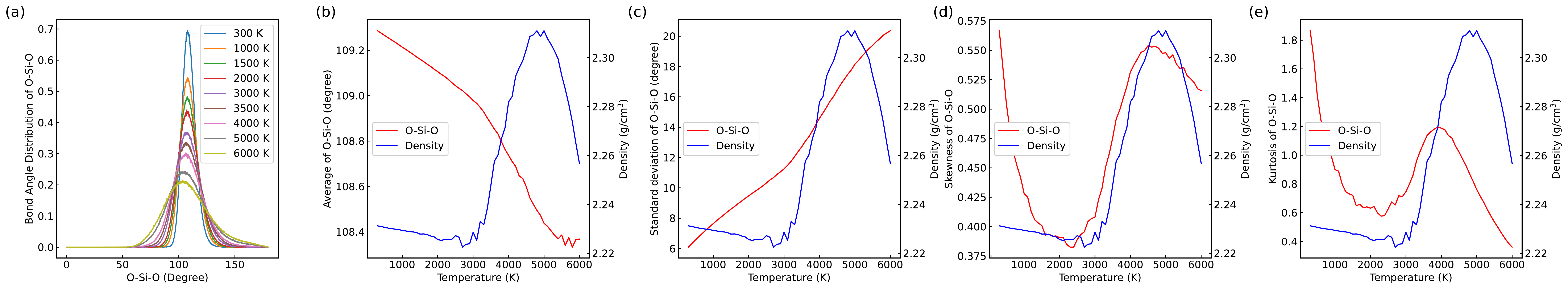}
  \caption{
  Statistics of the O-Si-O bond-angle distributions. (a) Distribution, (b) mean, (c) standard deviation, (d) skewness, and (e) kurtosis.
  }
  \label{bad-O-Si-O}
\end{figure*}
\begin{figure*}[htbp]
  \centering
  \includegraphics[width=\columnwidth]{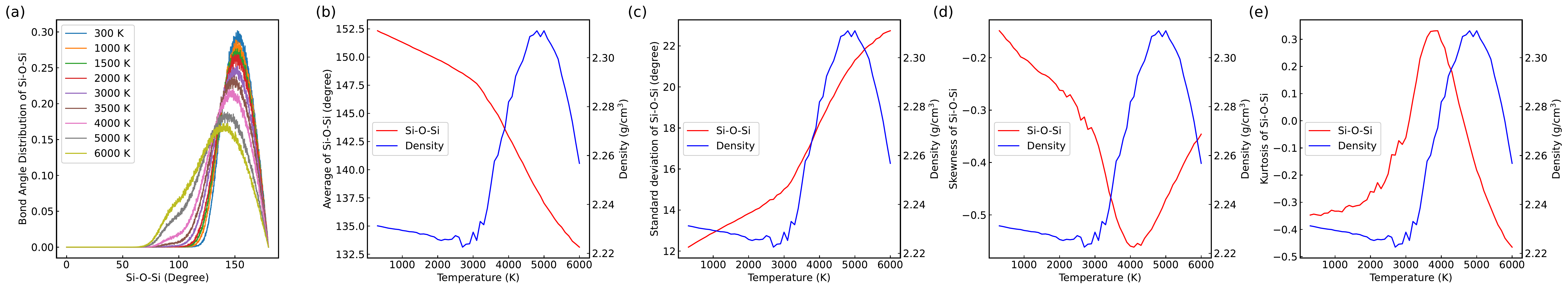}
  \caption{
  Statistics of the Si-O-Si bond-angle distributions. (a) Distribution, (b) mean, (c) standard deviation, (d) skewness, and (e) kurtosis.
  }
  \label{bad-Si-O-Si}
\end{figure*}
\begin{figure*}[htbp]
  \centering
  \includegraphics[width=\columnwidth]{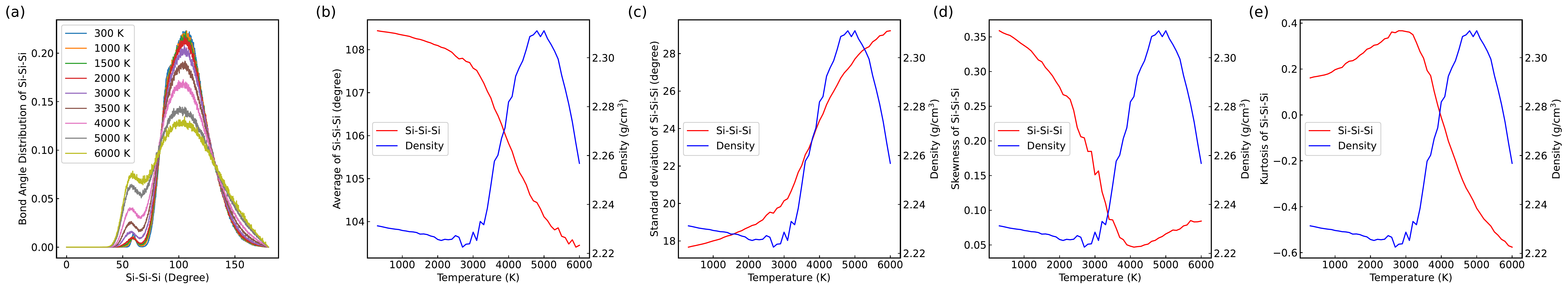}
  \caption{
  Statistics of the Si-Si-Si bond-angle distributions. (a) Distribution, (b) mean, (c) standard deviation, (d) skewness, and (e) kurtosis.
  }
  \label{bad-Si-Si-Si}
\end{figure*}
\begin{figure*}[htbp]
  \centering
  \includegraphics[width=\columnwidth]{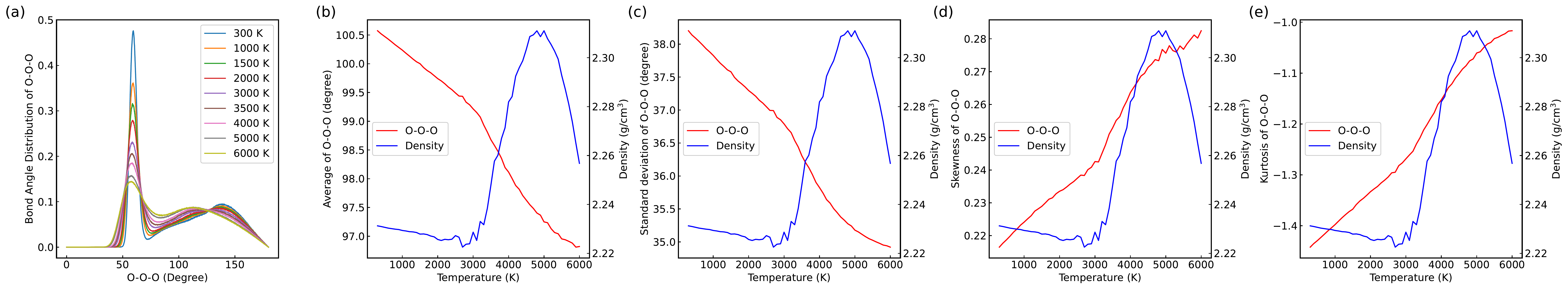}
  \caption{
  Statistics of the O-O-O bond angle distributions. (a) Distribution, (b) mean, (c) standard deviation, (d) skewness, and (e) kurtosis.
  }
  \label{bad-O-O-O}
\end{figure*}
\begin{figure*}[htbp]
  \centering
  \includegraphics[width=\columnwidth]{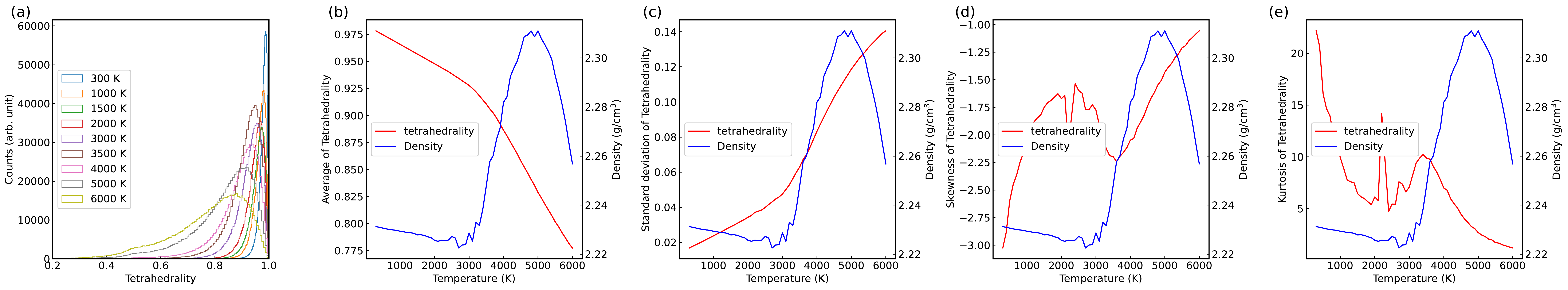}
  \caption{
  Statistics of the tetrahedral ordering parameter for SiO$_4$. (a) Distribution, (b) mean, (c) standard deviation, (d) skewness, and (e) kurtosis.
  }
  \label{tetra-SiO4}
\end{figure*}
\begin{figure*}[htbp]
  \centering
  \includegraphics[width=\columnwidth]{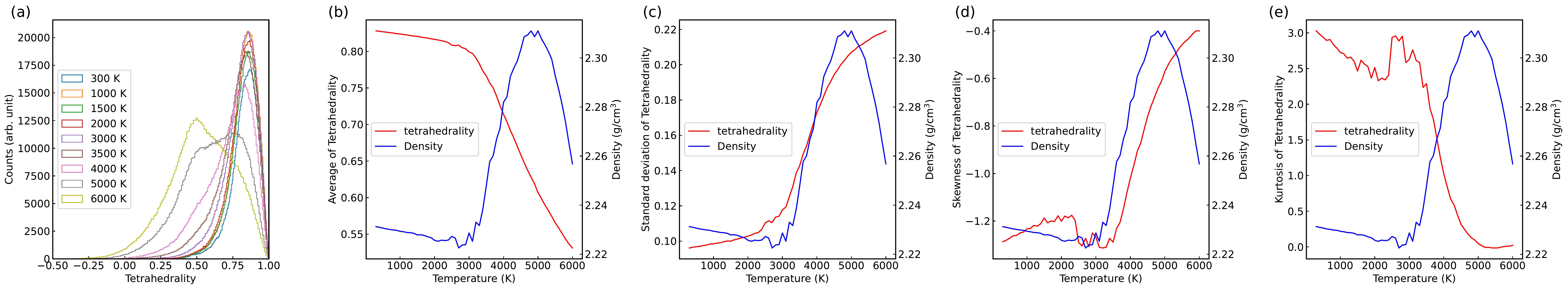}
  \caption{
  Statistics of the tetrahedral ordering parameter for SiSi$_4$. (a) Distribution, (b) mean, (c) standard deviation, (d) skewness, and (e) kurtosis.
  }
  \label{tetra-SiSi4}
\end{figure*}
\begin{figure*}[htbp]
  \centering
  \includegraphics[width=\columnwidth]{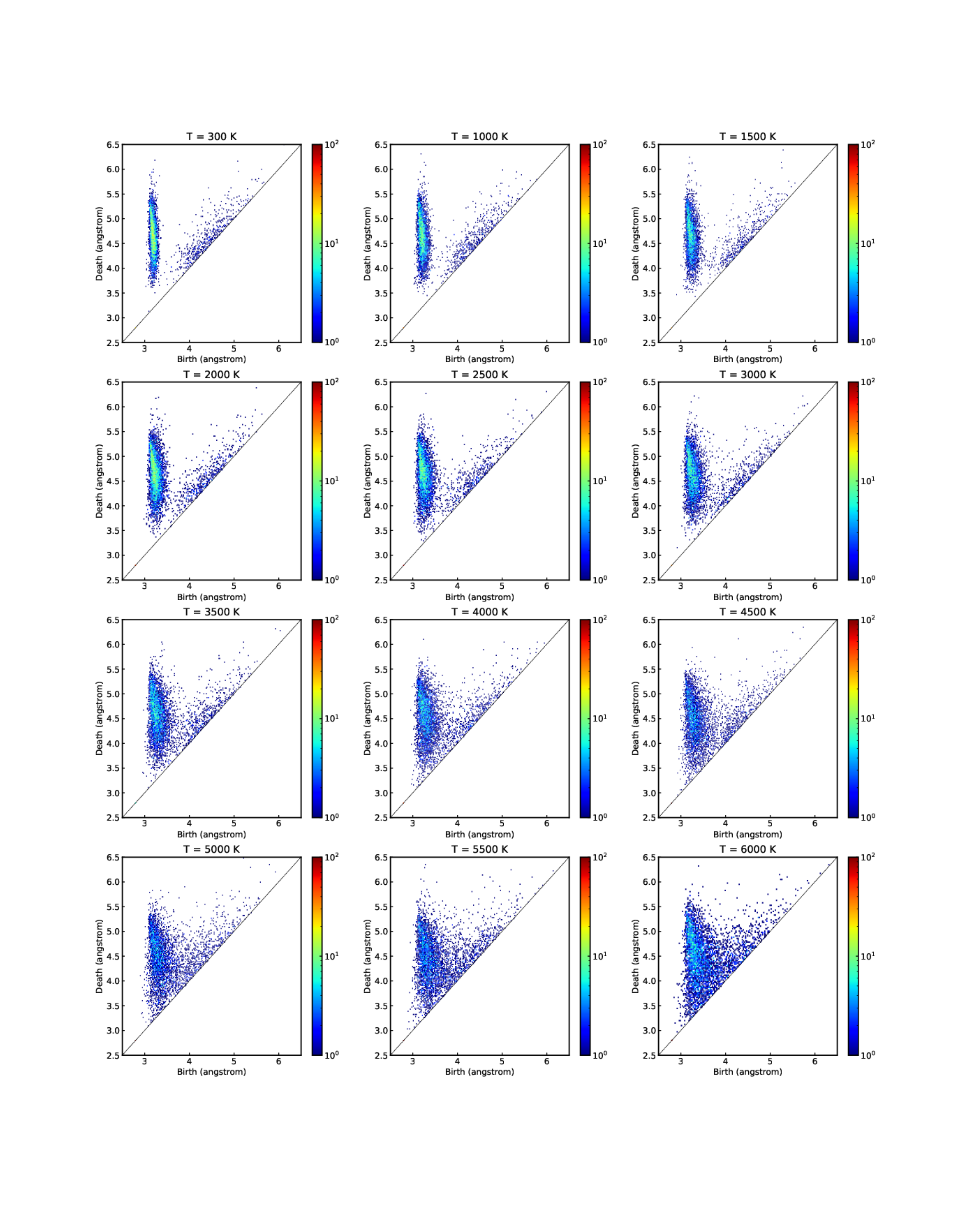}
  \caption{
  One-dimensional persistence diagrams of -Si-Si- network for each temperature.
  }
  \label{PD-Si}
\end{figure*}
\begin{figure*}[htbp]
  \centering
  \includegraphics[width=\columnwidth]{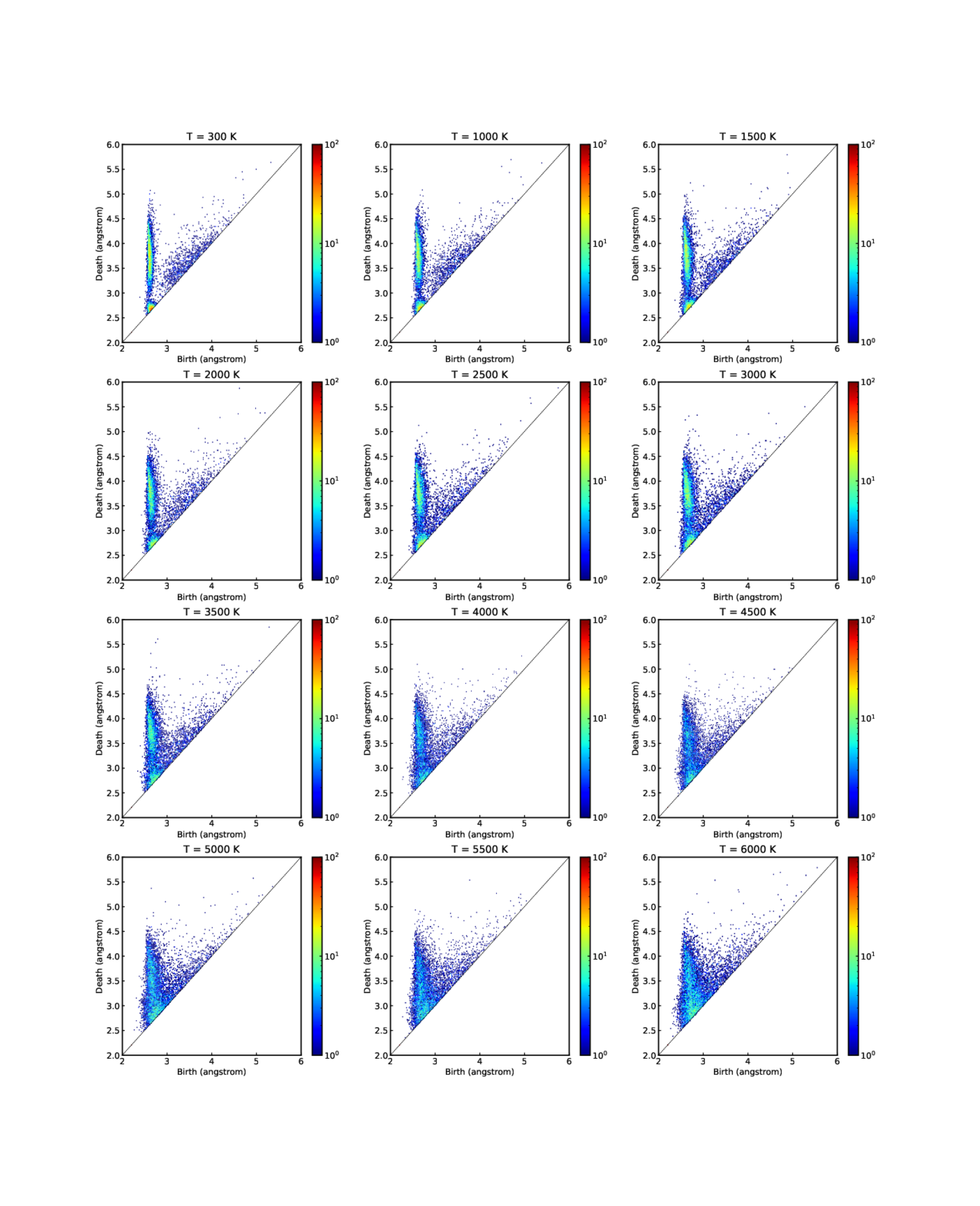}
  \caption{
  One-dimensional persistence diagrams of -O-O- network for each temperature.
  }
  \label{PD-O}
\end{figure*}
\begin{figure*}[htbp]
  \centering
  \includegraphics[width=\columnwidth]{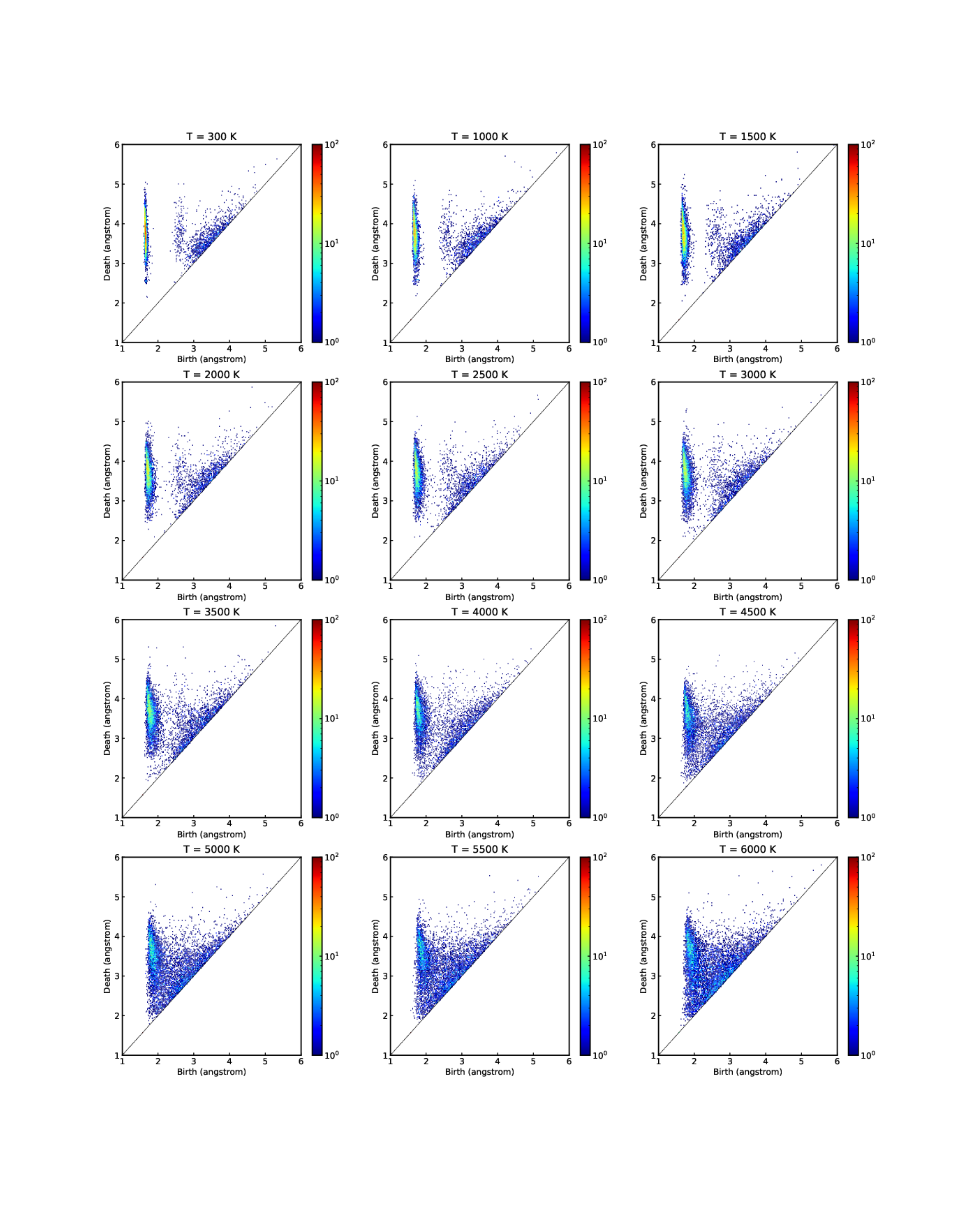}
  \caption{
  One-dimensional persistence diagrams of -Si-O- network for each temperature.
  }
  \label{PD-SiO}
\end{figure*}

This file includes the supplementary information for the paper titled ``Topological data analysis for revealing structural origin of density anomalies in silica glass."
Fig.~{\ref{temperature}} shows the temperature profile during the cooling process in the MD simulations. The glass was heated to 6100~K at a rate of 10~K/ps and then melted at 6100~K for 1.0~ns. The glass was cooled from 6100~K to 300~K at a rate of 1~K/ps, where it was equilibrated for 300 ps every 100~K. Further details about the MD simulations are provided in the main text. The glass structures were recorded every 1~ps during the cooling process, and the thermodynamic properties were recorded every 0.01 ps. Only the data from the last 150~ps were used to average the properties, such as $g(r)$, $S(q)$, bond-angle distribution, and tetrahedrality. PDs were computed using only the last configuration for each temperature, and all the other properties were averaged over the recorded values.
Fig.~{\ref{gr}} shows the averaged radial distribution functions, $g(r)$, of Si-O, Si-Si, and O-O.
Fig.~{\ref{gr-min}} shows the averaged first minimum, $r$, of $g(r)$ for Si-O, Si-Si, and O-O. The minimum values were obtained using R.I.N.G.S code~{\cite{2010ROU}}.
\red{Fig.~{\ref{population-sio2}} shows populations of Si and O atoms with different coordination numbers in the silica as a function of temperature, where the cutoff distance 2.21~$\AA$ was used. Fig.~{\ref{coord-num}} shows integrated coordination numbers of Si-O (1st NN), O-Si (1st NN), Si-Si (1st NN), Si-Si (2nd NN), O-O (1st NN), O-O (2nd NN) in the silica glass as a function of temperature. The employed cutoff values are written in the legends.}
Fig.~{\ref{sq}} depicts the averaged structure factors, $S(q)$, of Si-O, Si-Si, and O-O.
Figs.~{\ref{bad-O-Si-O}}, ~{\ref{bad-Si-O-Si}}, ~{\ref{bad-Si-Si-Si}}, and ~{\ref{bad-O-O-O}} show the bond angle distributions of O-Si-O, Si-O-Si, Si-Si-Si, and O-O-O, respectively.
Figs.~{\ref{tetra-SiO4}} and ~{\ref{tetra-SiSi4}} depict the statistics of the tetrahedral ordering parameter of SiO$_{4}$ and SiSi$_{4}$ tetrahedrons, respectively. The tetrahedral ordering parameter, $q_{\rm tet}$, was computed using the following equation:
\begin{equation*}
q_{\rm tet} = 1 - \frac{3}{8}\sum_{j=1}^{3}\sum_{k=j+1}^{4}(\cos{\theta_{j,k}}+\frac{1}{3}),
\end{equation*}
where $\theta_{j,k}$ is the angle between the four closest tetrahedral atoms and $j$ and $k$ are the indices of these atoms~{\cite{1998CHA, 2001JEF}}. Perfect tetrahedrality corresponds to $q_{\rm tet} = 1$, and the possible range of $q_{\rm tet}$ for a molecule is $-3 \le q_{\rm tet} \le 1$~{\cite{2001JEF}}.
%
Figs.~{\ref{PD-Si}}, {\ref{PD-O}}, and {\ref{PD-SiO}} show the one-dimensional PDs for the -Si-Si-, -O-O-, and -Si-O- networks, respectively.


\clearpage

\bibliography{references}